\documentclass[prd,aps,eqsecnum,floatfix,nofootinbib,preprint,tightenlines]{revtex4}
\usepackage{latexsym}
\usepackage{graphicx}
\usepackage{hyperref}
\def\bibi{\bibitem}

\def\a{\alpha}
\def\b{\beta}
\def\c{\chi}
\def\d{\delta}
\def\e{\epsilon}                % Also, \varepsilon
\def\g{\gamma}
\def\h{\eta}

\def\k{\kappa}

\def\m{\mu}
\def\n{\nu}

\def\p{\pi}                     % Also, \varpi
\def\r{\rho}                    %       \varrho
\def\s{\sigma}                  %       \varsigma
\def\t{\tau}

\def\x{\xi}
\def\z{\zeta}
\def\D{\Delta}

\def\G{\Gamma}

\def\L{\Lambda}
\def\O{\Omega}

\def\S{\Sigma}

\def\X{\Xi}

\def\ca{{\cal A}}

\def\cl{{\cal L}}
\def\cm{{\cal M}}
\def\cn{{\cal N}}
\def\co{{\cal O}}

\def\cu{{\cal U}}
\def\cv{{\cal V}}

\def\bo{\raisebox{-.4ex}{\large$\Box$}}                 % D'Alembertian
\def\cbo{{\,\raise-.15ex\Sc [\,}}                       % curly "
\def\ltap{\raisebox{-.4ex}{\rlap{$\sim$}} \raisebox{.4ex}{$<$}}   % < or ~
\def\Sl#1{\rlap{\hbox{$\mskip 3 mu /$}}#1}      % " upper
\def\ddt#1{{\buildrel {\hbox{\LARGE .\kern-2pt.}} \over {#1}}}% double dot-over
\def\ie{\mbox{\it i.e.} }

\def\etc{\mbox{etc.} }

\def\geqx{\,\raisebox{-1.0ex}{$\stackrel{\textstyle >}{\sim}$}\,}
\def\tr{{\rm tr}\,}

\def\det{{\rm det}}
\def\Det{{\rm Det}}

\def\floatcaption#1#2{ \caption{ #2 \label{#1}} }

\def\etc{{\it etc.}}
\def\seef{{\it cf.}}

\def\textit#1{{\it #1}\kern.1em }

\def\bq{\overline{q}}
\def\bu{\overline{u}}
\def\bd{\overline{d}}
\def\bs{\overline{s}}
\def\bK{\overline{K}}
\def\bchi{\overline{\c}}
\def\bM{\overline{M}}
\def\bQ{\overline{Q}}
\def\be{\overline{\eta}}
\def\bx{\overline{x}}
\def\bb{\overline{b}}
\def\bc{\overline{c}}
\def\bdelta{\overline{\d}}

\def\hM{\hat{M}}
\def\hm{\hat{m}}
\def\ha{\hat{a}}

\def\hA{\hat{A}}

\def\tm{\tilde{m}}
\def\tq{\tilde{q}}
\def\tphi{\tilde{\phi}}
\def\tX{\tilde{X}}
\def\tY{\tilde{Y}}
\def\tS{\tilde{\S}}

\def\tr{{\rm{tr}}}
\def\str{{\rm{str}}}
\def\exp{{\rm{exp}}}
\def\sdet{{\rm{sdet}}}

\DeclareSymbolFont{lettersA}{U}{txmia}{m}{it}
 \DeclareMathSymbol{\real}{\mathord}{lettersA}{"92}
 \DeclareMathSymbol{\field}{\mathord}{lettersA}{"83}

\begin{document}

\thispagestyle{empty}

\begin{center}
\vspace*{10mm}
{\large\bf Applications of chiral perturbation theory to lattice QCD}
\\[12mm]
Maarten Golterman
\\[8mm]
{\small\it
Department of Physics and Astronomy
\\San Francisco State University,
San Francisco, CA 94132, USA}
\\[10mm]
{ABSTRACT}
\\[2mm]
\end{center}
\begin{quotation}
These notes contain the written version of lectures given at the
2009 Les Houches Summer School ``Modern perspectives in lattice QCD:
Quantum field theory and high performance computing.''
The goal is to provide a pedagogical introduction to the subject, and not a
comprehensive review.  Topics covered include a general introduction,
the inclusion of scaling violations in chiral perturbation theory, partial quenching
and mixed actions, chiral perturbation theory with heavy kaons, and the
effects of finite volume, both in the $p$- and $\e$-regimes.
\end{quotation}

%%####%%
\newpage
\section{\label{intro} Why chiral perturbation theory for lattice QCD?}
%%####%%

It is often claimed that lattice QCD provides a tool for computing hadronic
quantities numerically, with fully controlled systematic errors,
from first principles.
This assertion is of course based on the fact that lattice QCD provides
a nonperturbative definition of QCD (in fact, the only one to date).
So, why not choose the parameters (in particular, the quark masses) at, or
very near, their
physical values, and compute all quantities of interest?

There are at least two major obstacles to this.  The first obstacle is that
lattice QCD is formulated in euclidean space, rather than in Minkowski
space. This is a necessary restriction if one wants to use Monte Carlo
methods in order to evaluate expectation values of operators from which one
extracts physical quantities.  In euclidean space
correlation functions do not give direct access to physical
scattering amplitudes; they first need to be continued to Minkowski space.
If one is interested in physics at some scale for which an effective field theory
(EFT) is available, one can use this EFT in order to match to euclidean
lattice correlation functions in the regime of validity of the EFT.
In the case of the chiral EFT for QCD, the form of the correlation
functions is predicted in terms of a finite number  of coupling
constants to a given finite order in a momentum expansion.
Once the value of these is known from lattice QCD
computations, the EFT can then be used to continue
to Minkowski space.  Chiral perturbation theory (ChPT) provides this EFT framework
for the low-energy physics of the (pseudo-) Nambu--Goldstone bosons of QCD.

A second obstacle is that in practice lattice QCD computations are carried
out at values of the up and down quark mass  larger than those
observed in nature,\footnote{The strange quark mass can be taken at its
physical value, although it is theoretically interesting to see what happens
if one varies the strange quark mass as well.} because of limits on the size
of the physical volumes
that can be handled with presently available computers, and because of
the rapid increase in algorithmic cost with decreasing quark masses.  If the quark
masses are nevertheless still small enough for ChPT to be applicable,
it can be used to extrapolate to the physical values for these light quark
masses, again using ChPT to connect an ``unphysical'' lattice computation
(namely, at values of the light quark masses larger than their real-world
values) to physical quantities (those
observed in nature).

It should be said that lattice QCD is moving toward the physical point, \ie,
that numerical computations are being done with the light (up and down)
quark masses at or very near the physical values.  If lattice quark masses
are very near the physical values, simple smooth extrapolations are in
principle enough to obtain hadronic quantities of interest at the physical
point, and the use of ChPT may become less important in this respect.
I will return to this point below.

Both of the obstacles described above are examples of how ChPT (and, in general,
EFTs) can be used to connect ``unphysical'' computations with physical
quantities.  This is particularly helpful in the case of lattice QCD,
where often computations are easier, or even only possible, in some
unphysical regime.   The use of ChPT is by no means limited to extrapolations
in quark masses, or continuation from euclidean space.
Other examples of the use of ChPT to bridge
the gap between the lattice and the real world are, in increasing
order of boldness:

\begin{itemize}
\item[1.] Nonzero lattice spacing.  While in state-of-the-art
lattice computations the lattice spacing is small (in units of
$\L_{QCD}$), scaling violations are still significant.  It is possible to
extend ChPT to parametrize scaling violations afflicting the physics
of Nambu--Goldstone bosons.  In particular, depending on what lattice
fermions one uses, chiral symmetry (and rotational invariance!) are
broken on the lattice, and such breakings can be parametrized in
ChPT through new coupling constants that vanish in the continuum limit.
This will be the topic of the second lecture, Sec.~\ref{lattice spacing}.
Another example in this category is the use of a finite volume
in numerical computations.  The dominant effects of this can also be
studied in ChPT, as we will see in Sec.~\ref{finite volume}.
\item[2.] Partial quenching and the use of mixed actions.  As we will
see, on the lattice one is free to choose the valence quark masses to be
different from sea quark masses.  This is of course completely
unphysical (it violates unitarity!), but it turns out to be very
practical -- it is (often much) cheaper to vary only valence quark masses
than to vary the sea (or ``dynamical'') quark mass, which affects
the ensemble of gauge configurations.  Moreover, as we will see, varying
valence and sea quark masses independently from each other
gives us extra handles on
the theory, making it easier to determine the couplings of ChPT
than it would be without this possibility.
This trick is known as
``partial quenching.''\footnote{``Quenched QCD'' corresponds to the
case in which the sea quark masses are taken to infinity, making the
fermion determinant constant.  This is of course unphysical
``beyond repair'': all effects of quark loops are irrevocably gone.}
One may even go further: one can use different discretizations
for sea quarks (typically a computationally less expensive method,
such as staggered or Wilson quarks) and for valence quarks
(a discretization using lattice fermions with very good chiral
and flavor symmetries --- operator mixing depends primarily on the
symmetries of those operators, which are constructed out of the
valence quarks).  This generalization of partially-quenched QCD
(PQQCD) is now often referred to as ``mixed-action'' QCD.  We will discuss
these methods in the third lecture, Sec.~\ref{PQmixed}.
\item[3.]
Other unphysical constructions used in lattice QCD; a prominent example
is the use of ``fourth-rooted'' staggered fermions.  Again, EFT
techniques provide much inside into the nature of this method,
and the comparison of EFT calculations with numerical results
can be used to test the validity of such methods.
\end{itemize}

Even if one is only interested in physical quantities which can directly
be computed on the lattice, such as (stable) hadron masses at the
physical values of the quark mass for a very small lattice spacing,
ChPT may still be used to test the results.  By varying the quark
mass away from the physical value, one might discover that the
numerical results do not match the predictions of continuum ChPT, for instance because
the lattice spacing is still too large, the volume too small,
or because of some other systematic effect.  In other words, ChPT provides
a useful tool for the validation of results obtained with lattice QCD.
%\ask{example of Oliver's lat2004 review?}

Extending this latter argument, it can also be very interesting to
use lattice QCD as a laboratory to explore what strong interaction
physics would look like at parameter values
different from those in the real world.  Again, interpreting the
results of numerical computations in lattice QCD through the help
of EFT techniques can be very useful in this respect.  An example
of this is the question about whether it is possible that the up-quark mass
$m_u$ could be zero.\footnote{Which would solve the strong CP problem.}
This question has now been answered with the help of lattice QCD
computations at {\em nonzero} $m_u$ (the answer is that $m_u=0$ is
now exluded with a very high confidence level).
This is only possible
with assistance of ChPT, which makes it possible to extrapolate to
what physics would have looked like at $m_u=0$, from computations
done at nonzero values of $m_u$.

Finally, with limited computational resources, there is the important
question as to how to use them most judiciously.  Does one use all
resources to do a few, very expensive, computations at physical
quark masses, in very large volume?  While this sounds very attractive,
restrictions on available computational resources
may force one to choose only
larger lattice spacings, and one thus pays with the increase of
one systematic error for the removal of other ones.  Or does one
try to strike a balance between the various systematic errors which
afflict lattice computations, by using larger-than-physical quark
masses, and less expensive fermions with symmetry properties that are
not as nice?  What choices to make depends very much on the physics one is
interested in, and they are to some extent subjective.  However,
ChPT and its extension to unphysical situations helps us with making
such choices, and it therefore is, and will likely remain, a very important tool for lattice
QCD practitioners.

The aim of these lectures is not to give a review of all the work that has
been done in ChPT and its applications to lattice QCD.
Rather, the aim is to provide a basic introduction to ChPT, with emphasis
on its applications to lattice QCD, geared at readers with a general knowledge
of quantum field theory and the basics of lattice QCD.  The list of references
included is heavily biased toward the papers from which I learned the subject.
It will not  be possible
to refer to all papers that have been written on this and closely related topics
and applications;
there are probably
at least a few hundred {\em more} than the references I did include.  The references
I did include are primarily those in which new concepts or methods are introduced
and explained, and I strongly recommend them for delving deeper into the many
aspects and applications of ChPT in its relation to QCD on the lattice.

It is not even possible to cover all the relevant and important
applications that have been
considered in the literature.  Notable examples of topics left out are applications
to baryons (including EFTs for two-nucleon systems), and to hadrons containing
heavy quarks,
although some aspects of these topics
are shared by ``heavy-kaon'' ChPT, which we
consider in Sec.~\ref{kaon}.
The topics I have chosen to include are a general introduction to ChPT in
Sec.~\ref{chpt}, the incorporation of scaling violations in ChPT in Sec.~\ref{lattice spacing},
partially quenched and mixed-action ChPT in Sec.~\ref{PQmixed}, two-flavor ChPT
with a heavy kaon in Sec.~\ref{kaon}, and ChPT in finite volume in Sec.~\ref{finite volume}.

Good reviews of continuum chiral perturbation theory can for instance be found in
Ref.~\cite{HLreview}, and, at the level of a textbook, Refs.~\cite{DGH,Georgi}.
For a more general introduction to EFT methods, including
ChPT, see Ref.~\cite{DBK}.  For applications of EFT methods to lattice QCD,
see Ref.~\cite{ASK}; for a more detailed review of ChPT in the context of lattice QCD,
see Ref.~\cite{SRSNara}.  Another lattice-oriented introduction, with applications of ChPT to lattice
calculations of weak matrix elements, is Ref.~\cite{CBtasi}.
For a recent review of applications of ChPT to hadron
phenomenology, see Ref.~\cite{Ecker}.

%%####%%
%\newpage
\section{\label{chpt} Continuum chiral perturbation theory}
%%####%%

We begin with a review of ChPT in the continuum, in infinite volume.  This is a vast subject
in itself, with many important applications to phenomenology.  I will
not review any of those here; the idea is to get an understanding of the
basics, which we will need for applications to lattice QCD.

%%####%%
\subsection{\label{chsym} Chiral symmetry}
%%####%%
Let us consider QCD with $N_f=3$ flavors, the up, down and strange quarks.
(We will also have reason to consider the $N_f=2$ theory, with only up and down
quarks, in situations where the strange quark can be considered heavy.)
Of course, heavier quarks exist, but we will be interested in physics
at energy scales well below the charm mass, so we may consider the theory
in which the charm, bottom and top quarks have been decoupled by integrating them
out.
In deference to the lattice, we will work in euclidean space
throughout these lectures, continuing back to Minkowski space when necessary.
The fermion part of the QCD lagrangian is
\begin{equation}
\label{QCD}
\cl=\bq_L\Sl{D}q_L+\bq_R\Sl{D}q_R+\bq_L Mq_R+\bq_R M^\dagger q_L\ ,
\end{equation}
in which $q_i$, $i=u,d,s$ is the three-flavor quark field.  The subscripts
$L$ and $R$ denote left- and right-handed projectors:
\begin{eqnarray}
\label{chproj}
q_L&=&\frac{1}{2}(1-\g_5)\,q\ ,\ \ \ \ \ q_R=\frac{1}{2}(1+\g_5)\,q\ ,\\
\bq_L&=&\bq\,\frac{1}{2}(1+\g_5)\ ,\ \ \ \ \ \bq_R=\bq\,\frac{1}{2}(1-\g_5)\ .
\nonumber
\end{eqnarray}
Note that we may define $\bq_{L,R}$ as we like, since the Grassmann fields $q$ and
$\bq$ are independent in euclidean space; we have chosen them to be consistent
with $\bq\to q^\dagger\g^0$ in the operator formalism.  $M$ is the quark mass matrix,
equal to
\begin{equation}
\label{M}
M=\pmatrix{m_u&0&0\cr 0&m_d&0\cr 0&0&m_s}\ .
\end{equation}
For massless QCD, with $M=0$, the lagrangian is invariant under the symmetry
group $U(3)_L\times U(3)_R$:
\begin{eqnarray}
\label{chsymm}
q_L&\to& U_Lq_L\ ,\ \ \ \ \ q_R\to U_Rq_R\ ,\\
\bq_L&\to&\bq_LU_L^\dagger\ , \ \ \ \ \ \bq_R\to\bq_RU_R^\dagger\ ,\nonumber
\end{eqnarray}
with $U_{L,R}\in U(3)_{L,R}$.  We observe that the full QCD lagrangian  is invariant
if we also let $M$ transform as
\begin{equation}
\label{spurion}
M\to U_LMU_R^\dagger\ .
\end{equation}
Of course, the quark masses do not transform, but this ``spurion''
trick, of letting some
parameters transform under the symmetry group, will be useful later,
for instance in Sec.~\ref{masses}.  Note that in euclidean space, $\cl$ (with
Eq.~(\ref{spurion})) is invariant under a larger group, $GL(3,\field)_L\times
GL(3,\field)_R$.  (If we consider this larger group, $U^\dagger_L$ and $U^\dagger_R$
in Eqs.~(\ref{chsymm}) and~(\ref{spurion}) have to be replaced by $U_L^{-1}$ and
$U_R^{-1}$, respectively.)  But if we identify
$\bq$ with $q^\dagger\g^0$, this reduces this larger group to $U(3)_L\times U(3)_R$,
from which it can be shown that the larger symmetry group in euclidean space has
no additional physical consequences.

It is well known, of course, that ``axial $U(1)$,'' \ie, transformations
with $U_L=U_R^\dagger=\exp{(i\theta)}{\bf 1}$, are not preserved when QCD is
quantized, even when $M=0$.  The actual symmetry group is therefore
$SU(3)_L\times SU(3)_R\times U(1)$, where the $U(1)$ factor is quark
number.  On mesons, the topic of these lectures, quark-number $U(1)$ is
trivially realized, and we may drop it from consideration.

In the real world, we find that there exists an octet of pseudoscalar
mesons with masses smaller than any other hadron masses.  Moreover, one
observes that relations between their masses and interactions are
reasonably well described by assuming an approximate $SU(3)$ flavor
symmetry.\footnote{We will return to the issue that the strange quark mass
is much larger than the up and down quark masses, see Sec.~\ref{kaon}.}
Under this $SU(3)$, the pseudoscalar mesons form an octet
\begin{equation}
\label{phi}
\phi=\pmatrix{\frac{\p_0}{\sqrt{2}}+\frac{\eta}{\sqrt{6}}&\p^+&K^+\cr
\p^-&-\frac{\p_0}{\sqrt{2}}+\frac{\eta}{\sqrt{6}}&K^0\cr
K^-&\bK^0&-\frac{2\eta}{\sqrt{6}}}\sim
\pmatrix{u\bu&u\bd&u\bs\cr d\bu&d\bd&d\bs\cr s\bu&s\bd&s\bs}\ ,
\end{equation}
where we also indicated the (valence) quark content.\footnote{We ignore
isospin breaking, which causes the fields $\p^0$ and $\eta$
of Eq.~(\ref{phi}) to mix, in most of these lectures.}
(In the theory with only two light flavors, $\phi$ reduces
to the two-by-two upper left-hand block, with the $\eta$ omitted.)
Under $SU(3)$ this
octet transforms as
\begin{equation}
\label{vsu3}
\phi\to U\phi U^\dagger\ ,
\end{equation}
and it is clear that this $SU(3)$ can be identified with the diagonal
subgroup $SU(3)_V$ of $SU(3)_L\times SU(3)_R$, for which $U_L=U_R=U$
in Eq.~(\ref{chsymm}).

If the full symmetry group $SU(3)_L\times SU(3)_R$ were realized
``manifestly'' in nature, we would observe larger hadronic multiplets,
and in particular, we would observe ``parity partners,'' \ie, pairs of hadrons with
the same mass but opposite parity, since parity
takes left-handed quarks into right-handed ones and {\it vice versa}.%
\footnote{In mathematical terms, parity is an automorphism of the group $SU(3)_L\times SU(3)_R$.}
Such parity partners are in general not observed, and specifically, there
are no scalar mesons with masses near those of the pseudoscalar multiplet
of Eq.~(\ref{phi}).  Instead, it is universally believed (with very strong
evidence from both the real world and lattice computations) that, in
a (hypothetical!) world with massless quarks, the chiral group is
spontaneously broken to the diagonal subgroup, $SU(3)_L\times SU(3)_R
\to SU(3)_V$.  This requires eight Nambu--Goldstone bosons (NGBs), which are identified
with the pions, eta and kaons of Eq.~(\ref{phi}).  Since the broken
generators distinguish between left- and right-handed quarks, these NGBs
have to be pseudoscalars.  In the real
world, quarks are not massless, but if the quark masses $m_{u,d,s}$ are
small compared to $\L_{QCD}$, they can be
treated as a perturbation (as we will see below), giving the NGBs a small
mass.  Indeed, the pions have a very small mass of about 140~MeV,
while the kaons and eta have a mass of about 500~MeV, which is at least
smallish compared to other hadron masses.  For a proof that $SU(3)_V$
does not undergo spontaneous symmetry breaking in continuum QCD, see Ref.~\cite{VW}.

Our interest in these lectures will be in the physics of these NGBs, at
energies below those at which any other hadrons can be produced.
We therefore expect that it should be possible to find an EFT for the
physics of NGBs.  What that means is that it should be possible to write
down a local lagrangian in terms of the field $\phi$ that, in some
systematic
approximation, reproduces correlation functions involving only NGBs,
with restrictions imposed by the principles of quantum field theory.
The chiral lagrangian is precisely this EFT.

%%####%%
\subsection{\label{lagr} The chiral lagrangian}
%%####%%
Let us therefore start with a closer look at spontaneous symmetry breaking
of chiral symmetry, for the case that $M=0$.  An order parameter for this
breaking is the renormalized condensate
\begin{equation}
\label{condensate}
\langle\bq_{Ri} q_{Lj}\rangle=\langle\bq_{Li} q_{Rj}\rangle\propto\L_{QCD}^3
\d_{ij}\ ,
\end{equation}
where the power of $\L_{QCD}$ estimates the magnitude of the condensate.
As is always the case with spontaneous symmetry breaking,
there is a manifold of equivalent
vacua, and indeed, one may rotate this condensate as
\begin{equation}
\label{gencond}
\O_{ij}=\langle q_{Li}\bq_{Rj}\rangle\to U_L\O U_R^\dagger\ ,
\end{equation}
where I indicated how $\O$ transforms under the chiral group
(here I sum over spin and color indices, so that $\O$ is a color and spin
singlet).
These vacua are all equivalent, and they are rotated into each other by
elements of the coset $SU(3)_L\times SU(3)_R/SU(3)_V$, which happens to be isomorphic
to the group $SU(3)$.  The standard choice of Eq.~(\ref{condensate})
leaves the diagonal subgroup $SU(3)_V$ invariant, but this does not mean that with
another choice of condensate there would be no $SU(3)$ invariance: the
unbroken $SU(3)$ would simply be differently embedded in $SU(3)_L\times SU(3)_R$.%
\footnote{If we choose some other value $\O$ for the condensate, the subgroup
leaving this invariant is the subgroup for which $U_L=\O U_R\O^{-1}$, which is
isomorphic to $SU(3)_V$.  Note that for a different value of $\Omega$ the
definition of parity would also need to be modified.}

The color-singlet
operators $\tr(\G q_{Li}\bq_{Rj})$ and their parity conjugates, when acting
on the vacuum, create mesons with flavor quantum numbers corresponding
to the flavor indices $i$ and $j$, and spin-parity corresponding to the
matrix $\G$, which is some product of Dirac gamma matrices.   Since we are
interested here in
scalar and pseudoscalar mesons, we choose $\G=1$.
Introducing a complex scalar field $H_{ij}$ with the same quantum numbers,
we can write down an effective lagrangian of the form
\begin{equation}
\label{eft}
\cl_{eff}=\tr(\partial_\m H^\dagger\partial_\m H)+\cl_{int}(H,H^\dagger)
+\cl_{other}(H,H^\dagger,{\rm other\ hadrons})\ .
\end{equation}
This lagrangian should obey the same symmmetries as the QCD lagrangian,
and in particular, it should be invariant under $SU(3)_L\times SU(3)_R$,
with $H\to U_LHU_R^\dagger$.  We can now decompose $H=R\S$ with $R$
hermitian and positive, and $\S$ unitary.  Symmetry breaking with the
pattern $SU(3)_L\times SU(3)_R\to SU(3)_V$ implies that $R$ picks up
an expectation value, which again we can take to be proportional to the
unit matrix, as in Eq.~(\ref{condensate}).  $\S$ parametrizes the vacuum
manifold, and thus describes the NGBs predicted by chiral symmetry breaking.
If all other hadrons are massive, and we are interested only in the
physics of NGBs below the typical hadronic scale, we can integrate out all
non-Goldstone fields, leaving us with a low-energy effective lagrangian
in terms of only $\S$.  This lagrangian is local at scales below the hadronic scale.
Because of the anomalous axial $U(1)$, the meson
associated with this symmetry is also heavy, and we may take
$\S$ in $SU(3)$ rather than $U(3)$, and parametrize it as
\begin{equation}
\label{sigma}
\S=\exp\left(2i\phi/f\right)\to U_L\S U_R^\dagger\ ,
\end{equation}
with $\phi$ the field of Eq.~(\ref{phi}). The field
$\S$ inherits its transformation under the chiral group from the chiral transformation of
$H$, as indicated above.  We inserted a dimensionful
parameter $f\propto\L_{QCD}$ so that the field $\phi$ has the canonical mass dimension
of a scalar field.  Observe that $\S$ precisely parametrizes the vacuum
manifold as in Eq.~(\ref{gencond}) for the choice $\O_{ij}=-\langle\bq q\rangle
\d_{ij}/(2N_f)$.  The unbroken group $SU(3)_V$ is linearly
realized: Eq.~(\ref{sigma}) implies indeed that $\phi$ transforms as in
Eq.~(\ref{vsu3}) under $SU(3)_V$.  This is not true for the full chiral
group: the field $\phi$ transforms nonlinearly under all symmetries
outside the subgroup $SU(3)_V$.  For more discussion, see Sec.~\ref{kaon}.

Our lagrangian thus simplifies to
\begin{equation}
\label{eft1}
\cl_{eff}=\frac{1}{8}f^2\tr(\partial_\m\S^\dagger\partial_\m\S)
+\cl_{int}(\S,\S^\dagger)
\end{equation}
where I have normalized the first term such that, upon expanding $\S$ in terms of $\phi$,
the meson fields of Eq.~(\ref{phi}) have properly normalized kinetic terms.
While this is progress,  we clearly need more input to turn this into any practical use.

The EFT does not have to be renormalizable.  In fact, one expects the
presence of a cutoff of order 1~GeV, because we integrated out all
other hadrons, with masses of order 1~GeV and up.  But we
do expect that pion\footnote{I will often use ``pion'' to refer
to all (pseudo-) Nambu--Goldstone bosons, kaons and eta included.} interactions
obey all fundamental properties of quantum field theory:
unitarity and causality, crossing symmetry, clustering and Lorentz invariance
\cite{SWphysica}, of course
all to the precision with which the EFT reproduces the physics predicted
by QCD.  We will assume\footnote{I am not aware of any general
proof from the underlying theory, QCD.} that all these
properties are satisfied if we take $\cl_{eff}$ to be the most general
local, Lorentz (or euclidean) invariant function of the field $\S$
and its derivatives \cite{SWphysica}.\footnote{A more constructive argument, in which
clustering plays a central role, was given in Ref.~\cite{HLannphys}.}
  Moreover, since the EFT should obey all symmetries
of the underlying theory,  it should be invariant under
$SU(3)_L\times SU(3)_R$.

Our task is therefore to construct local invariants out of the field $\S$.
Because of the transformation rule for $\S$, \seef\ Eq.~(\ref{sigma}),
such invariants can only be formed by alternating $\S$ with $\S^\dagger$,
and taking a trace of such a product.  One may also multiply such
traces together to form new invariants.  However, all such invariants
collapse to a constant, because $\S\S^\dagger={\bf 1}$!  The only way
out is to allow for derivatives on the fields, as in the first term
of Eq.~(\ref{eft1}).  Because of Lorentz invariance, we need at least two
such derivatives, and we can organize terms in our lagrangian into groups
of terms with the same number of derivatives.  The unique term with
just two derivatives is the one shown as the first term in Eq.~(\ref{eft1}).

This construction leads to several fundamental consequences.
First, since we need at least two derivatives in all terms,
pion interactions vanish when their momenta vanish, and thus they
become weak for small momenta (and masses, as we will see below).
This, in turn, implies that we may organize our EFT in terms of
a derivative expansion: pion correlation functions with small
momenta on the external legs can be expanded in terms of a small parameter
$p/\L_\c$, where $p$ is a typical pion momentum, and $\L_\c$ is the
typical hadronic scale, of order 1~GeV.\footnote{This assumes that there
are no other ``accidentally'' massless (or very light) hadrons other than
the pions.}

These observations give us a very powerful method for constructing the
chiral lagrangian.  If our aim is to work only to leading order in the
derivative expansion, the chiral lagrangian is simply given by the first
term in Eq.~(\ref{eft1}):
\begin{equation}
\label{l2}
\cl^{(2)}=\frac{1}{8}f^2\tr(\partial_\m\S^\dagger\partial_\m\S)\ .
\end{equation}
Note that this is an interacting theory, because $\S$ is nonlinear
in the pion fields $\phi$.  It therefore describes all pion physics
to leading order in pion momenta.  For example, writing $\phi$ as $\phi=\phi_a T_a$,
with $T_a$ the generators of $SU(3)$ obeying
\begin{equation}
\label{norml}
\tr(T_a T_b)=\d_{ab}\ ,\ \ \ \ \ [T_a,T_b]=\sqrt{2}if_{abc}T_c\ ,
\end{equation}
one finds the leading-order prediction for the pion
scattering amplitude from a tree-level calculation using Eq.~(\ref{l2}):
\begin{equation}
\label{piscat}
\ca(ab\to cd)=\frac{1}{3f^2}\left\{(t-u)f_{abe}f_{cde}+(s-u)f_{ace}f_{bde}
+(s-t)f_{ade}f_{bce}\right\}\ ,
\end{equation}
in which $s$, $t$ and $u$ are the Mandelstam variables.  This is our first
prediction from ChPT.

Beyond leading order, we need to worry about two different (but, as we will
see, intricately related) issues.  We need to construct the most
general term of order $p^4$, \ie, with four derivatives, and we
also need to worry about unitarity (up to a given order).  A nice way
of building terms with higher derivatives starts from the observation that
we can define an object with one derivative that transforms only under
$SU(3)_L$, or, analogously, an object which transforms only under $SU(3)_R$:
\begin{eqnarray}
\label{LR}
L_\m&=&\S\partial_\m\S^\dagger = -\partial_\m\S\S^\dagger\ ,\\
R_\m&=&\S^\dagger\partial_\m\S = -\partial_\m\S^\dagger\S=-\S^\dagger L_\m\S
\ .\nonumber
\end{eqnarray}
The most general lagrangian with four derivatives is \cite{GL}
\begin{equation}
\label{l4}
\cl^{(4)}=-L_1\,\left(\tr(L_\m L_\m)\right)^2-L_2\,\tr(L_\m L_\n)\,\tr(L_\m L_\n)
-L_3\,\tr(L_\m L_\m L_\n L_\n)\ ,
\end{equation}
in which $L_{1,2,3}$ are new (dimensionless) coupling constants.
For general $N_f$, there is an additional term of the form
$\tr(L_\m L_\n L_\m L_\n)$, but for $N_f=3$ this can be written in terms
of the three terms appearing in Eq.~(\ref{l4}) \cite{GL}.  Even with only three
light quarks, this is not true in the partially-quenched case, see Sec.~\ref{PQmixed}.

We see that the new terms in $\cl^{(4)}$
will contribute to pion scattering at tree level:
each of these terms starts off with a $(\partial\phi/f)^4$ term when we
expand $\S$ in terms of $\phi$.  Thus, indeed, such contributions will be
of order $(p/f)^4$, \ie, of order $(p/f)^2$ relative to Eq.~(\ref{piscat}) --
apparently $f$ plays
the role of the typical hadronic scale $\sim$~1~GeV, and $(p/f)^2$ is the
expansion parameter of ChPT -- see below.

In order to preserve unitarity, we should also calculate loop contributions,
insofar as they contribute to order $p^4$.  They arise
when we consider the one-loop scattering diagrams with two
four-pion vertices from $\cl^{(2)}$.  This leads to contributions of the
form $(s/4\p f^2)^2\log{(-s/\L^2)}$ (and more such terms also involving
$t$ and $u$, as dictated by crossing symmetry),
where $\L$ is the cutoff needed in order to define the theory, and the
factor $1/(4\p)^2$ comes from the loop integral.
The appearance of the logarithm is interesting for two reasons:
it shows that, as mandated by unitarity, there is a two-particle
cut  in the pion scattering amplitude.  Second, it is clear that a cutoff
is needed in order to define theory.

Physical quantities, such as the scattering amplitude, cannot depend
on this cutoff, which has only been introduced in order to define the effective theory.
But indeed, since also polynomial terms from $\cl^{(4)}$
appear at this order (order $p^4$), the cutoff dependence can be removed
by taking $L_{1,2,3}$ to be dependent on $\L$, such that the scattering
amplitudes (and other physical quantities) are not.  We see that in order
to renormalize the theory defined by $\cl^{(2)}$, we need to introduce
$\cl^{(4)}$.  The new constants $L_{1,2,3}$ represent  the effect of the
underlying physics that has been integrated out in the EFT.
This is the beginning of a general pattern, as we will argue in
Sec.~\ref{power} below.

Before we discuss higher orders, let us consider order $p^4$ in some more
detail.  First, note that the one-loop contribution is in fact
of order $p^2/(4\pi f)^2$ relative to Eq.~(\ref{piscat}).   Any EFT involves an
expansion in a ratio of scales, and in ChPT the higher scale in this ratio
appears in the form $4\p f$.   The size of $f$ can be estimated from
pion scattering (using Eq.~(\ref{piscat})); for another method, see Sec.~\ref{currents}.
At the level of ChPT, the ``low-energy'' constants (LECs) $f$ and $L_i$
are free parameters, which can only be determined either by comparison
with experiment, or by matching ChPT to a lattice QCD computation.  We can say something
about the ``generically expected'' values of the $L_i$.
Since the $L_i$
absorb the scale dependence, we expect that their magnitude changes by an amount of
order $1/(4\p)^2\,\log(\L/\L')$ when we change the scale $\L\to\L'$.
Since the cutoff $\L$ appears because we integrated out all heavier
hadrons, a physically sensible choice for $\L$ is the typical hadronic
scale of 1~GeV.  With this interpretation, varying the cutoff within an order
of magnitude is reasonable.  This shifts the $L_i$ by an amount of order
 $1/(4\p)^2$, and gives us a sense of what one expects the values of these
 LECs to be.

%%####%%
\subsection{\label{power} Power counting}
%%####%%
Now, let us consider the derivative expansion more systematically, and show that indeed
the chiral theory is a proper EFT.   What this means is that there exists a systematic
power counting, and that to any order in the expansion we need only a finite number
of coupling constants to define the theory.

Consider an amputated connected diagram with $V_d$ vertices with $d$ derivatives.  Collectively
denoting the external momenta by $p$, this diagram is of order $\sum_d dV_d-2I+4L$ in $p$,
where $I$ is the number of internal lines, and $L$ is the number of loops.  This follows from
simply counting powers of $p$.
Using, as usual, that the number of loops
$L=I-\sum_d V_d+1$, we can rewrite this as $\sum_d(d-2)V_d+2L+2$.  Our diagram
is thus of order
\begin{equation}
\label{pc}
f^2p^2\left(\frac{p^2}{f^2}\right)^N\left(\frac{1}{f}\right)^E\ ,
\end{equation}
with
\begin{equation}
\label{pc1}
N=\sum_d\frac{1}{2}(d-2)V_d+L\ ,
\end{equation}
and $E$ the number of external legs.  The powers of $f$ in this result follow
from dimensional analysis.  Ignoring the cutoff for now, this is the only other
scale in the problem, if we express all other LECs as products of dimensionless
constants times the appropriate power of $f$.

We see that all contributions to a certain amplitude of a fixed order in external momenta
correspond to a fixed value of $N$.  For instance, for $N=0$, only tree-level
diagrams, calculated from $\cl^{(2)}$ contribute, because $N=0$ requires that
$L=0$ and $V_d=0$ for $d>2$.  For $N=1$, one-loop
diagrams coming from $\cl^{(2)}$ combine with tree-level diagrams coming from
$\cl^{(4)}$, consistent with our discussion in the previous section.  In general,
if we calculate to some fixed order in $N$, we need the chiral lagrangian only
up to $\cl^{(2N+2)}$.  In other words, while our EFT is not renormalizable, it is
nevertheless predictive if we work to a fixed order in the derivative expansion,
since to that order only a finite number of LECs occur in the chiral lagrangian
\cite{SWphysica}.%
\footnote{The number of LECs grows rapidly with $N$, rendering our
EFT practically useless beyond $N=2$.}

Each loop provides a factor $1/(4\p)^2$.  If we assume (\seef\ Sec.~\ref{lagr}) that
this number also sets the natural size of all dimensionless LECs, that turns the
momentum expansion into an expansion in powers of $p^2/(4\p f)^2$.
Then, so far we have ignored the fact that loops lead to divergences.  That means that
there is another scale, the cutoff, that can appear in our result.  Putting everything
together, we therefore amend our power-counting result: our diagram takes the
schematic form (restoring a delta function for momentum conservation)
\begin{equation}
\label{pc2}
(2\p)^4\delta\left(\sum_{i=1}^E p_i\right)\,f^2p^2
\left(\frac{p^2}{(4\p f)^2}\right)^N\left(\frac{1}{f}\right)^E F(p^2/\L^2)\ ,
\end{equation}
with $F$ a dimensionless function.  Contributions to $F$ can be of three types.
First, any positive powers of $p^2/\L^2$ can be ignored, as they correspond to
higher-order contributions.  Then, depending on the regulator, negative powers may occur
(\ie, positive powers of the cutoff);
those correspond to power divergences.  Since all divergences have to be local \cite{Collins}, such divergences can be absorbed into lower-order LECs.  Finally, there can be
logarithmic divergences, which, as we have seen, can be renormalized by
LECs in $\cl^{(2N+2)}$ if they occur at $N$ loops.  In dimensional regularization,
power divergences do not occur, making this regulator the most practical one
for calculations in ChPT.\footnote{For a different chirally invariant regularization
using a lattice cutoff, see Ref.~\cite{LO}.}

The conclusion of this section is that a well-defined power-counting
scheme exists, turning our EFT, \ie, ChPT, into a systematic and practically
useful tool.  This observation will play an important role when we start using
ChPT in applications to lattice QCD.

%%####%%
\subsection{\label{currents} Conserved currents}
%%####%%
It follows from Noether's theorem that there are sixteen conserved currents,
associated with the generators of the group $SU(3)_L\times SU(3)_R$.
At the level of QCD, these currents are fixed by the lagrangian and the symmetry
group, and the same should thus be true at the level of the chiral lagrangian.
In both cases, these currents can be calculated following Noether's procedure.
Equivalently, one may couple the QCD lagrangian to external gauge fields,
$\ell_\m$ and $r_\m$ (taken to be hermitian), which transform under local chiral transformations
as
\begin{eqnarray}
\label{sourcetransf}
\ell_\m\to U_L\ell_\m U_L^\dagger-i\partial_\m U_L U_L^\dagger\ ,\\
r_\m\to U_R r_\m U_R^\dagger-i\partial_\m U_R U_R^\dagger\ ,\nonumber
\end{eqnarray}
with the covariant derivatives in Eq.~(\ref{QCD}) turning into
$D_\m\to D_\m-i\ell_\m$ when acting on left-handed quarks, and
$D_\m\to D_\m-ir_\m$ when acting on right-handed quarks.  This substitution
makes the QCD lagrangian (we still are considering the case that $M=0$!)
invariant under local chiral transformations, and thus the chiral lagrangian
should also be made invariant under {\em local} chiral transformations (in
addition to Lorentz (or euclidean) invariance and parity).  This can
be accomplished by replacing
\begin{equation}
\label{Scovder}
\partial_\m\S\to D_\m\S=\partial_\m\S-i\ell_\m\S+i\S r_\m\to U_L D_\m\S U_R^\dagger
\end{equation}
everywhere in the chiral lagrangian, given to order $p^4$ by the sum of
Eqs.~(\ref{l2}) and~(\ref{l4}).   The second arrow in Eq.~(\ref{Scovder}) indicates how the
covariant derivative of $\S$ transforms; $U_{L,R}$ are now local transformations
in $SU(3)_{L,R}$.
In addition, more invariant terms can be
constructed, if we use also the building blocks
\begin{eqnarray}
\label{Fs}
L_{\m\n}&=&\partial_\m\ell_\n-\partial_\n\ell_\m-i[\ell_\m,\ell_\n]\ ,\\
R_{\m\n}&=&\partial_\m r_\n-\partial_\n r_\m-i[r_\m,r_\n]\ .\nonumber
\end{eqnarray}
At order $p^2$ there are no new terms (because these field strengths
have two Lorentz indices), but at order $p^4$ several new terms arise:
\begin{eqnarray}
\label{deltal4}
&&\hspace{-0.5cm}
iL_9\,\tr\left(L_{\m\n}D_\m\S(D_\n\S)^\dagger+R_{\m\n}(D_\m\S)^\dagger D_\n\S\right)
-L_{10}\,\tr(L_{\m\n}\S R_{\m\n}\S^\dagger)\\
&&\hspace{2.5cm}-H_1\tr(L_{\m\n}L_{\m\n}+R_{\m\n}R_{\m\n})\ .\nonumber
\end{eqnarray}
With the chiral lagrangian as constructed above, we have that
\begin{equation}
\label{zeq}
\log Z_{ChPT}(\ell_\m,r_\m)=\log Z_{QCD}(\ell_\m,r_\m)+{\rm constant}\ ,
\end{equation}
up to the order at which we work in the chiral expansion.
Since $\ell_\m$ and $r_\m$ couple to the left- and right-handed Noether currents,
(connected) correlation functions of these currents are generated by taking derivatives
with respect to $\ell_\m$ and $r_\m$.  In words, Eq.~(\ref{zeq}) states that these correlation
functions, as calculated in ChPT, equal those of QCD, to some given order in the derivative
expansion.

By taking one derivative of the chiral lagrangian
with respect to $\ell_\m$ or $r_\m$, and setting
these sources equal to zero, we find the
conserved left- and right-handed currents. Showing only the lowest-order
explicitly:
\begin{eqnarray}
\label{conscurrents}
J^L_\m&=&\frac{i}{4}f^2\partial_\m\S\S^\dagger+\dots
=-\frac{1}{2}f\partial_\m\phi+\dots\ ,\\
J^R_\m&=&-\frac{i}{4}f^2\S^\dagger\partial_\m\S+\dots
=\frac{1}{2}f\partial_\m\phi+\dots
\ .\nonumber
\end{eqnarray}
Therefore, to lowest order in the derivative expansion, and writing
\begin{eqnarray}
\label{gener}
\phi&=&\phi_a T_a\ ,\\
J^{L,R}_\m&=&J^{L,R}_{a\m}T_a\ ,\nonumber
\end{eqnarray}
with $T^a$ the $SU(3)$ generators normalized, as before, through $\tr(T_a T_b)=\d_{ab}$,
we get for the pion
to vacuum matrix element of the axial current
\begin{equation}
\label{vacpion}
\langle 0|J^R_{a\m}(x)-J^L_{a\m}(x)|\phi_b(p)\rangle=-ip_\m f\d_{ab}\,e^{-ipx}\ ,
\end{equation}
and we conclude that $f$ is equal to the pion decay constant, $f_\p$, in the
chiral limit.\footnote{Our normalization is such that in the real world $f_\p=130.4$~MeV.
Another common convention uses a value smaller by a factor $\sqrt{2}$.}
We also find another prediction: to leading order in the chiral expansion
$f_\p=f_K=f_\h$.

We end this section with a number of remarks:
\begin{itemize}
\item[1.] The observation that we can obtain conserved currents in the effective theory
just as well from the Noether procedure or from the ``source method'' that I described above
 is correct, but both methods do not in general lead to the same current.  At order $p^2$
the currents obtained using either of these methods are the same, but at order $p^4$
they differ by a term proportional to $L_9$.  Clearly, if we do not introduce the sources
$\ell_\m$ and $r_\m$ at all, the $L_9$ term in Eq.~(\ref{deltal4}) never appears, and therefore
the Noether current has no term proportional to $L_9$.  But if we use the source method
described above Eq.~(\ref{conscurrents}) and apply it to the terms in Eq.~(\ref{deltal4}), we
find additional terms
\begin{eqnarray}
\label{addcurrents}
\D J_\m^L&=&iL_9\;\partial_\n[L_\m,L_\n]\ ,\\
\D J_\m^R&=&iL_9\;\partial_\n[R_\m,R_\n]\ .\nonumber
\end{eqnarray}
However, these extra terms are automatically conserved, $\partial_\m\D J_\m^{L,R}=0$
identically.  Therefore, the most general form of the conserved current is that provided
by the source method.
\item[2.] When we gauge the group $SU(3)_L\times SU(3)_R$, as we did above,
it is in fact anomalous.  In order to reproduce the ``nonabelian'' anomaly in the
chiral theory, we need to add the gauged Wess--Zumino--Witten term \cite{WZ}.  Since this
part of the chiral lagrangian will play no role in the rest of these lectures, we will
not discuss this any further.
\item[3.] While (apart from the anomaly), the derivation given above looks very
straightforward, in fact the freedom to perform field redefinitions and to add
total-derivative terms is needed to complete the proof that the correct recipe
is to just choose the chiral lagrangian to be locally invariant \cite{HLannphys}.
\item[4.] A new term proportional to $H_1$,
not containing the pion fields $\S$, shows up.  This term
can only contribute contact terms to correlation functions and is thus not
physical.  For instance, if one considers the two-point function
$\langle J^L_\m(x) J^L_\n(y)\rangle$ at one loop, this has a logarithmic divergence,
which can be absorbed into $H_1$.  Such parameters in the chiral
lagrangian are sometimes referred to as ``high-energy constants.''
\item[5.] The coupling of the chiral lagrangian to the sources $\ell_\m$ and
$r_\m$ tells us how to couple pions to the photon and to (virtual) $W$ and $Z$
mesons.  For instance, one obtains the photon coupling by taking
\begin{equation}
\label{photon}
\ell_\m=r_\mu=-\frac{1}{3}ea_\m\pmatrix{2&0&0\cr 0&-1&0\cr 0&0&-1
}\ ,
\end{equation}
with $a_\m$ the photon field and $e$ the charge of the electron.
\end{itemize}
Generally, electromagnetic effects in hadronic quantities are of approximately the
same size as the isospin breaking coming from the fact that in nature $m_u\ne m_d$.
This implies that ultimately electromagnetic effects will have to be taken into
account in lattice QCD computations.  For explorations in this direction, see
Ref.~\cite{MILCEM}.

%%####%%
\subsection{\label{masses} Quark masses}
%%####%%
It is time to remember that in the real world the masses of the up, down
and strange quark do not vanish. Generalizing Eq.~(\ref{spurion}), the mass terms in Eq.~(\ref{QCD}) can be
replaced by source terms, with hermitian scalar and pseudoscalar
sources $s(x)$ and $p(x)$:
\begin{equation}
\label{scalarsources}
\bq_L(s+ip)q_R+\bq_R(s-ip)q_L\ ,
\end{equation}
with $SU(3)_L\times SU(3)_R$ transformation rules
\begin{equation}
\label{stransf}
s+ip\to U_L(s+ip)U_R^\dagger\ ,\ \ \ \ \ s-ip\to U_R(s-ip)U_L^\dagger\ .
\end{equation}
We recover the quark mass terms by setting $s=M$ with $M$ as in Eq.~(\ref{M})
and $p=0$.

This gives us a new building block for constructing terms in the
chiral lagrangian.  To lowest order in $s$ and $p$, using parity symmetry,
there is a unique operator that can be added to $\cl^{(2)}$, which now becomes
\begin{eqnarray}
\label{l2again}
\cl^{(2)}&=&\frac{1}{8}f^2\,\tr\left((D_\m\S)^\dagger D_\m\S\right)-\frac{1}{8}f^2\,\tr\left(\c^\dagger\S+\S^\dagger\c\right)
\ ,\\
\c&\equiv&2B_0(s+ip)\ ,\nonumber
\end{eqnarray}
with $B_0$ a new LEC.  Setting $\c=\c^\dagger=2B_0M$, and expanding
$\S$ to quadratic order in $\phi$, we can read off the (tree-level) pion masses
in terms of the quark masses:
\begin{eqnarray}
\label{mesonmasses}
m_{{\p^+}}^2&=&B_0(m_u+m_d)\ ,\\
m_{K^+}^2&=&B_0(m_u+m_s)\ ,\nonumber\\
m_{K^0}^2&=&B_0(m_d+m_s)\ ,\nonumber\\
m_{{\p^0}}^2&=&B_0\left(m_u+m_d+O\left(\frac{(m_u-m_d)^2}{m_s}\right)\right)\ ,\nonumber\\
m_\eta^2&=&\frac{1}{3}B_0\left(m_u+m_d+4m_s+O\left(\frac{(m_u-m_d)^2}{m_s}
\right)\right)\ .\nonumber
\end{eqnarray}
This gives our third prediction from chiral symmetry: five meson masses are
expressed in terms of three parameters.  If we ignore isospin breaking
(\ie, set $m_\ell\equiv m_u=m_d$), we find that $m_{{\p^0}}=m_{{\p^+}}$
and $m_\eta^2=(2(m_{K^+}^2+m_{K^0}^2)-m_{{\p^+}}^2)/3$, relations which
agree with experiment at the few percent level.  The latter relation is the
well-known Gell-Mann--Okubo relation.

Since the masses on the left-hand side
are physical quantities, also the expressions on the right-hand side should be
physical; in particular, they should not depend on the renormalization scale
of the underlying theory.
Indeed, by taking a derivative with respect to $s$ and then setting all sources
equal to zero, we find that in the chiral limit
\begin{equation}
\label{psibarpsi}
\langle\bu u\rangle=\langle\bd d\rangle=\langle\bs s\rangle=-f^2B_0\ ,
\end{equation}
and we then use that $m_u\langle\bu u\rangle$, \etc, are scale independent.  The
quark masses appear in the chiral lagrangian only in the scale-independent combinations
\begin{equation}
\label{combmass}
\c_u\equiv 2B_0m_u\ ,\ \ \ \ \ \c_d\equiv 2B_0m_d\ ,\ \ \ \ \  \c_s\equiv 2B_0m_s\ .
\end{equation}

{}From this tree-level exercise, we see that for onshell momenta, one power
of the quark mass should be counted as order $p^2$, because for an onshell
pion we have that $p^2=-m_{{\p^+}}^2$, \etc\ With this power counting for the
quark masses, the whole discussion of Sec.~\ref{power} applies, with the proviso
that $p^2$ in formulas like Eq.~(\ref{pc2}) can now also stand for any of the
squared meson masses.  With $f=130$~MeV, we note that the expansion
parameters of the chiral expansion are
\begin{equation}
\label{expansionpars}
\frac{m_\p^2}{(4\p f)^2}\approx 0.007\ ,\ \ \ \ \ \frac{m_K^2}{(4\p f)^2}\approx 0.09\ ,
\end{equation}
which gives one hope that even for kaons the chiral expansion may be well
behaved.  We will return to this issue in Sec.~\ref{kaon}.   As an aside, we also see
why the heavy quark masses (charm \etc) cannot be accounted for in ChPT:
The corresponding ``expansion'' parameters would not be small.  In fact,
the situation is ``upside down'': one may instead consider expansions in
$\L_{QCD}/m_{heavy}$, leading to Heavy-Quark EFT.\footnote{For
an introduction to heavy quarks, see the lectures by Rainer Sommer at this
school.}

In order to find the meson masses, we expanded the nonlinear $\S$ around
${\bf 1}$.  Indeed, when all masses in Eq.~(\ref{M}) are positive, $\S={\bf 1}$ is
the minimum of the potential, which is given by the second term in Eq.~(\ref{l2again}).  When we
allow two of the masses to be negative, the situation is essentially the same.
For instance, if $m_{u,d}<0$ while $m_s>0$, the vacuum is
\begin{equation}
\label{negvac}
\S_{vac}={\rm diag}(-1,-1,1)\ ,
\end{equation}
which is equivalent to the trivial vacuum by an $SU(3)$ rotation.  When, however,
an odd number of quark masses are negative, there is no $SU(3)$ rotation
relating the vacuum to the trivial vacuum, and the theory can be in a different
phase, in which $CP$ is broken because of the appearance of a $\theta$-term
with $\theta=\p$ \cite{Witten}.  Since in nature all quark masses are positive
(or equivalently, $\theta=0$), we will only consider the case of positive quark
masses throughout this talk.\footnote{As mentioned already in the first section,
the scenario with $m_u=0$ is ruled out
\cite{MILC2004,MILC2007}.}  However, we will see in Sec.~\ref{phase}
that a nontrivial vacuum
structure is nevertheless possible for some discretizations of QCD at nonzero
lattice spacing.

Of course, with the scalar source $\c$ as a new building block, more terms
can appear in $\cl^{(4)}$ as well.  Gauging Eq.~(\ref{l4}) and including Eq.~(\ref{deltal4}),
the most general form becomes\footnote{Some terms in Eq.~(\ref{l4total})
have been removed by field redefinitions.}
\begin{eqnarray}
\label{l4total}
\cl^{(4)}&=&-L_1\,\left(\tr\left(D_\m\S(D_\m\S)^\dagger\right)\right)^2
-L_2\,\tr\left(D_\m\S(D_\n\S)^\dagger\right)\,\tr\left(D_\m\S(D_\n\S)^\dagger\right)\\
&&-L_3\,\tr\left(D_\m\S(D_\m\S)^\dagger D_\n\S(D_\n\S)^\dagger\right)\nonumber\\
&&+L_4\,\tr\left(D_\m\S(D_\m\S)^\dagger\right)\,\tr(\c^\dagger\S+\S^\dagger\c)
+L_5\,\tr\left(D_\m\S(D_\m\S)^\dagger(\c^\dagger\S+\S^\dagger\c)\right)\nonumber\\
&&-L_6\,\left(\tr(\c^\dagger\S+\S^\dagger\c)\right)^2-L_7\,\left(\tr(\c^\dagger\S-\S^\dagger\c)\right)^2-L_8\,\tr(\c^\dagger\S\c^\dagger\S+\S^\dagger\c\S^\dagger\c)\nonumber\\
&&+iL_9\,\tr\left(L_{\m\n}D_\m\S(D_\n\S)^\dagger+R_{\m\n}(D_\m\S)^\dagger D_\n\S\right)
-L_{10}\,\tr(L_{\m\n}\S R_{\m\n}\S^\dagger)\nonumber\\
&&-H_1\tr(L_{\m\n}L_{\m\n}+R_{\m\n}R_{\m\n})
-H_2\,\tr(\c^\dagger\c)\ .\nonumber
\end{eqnarray}
The full lagrangian $\cl^{(2)}+\cl^{(4)}$ of Eqs.~(\ref{l2again}) and~(\ref{l4total})
is invariant under the local group $SU(3)_L\times SU(3)_R$.  We note that
five new LECs and one new ``high-energy'' constant have been added.
The term proportional to $H_2$ contributes to the mass dependence of
the chiral condensate, and is needed for instance in order to match lattice QCD, where
the condensate includes a term proportional to $m/a^2$ (with $m$ the
relevant quark mass).

As an application, we quote some results of the quark-mass dependence of
various physical quantities to order $p^4$ \cite{GL}, in the isospin limit $m_u=m_d$:
\begin{eqnarray}
\label{oneloopresults}
f_\p&\!\!\!=\!\!\!&f\left\{1+\frac{8L_5}{f^2}\hm_\ell+\frac{8L_4}{f^2}(2\hm_\ell+\hm_s)
-2L(m_\p^2)-L(m_K^2)\right\}\ ,\\
\frac{f_K}{f_\p}&\!\!\!=\!\!\!&1+\frac{4L_5}{f^2}(\hm_s-\hm_\ell)+\frac{5}{4}L(m_\p^2)-\frac{1}{2}L(m_K^2)
-\frac{3}{4}L(m_\eta^2)\ ,\nonumber\\
\frac{f_\eta}{f_\p}&\!\!\!=\!\!\!&\left(\frac{f_K}{f_\p}\right)^{4/3}\left\{1+
\frac{1}{48\p^2 f^2}\left(3m_\eta^2\log\frac{m_\eta^2}{m_K^2}
+m_\p^2\log\frac{m_\p^2}{m_K^2}\right)\right\}\ ,\nonumber\\
\frac{m_\p^2}{2\hm_\ell}&=&1+\frac{8(2L_8-L_5)}{f^2}\,2\hm_\ell+
\frac{16(2L_6-L_4)}{f^2}(2\hm_\ell+\hm_s)+\!L(m_\p^2)\!-\!\frac{1}{3}L(m_\eta^2)\ ,\nonumber\\
\frac{m_K^2}{\hm_\ell+\hm_s}&=&1+\frac{8(2L_8-L_5)}{f^2}(\hm_\ell+\hm_s)+
\frac{16(2L_6-L_4)}{f^2}(2\hm_\ell+\hm_s)+\frac{2}{3}L(m_\eta^2)\ ,\nonumber
\end{eqnarray}
in which
\begin{eqnarray}
\label{abbrev}
\hm_\ell&=&2B_0m_u=2B_0m_d\ ,\ \ \ \ \ \hm_s=2B_0m_s\ ,\\
L(m^2)&=&\frac{m^2}{(4\p f)^2}\,\log\left(\frac{m^2}{\L^2}\right)\ ,\nonumber
\end{eqnarray}
and where we can use the tree-level expressions~(\ref{mesonmasses})
inside the logarithms.
Because all meson interactions that follow from the chiral lagrangian are even
in the meson fields,\footnote{This is not true for the Wess--Zumino--Witten term, which
however at this order does not contribute to the quantities of Eq.~(\ref{oneloopresults}) to
order $p^4$.} and the calculation of both masses and decay constants
involves two-point functions, the only diagrams that appear at one loop are
tadpole diagrams.

%%%%%%%%%%%%%%%%%%%
%%% chiral fits
\begin{figure}[t!]
\vspace*{10ex}
\hspace*{5ex}
%
%\begin{center}
\begin{picture}(75,55)(0,0)
\put(-17,0){\includegraphics*[height=6.5cm]{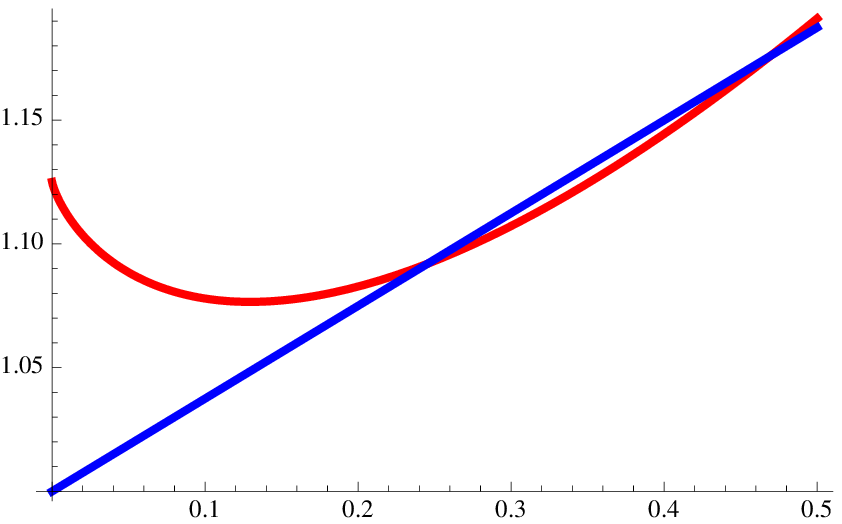}}
\end{picture}
%\end{center}
\vspace*{0ex}
\begin{quotation}
\floatcaption{fig:chiralfits}{The functions $1.125+L(m_\p^2)$ (red curve) and
$1+m_\p^2/(4\p f)^2$ (blue line), as a function of $m_\p^2$ in GeV$^2$.  I took
$\L=m_\r=770$~MeV and $f=130$~MeV.}
\end{quotation}
\vspace*{-1ex}
\end{figure}
%%%%%%%%%%%%%%%%%%%

Again, ChPT makes a prediction: if we use $f_K/f_\p=1.2$ as input, one finds
that (to this order in ChPT) $f_\eta/f_\p=1.3$.  With these values
for $f_K/f_\p$ and $f_\eta/f_\p$ we see that ChPT to order $p^4$ works reasonably
well for quantities involving the strange quark, with order-$p^4$
corrections of about 20--30\%.\footnote{For a recent
discussion of the experimental value of $f_\eta$, including mixing because
of $SU(3)$ breaking, see Ref.~\cite{RE}.}
For a discussion
of meson masses and decay constants to order $p^6$ in ChPT, see Ref.~\cite{Betal}.
For a recent review of the phenomenological status of ChPT, and many references,
see Ref.~\cite{Ecker}.

An important observation is that all LECs, $f$, $B_0$ and the $L_i$'s, are
independent of the quark mass; they only depend on the number of flavors,
$N_f$.\footnote{For the relation between $N_f=3$ LECs and $N_f=2$ LECs,
see Ref.~\cite{GL}.}  All quark-mass dependence in Eq.~(\ref{oneloopresults}) is explicit.
It follows from this that the values of the LECs can, in principle, be obtained from
lattice QCD with unphysical values of the quark masses (as long as they are
small enough for ChPT to be valid --- an important restriction).  Thus, if one
finds a set of quantities through which all LECs can be determined from the
lattice, one can then use these values to calculate other quantities that are
less easily accessible on the lattice.  A simple example of this is that one can
extrapolate physical quantities to the physical values of the light quark masses,
from lattice computations at larger values of $m_\ell$ in the isospin limit.\footnote{Since
also EM interactions break isospin, they also will have to be taken into account
if one reaches a precision at which isospin breaking becomes significant.}
Unfortunately, it is not possible to extract all the $L_i$ from two-point
functions, since the constants $L_{1,2,3}$ do not show up in masses and
decay constants.

%%%%%%%%%%%%%%%%%%%
%%% chiral fits
\begin{figure}[t!]
\vspace*{10ex}
\hspace*{5ex}
\begin{picture}(75,55)(0,0)
\put(-12,0){\includegraphics*[height=7.5cm]{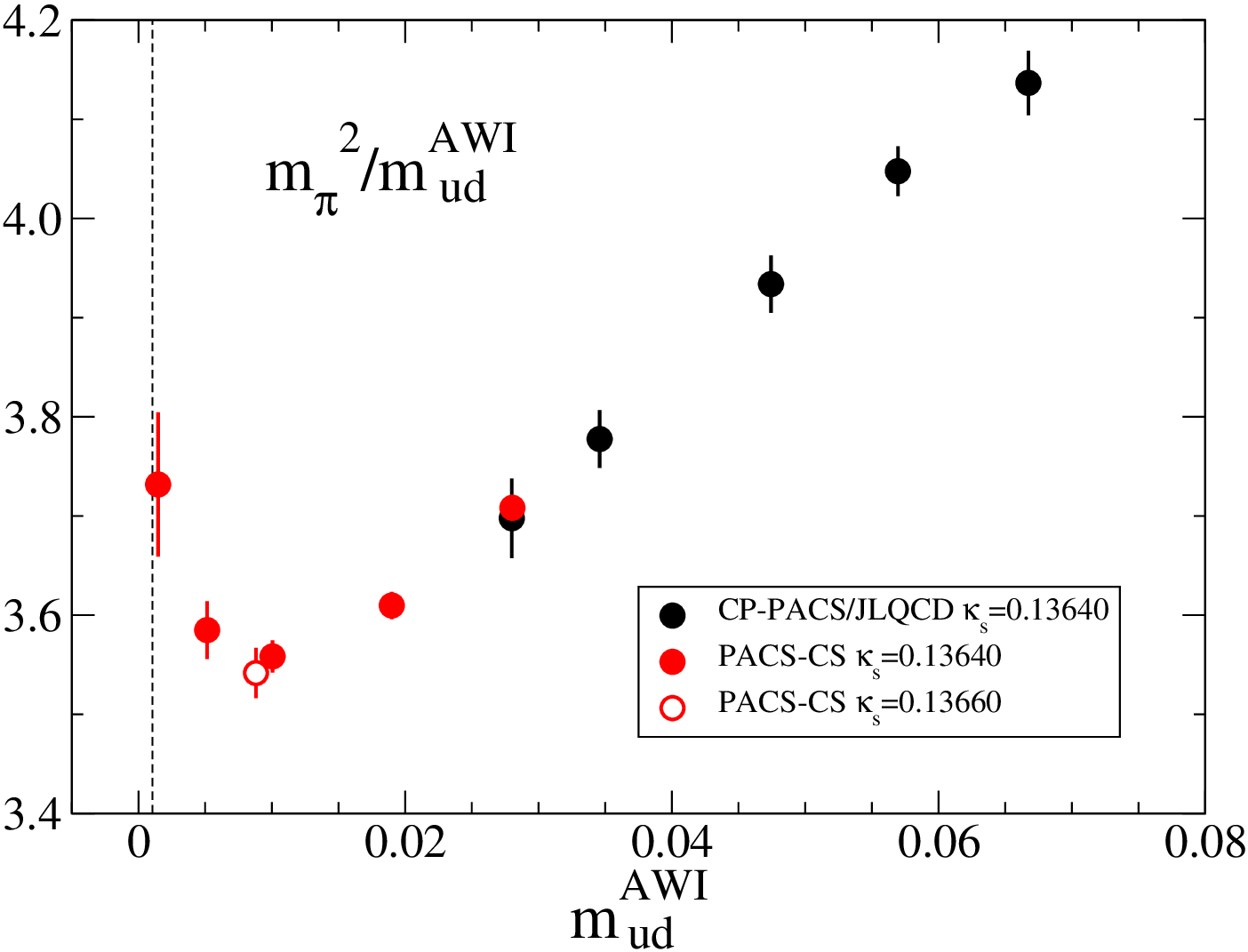}}
\end{picture}
\vspace*{0ex}
\begin{quotation}
\floatcaption{fig:PACS-CS}{Comparison of the PACS-CS (red) and the CP-PACS/JLQCD (black)
 results for $m_\pi^2/m_{\rm ud}^{\rm AWI}$ as a function of the light quark mass
 $m_{\rm ud}^{\rm AWI}$ (the so-called ``axial Ward identity'' definition of the quark mass).
 The vertical line denotes the physical point.
 {}From Ref.~\cite{PACS-CS}.}
\end{quotation}
\vspace*{-1ex}
\end{figure}
%%%%%%%%%%%%%%%%%%%

We conclude this section with an important lesson about the use of ChPT
for fitting the quark mass dependence of a hadronic quantity on the lattice.
As we see in Eq.~(\ref{oneloopresults}), nonanalytic terms show up, here in the
form of so-called ``chiral logarithms.''   Fig.~\ref{fig:chiralfits} shows how important
it can be to include such nonanalytic terms in chiral fits, by comparing the
functions $1.125+L(m_\p^2)$ with $1+m_\p^2/(4\p f)^2$.  The logarithm has a
dramatic effect: it looks nothing like the linear
curve in the region around the physical pion mass, $m_\p^2=0.02$~GeV$^2$.
If one would only have lattice data points in the region $m_\p^2\geqx 0.2$~GeV$^2$
with error bars that would more or less overlap with both curves,
and one would perform a linear fit,\footnote{Both functions can be thought of as
two-parameter fits.  In the case of the logarithm, the two parameters are the constant
and $\L$, in the case of the linear function, they are the constant and the slope.}
we see that this might lead to errors of order
10\% in the value at the physical pion mass.  In order to confirm the existence
of the chiral logarithm in the data, clearly data points in the region of the
curvature of the logarithm are needed.  In the cartoon example of Fig.~\ref{fig:chiralfits}
this means that we need data points in the region down to $m_\p\sim 200$~MeV.
In addition, data points with very good statistics help.

For a realistic example, see Fig.~\ref{fig:PACS-CS}, which shows $m_\p^2/m_\ell$
as a function of the light quark mass, $m_\ell$, for fixed (physical) strange mass, from
Ref.~\cite{PACS-CS} (to which I refer for details).   If one would only consider the
large $m_\p$ points (the black points), one would not be able to reliably fit
the chiral logarithm in Eq.~(\ref{oneloopresults}), and thus also not the linear combination of
LECs that accompany the logarithm.  Only with the red points, which were obtained
for small values of $m_\p$, or equivalently $m_\ell$, can one hope to perform
sensible chiral fits.  For a discussion of such fits, see Ref.~\cite{PACS-CS}.   Here we
only note that both $O(a^2)$ effects and finite volume can in principle modify the
chiral logarithms.   These issues can be systematically studied with the help of
ChPT, as we will see in Secs.~\ref{lattice spacing} and \ref{finite volume}.

%%####%%
\section{\label{lattice spacing} ChPT at nonzero lattice spacing}
%%####%%

If we compute hadronic quantities on the lattice, the values we
obtain will differ from their continuum values by scaling violations ---
terms of order $a^n$ with some positive integer power $n$ (possibly modified
by logarithms), where $a$ is the lattice spacing.  For example, for
the mass of some hadron evaluated on the lattice at some nonzero
$a$ we expect that
\begin{equation}
\label{sv}
M_{lattice}=M_{continuum}(m_{quark})+c_1a\L_{QCD}^2+O(a^2,am_{quark})\ ,
\end{equation}
with $c_1$ at most logarithmically dependent on $a$.\footnote{For lattice regulators with an exact
chiral symmetry, one expects scaling violations of order $a^2$ instead of
order $a$.  See for example Sec.~\ref{staggered}.}   In Eq.~(\ref{sv}) I indicated schematically that the hadron mass also
depends on the quark masses in the theory.

Evidently, at nonzero lattice spacing, there is a ``new'' infrared
scale $a\L_{QCD}^2$ that emerges from combining the physical infrared
scale $\L_{QCD}$ with the ultraviolet scale $a$.   Physical quantities
thus depend not only on the quark masses, but also on this new scale,
and one expects that a combined expansion in terms of $m_{quark}$
and $a\L_{QCD}^2$ should be possible when these parameters are both
small enough.  The goal of this section is to see how ChPT can be extended
to provide this combined expansion.

Before we begin doing that, there are two important observations.  First,
ChPT is constructed by making use of the chiral symmetry of the underlying
theory --- it is nothing more or less than a very efficient way of implementing
the Ward identities for chiral symmetry.  This implies that if we use a
chirally invariant regulator on the lattice, such as domain-wall fermions%
\footnote{At very small $m_{residual}$; see David Kaplan's lectures at
this school.} or overlap fermions, there will be almost no change in the form of
the chiral lagrangian.  Almost no change, because for any lattice regulator continuous
rotational invariance is broken, so one expects that new operators
will appear in the chiral lagrangian that break continuous rotational invariance,
but respect the lattice subgroup (usually the group of hypercubic rotations).
Quite generally, in the construction of the chiral lagrangian for Nambu--Goldstone
bosons, the only quantity transforming as a four-vector is the derivative,
so such terms will have to have at least four derivatives.  Combined with the
fact that they need to carry a positive power of the lattice spacing, they are
usually of quite high order in the chiral expansion, as we will see.
The only other difference is that the continuum LECs themselves are afflicted
by scaling violations, which can in principle be removed by an extrapolation
to $a=0$.

It is only for regulators that break some of the continuum chiral
symmetries that the form of the chiral lagrangian will be different.  The two
most important examples are Wilson-like fermions (including also
clover fermions and twisted-mass QCD), for which vectorlike flavor symmetries
are preserved but chiral symmetries are broken, and staggered fermions,
for which flavor symmetries are broken, and only one chiral symmetry is
preserved.  Here I will primarily discuss Wilson fermions, since they are
conceptually the simplest.  Section~\ref{staggered} is devoted to a brief
overview of ChPT for QCD with staggered quarks.

A second point is that we will be dealing with an expansion in two small
parameters, $m_{quark}$ and $a\L_{QCD}^2$, and we thus have to
define a power counting --- we have to compare their relative size in order
to know to what relative order in each parameter we have to expand.
Typical quark masses in lattice computations
range between 10 and 100~MeV, and typical
lattice spacings vary between $a^{-1}\sim 2-3$~GeV.  With $\L_{QCD}\sim
300$~MeV, this means that $m_{quark}$ and $a\L_{QCD}^2$ are of the
same order, and this is what I will assume to get the discussion started,
until Sec.~\ref{phase}.

QCD with Wilson fermions has no chiral symmetry, and it thus seems that
our guiding principle for constructing the chiral effective theory is completely
lost: we should allow for all kinds of operators in the chiral lagrangian
that are only invariant under $SU(3)_V$, and the only thing we know is
that any such operator that is forbidden in the continuum theory is multiplied
by some positive power of the lattice spacing.  It thus seems that, while
an EFT at low energies should exist for small enough lattice spacing, it will
not be of much practical use.  It turns out, however, that this is not true.

The solution to this apparent roadblock consists of first considering a ``low-energy''
continuum EFT of quarks and gluons, where ``low energy'' means that we
want to consider correlation functions of quarks and gluons at momenta
$\L_{QCD}\ll p\ll 1/a$.
Since at such momenta the degrees of freedom in this EFT are the same as in the
lattice theory, no powers of $\L_{QCD}$ can appear in this EFT, and any
operator of order $a^n$ ($n\ge 0$) thus has to be multiplied by an operator of
mass dimension $4+n$, so that the full lagrangian always has dimension four.
This means that, for small values of $n$ the operators that can appear in this
EFT are very constrained, and, as we will see, this allows us to also
restrict the form of the operators for small values of $n$ in the chiral theory.
The idea of considering the quark-gluon continuum EFT for the lattice theory
is due to Symanzik \cite{Sym}, and we will refer to this EFT as the Symanzik effective
theory (SET).  The insight that this can be used to develop ChPT including
scaling violations systematically is due to Sharpe and Singleton \cite{ShSi}.

%%####%%
\subsection{\label{SET} Symanzik effective theory}
%%####%%
Let us begin with a brief review of the construction of the SET for lattice QCD
with Wilson fermions.\footnote{For a more detailed discussion of Symanzik
effective theories in the context of improvement, see the lectures of Peter
Weisz at this school.}  The Symanzik expansion is an expansion in powers
of $a$,
\begin{equation}
\label{Sym}
\cl_{SET}=\cl^{(4)}_S+a\cl^{(5)}_S+a^2\cl^{(6)}_S+\dots\ .
\end{equation}
The first term, $\cl^{(4)}_S$, is the continuum theory,
\begin{equation}
\label{cl4}
\cl^{(4)}_S=\cl_{gluons}+\bq\Sl{D}q+\bq M q\ ,
\end{equation}
with $M$ the quark mass matrix given in Eq.~(\ref{M}).\footnote{For the application
of these ideas to twisted-mass QCD, see the review of Ref.~\cite{SRSNara}.}  In fact, because of the
lack of chiral symmetry of Wilson fermions, the first term in the Symanzik
expansion is a term of dimension three of the form $(c/a)\bq q$ with $c$
a numerical constant.  The quark masses in $M$  thus have to be defined as the
difference between the lattice quark masses and the power divergence
$c/a$:
\begin{equation}
\label{massrel}
m_i=Z_S^{-1}(m_{0,i}-c)/a\ ,\ \ \ i=u,d,s\ ,
\end{equation}
where $m_{0,i}$ is the bare lattice mass for flavor $i$ in lattice units, and
$Z_S$ is the multiplicative renormalization factor needed to relate lattice
masses to the continuum regulator that we use to define the SET.  It is
conceptually easiest to think of the continuum regulator as dimensional
regularization (in some minimal subtraction scheme).  This brings us to an
important point: the fact that, for momenta $\L_{QCD}\ll p\ll 1/a$, an effective theory
of the form~(\ref{SET}) can be used to represent the lattice theory was only
argued to exist using perturbation theory.\footnote{And, in the original
work by Symanzik, only for $\phi^4$ theory \cite{Sym} and the two-dimensional
nonlinear sigma model \cite{Symsigma}, although there is little doubt that
the ideas carry over to gauge theories as well.}   It is an assumption, essentially
based on locality, that the SET is also valid nonperturbatively: the claim is
that discretization effects in correlation functions with all momenta $\L_{QCD}\ll p\ll 1/a$
are given by insertions of local operators, to any finite order in an expansion
in $ap$.   We note that this is nothing else than the basic assumption underlying the
construction of all EFTs, if we replace $ap$ by the ratio of scales that provides the
small parameter for the EFT.

Of course,
in order to ``think'' about the SET outside perturbation theory, one would have
to use a nonperturbative regulator --- such as the lattice!  One can imagine
using a lattice regulator to define the SET with a lattice spacing $a'$ much
smaller than $a$, so that $p\ll 1/a\ll 1/a'$, and since $a'\ll a$ one does not have
to include powers of $a'$ in the expansion~(\ref{Sym}).  This ignores the
possibility of power divergences in the smaller lattice spacing of the form
$1/a'$ or $1/{a'}^2$, which one might have to subtract in order to define the
SET.  Since I will only use the form of the SET, without ever doing any
calculations with it, I will sidestep this issue, and assume that we can work with
the expansion~(\ref{Sym}).\footnote{Since we are considering a renormalizable
theory, there should only be a finite number of power divergences in $a'$.
For more discussion of related issues in the context of staggered fermions,
see Ref.~\cite{BGSset}.}

Next, let us consider $\cl^{(5)}_S$.  This consists of all dimension five operators
consistent with the exact symmetries of the underlying lattice theory.  For Wilson
fermions
\begin{equation}
\label{cl5}
\cl^{(5)}_S=b_1\;\bq i\s_{\m\n}G_{\m\n}q+b_2\;\bq D_\m D_\m q+b_3\;\bq M\Sl{D}q
+b_4\;\bq M^2 q+b_5\;\tr(M)\cl_{gluons}\ ,
\end{equation}
in which the $b_i$ are real dimensionless coefficients, independent of quark
masses, that can be calculated in perturbation theory, or, nonperturbatively,
on the lattice.   $G_{\m\n}$ is the gluon field-strength
tensor.  Note that the $b_i$ are not really constant: they depend on logarithms
of the ratio of the scales used to define the lattice theory and the SET.
If dimensional regularization is used for the SET, $b_i=b_i(g^2(\m),\log{a\m})$.
We will ignore this additional logarithmic dependence on $a$, as it should be
much milder than the explicit power dependence on $a$ in Eq.~(\ref{Sym}).

Since the $b_3$ and $b_5$ terms are of the form quark mass times terms which
already occur
in $\cl^{(4)}_S$, they will translate into $O(p^6)$ terms in ChPT.  The counting
here works as follows.  Terms in $\cl^{(4)}$ translate into $O(p^2)$ terms in
ChPT, \seef\ Eq.~(\ref{l2}).  The explicit factors of quark mass and $a$ add
two powers of $p^2$, since we chose a power counting scheme in which
$p^2\sim m_{quark}\L_{QCD}\sim a\L_{QCD}^3$.  Since here we will work
to order $p^4$, we may drop these terms from consideration.  The same is
true for the $b_4$ term, which is of order $a(m_{quark})^2$.

At the next order in the Symanzik expansion, the number of terms of dimension
six, consistent with all exact lattice symmetries, proliferates, and I will not
list them all \cite{ShWo}.  There are purely gluonic operators, that do not break chiral
symmetry, but they may break Lorentz (or, rather, euclidean) invariance,
such as
\begin{equation}
\label{gluonl6}
\sum_{\m,\k}\tr(D_\m G_{\m\k}D_\m G_{\m\k})\ ,
\end{equation}
where for clarity I showed index sums explicitly.
While this term would thus lead to new terms in the chiral lagrangian,
it is easy to see that these are higher order than $p^4$.   As we already noted before,
in order to construct
a chiral operator that breaks euclidean invariance, one needs at least four
derivatives.  With the additional two powers of $a$ that multiply terms in
$\cl^{(6)}_S$, this makes such terms $O(p^8)$ in our power counting.\footnote{
Such terms would be $O(p^6)$ in a power counting in which
$m_{quark}\sim a^2\L_{QCD}^3$.}  Note that this observation hinges on the fact that
in pion ChPT the derivative is the only object with a Lorentz index.
For instance, in baryon ChPT this is not true (there is also the four-velocity
of the heavy baryon, which is order one in power counting), and the effects
of the breaking of continuous rotational symmetry already shows up at order
$p^4$ \cite{BT}.

Operators involving quarks are of two types: bilinear and quartic in the
quark fields.  In order to generate new terms in the chiral lagrangian, they
have to break one of the symmetries of the continuum theory.   There are
bilinears which break euclidean invariance, which, as we argued above, we do not
have to consider at order $p^4$.   Dimension-six operators that break chiral
symmetry (and are not of the form $\tr(M)\cl^{(5)}_S$) are, in the basis of Ref.~\cite{Baeretal},
\begin{eqnarray}
\label{cl6}
&O^{(6)}_5=(\bq q)^2\ , &O^{(6)}_{10}=(\bq t_r q)^2\ ,\\
&O^{(6)}_6=(\bq \g_5 q)^2\ , &O^{(6)}_{11}=(\bq t_r \g_5 q)^2\ ,\nonumber\\
&O^{(6)}_9=(\bq \s_{\m\n}q)^2\ , &O^{(6)}_{14}=(\bq t_r \s_{\m\n}q)^2\ ,
\nonumber
\end{eqnarray}
where $t_r$ are the $SU(3)_{color}$ generators.

%%####%%
\subsection{\label{WChPT} Transition to the chiral theory}
%%####%%
Now we are ready to make the transition to the chiral theory.  First, consider
terms of order $a$, corresponding to $\cl^{(5)}_S$ in Eq.~(\ref{Sym}).  Using a trick
similar to the one we used translating quark-mass terms to ChPT, we introduce
a spurion field $A$, and write the first term in Eq.~(\ref{cl5}) as
\begin{equation}
\label{spurioncl5}
ab_1\;\bq i\s_{\m\n}G_{\m\n}q\to b_1\left(\bq_L i\s_{\m\n}G_{\m\n}A q_R
+\;\bq_R i\s_{\m\n}G_{\m\n}A^\dagger q_L\right)\ ,
\end{equation}
which is invariant under $SU(3)_L\times SU(3)_R$ if we let $A$ transform
as (\seef\ Eqs.~(\ref{spurion}) and~(\ref{stransf}))
\begin{equation}
\label{Aspurion}
A\to U_L A U_R^\dagger\ .
\end{equation}
Note that we recover the term on the left of Eq.~(\ref{spurioncl5}) by setting
$A=a{\bf 1}$.

We see that the $b_1$ term, from the point of view of chiral symmetry,
has exactly the same structure as the quark mass term.  At lowest order
in ChPT, we thus find a new term at leading order in the
chiral lagrangian of the form
\begin{equation}
\label{l2a}
-\frac{1}{4}f^2W_0\;\tr(A^\dagger\S+\S^\dagger A)\ ,
\end{equation}
in which $W_0$ is a new LEC analogous to $B_0$.  The $b_2$ term in
Eq.~(\ref{cl5}) has exactly the same chiral structure, and thus does not lead
to any new operator in ChPT.    This is an example of the fact that there
is in general not a one-to-one correspondence between operators in the
SET and in ChPT.

Once we set $A=a{\bf 1}$,  the new term can actually be
completely absorbed into the mass term in Eq.~(\ref{l2again}), if we shift
the quark mass matrix by
\begin{eqnarray}
\label{shift}
M&\to &M' \equiv M+\frac{W_0}{B_0}a{\bf 1}\ ,\\
{\rm or}\ \ m_i&\to & m'_i=m_i+\frac{W_0}{B_0}a\ ,\ \ \ i=u,d,s\ .\nonumber
\end{eqnarray}
If we determine, nonperturbatively, the lattice quark masses by requiring the
pion mass to vanish,\footnote{Or the PCAC quark mass to go to zero.}
then this determines $M'$, rather than $M$ --- in other words, the shift
of order $a$ in Eq.~(\ref{shift}) is automatically taken into account in the
determination of the critical quark mass for Wilson fermions.   Note that this
does not remove all $O(a)$ effects from the theory when we extend it to
include other hadrons,\footnote{For instance, the nucleon mass has $O(a)$ corrections
which are not removed by the shift~(\ref{shift}) \cite{nucleonmass,BT}.}
or operators other than the action, \seef\ Sec.~\ref{axial}.

Next, we consider terms of order $ap^2\sim am_{quark}\sim a^2$, all of
order $p^4$ in our power counting (appropriate powers of $\L_{QCD}$
are implicit).  These come from two sources: higher order terms in the
spurion $A$, as well as terms arising from $\cl^{(6)}_S$.  However, both of these
lead to the same new $O(p^4)$ terms in the chiral lagrangian.  The point
is that the same spurion field can be used to also make the operators in
Eq.~(\ref{cl6}) invariant under $SU(3)_L\times SU(3)_R$:
\begin{equation}
\label{spurioncl6}
a^2(\bq q)^2\to (\bq_L A q_R+\bq_R A^\dagger q_L)^2\ ,
\end{equation}
and similar for the other operators in Eq.~(\ref{cl6}).  All these operators
can thus be made invariant using the ``composite'' spurions
\begin{eqnarray}
\label{cspurions}
A\otimes A&\to &U_LAU_R^\dagger\otimes U_LAU_R^\dagger\ ,\\
A\otimes A^\dagger&\to &U_LAU_R^\dagger\otimes U_RA^\dagger
U_L^\dagger\ ,\nonumber\\
A^\dagger\otimes A&\to &U_RA^\dagger
U_L^\dagger\otimes U_LAU_R^\dagger\ ,\nonumber\\
A^\dagger\otimes A^\dagger&\to &U_RA^\dagger
U_L^\dagger\otimes U_RA^\dagger
U_L^\dagger\ .\nonumber
\end{eqnarray}
Using these spurions, we find the following new terms in the chiral lagrangian
\begin{eqnarray}
\label{newasq}
\tr(A\S^\dagger)\;\tr(A\S^\dagger)+{\rm h.c.}&\to &a^2\left(\tr(\S^\dagger)\right)^2+{\rm h.c.}\ ,\\
\tr(A\S^\dagger A\S^\dagger)+{\rm h.c.}&\to &a^2\;\tr(\S^\dagger\S^\dagger)+{\rm h.c.}\ ,
\nonumber\\
\tr(A\S^\dagger)\;\tr(A^\dagger\S)&\to &a^2\;\tr(\S^\dagger)\;\tr(\S)\ .\nonumber
\end{eqnarray}
Setting sources
$\ell_\m$ and $r_\m$ equal to zero,\footnote{For the case including all sources and new ``high-energy'' constants, see Ref.~\cite{SW}.}
the complete set of additional terms in the chiral lagrangian at order $p^4$,
including new LECs multiplying each of the new operators is \cite{Baeretal}
\begin{eqnarray}
\label{newp4}
\D\cl^{(4)}&=&\ha W_4\;\tr(\partial_\m\S\partial_\m\S^\dagger)\;\tr(\S+\S^\dagger)
+\ha W_5\;\tr\left(\partial_\m\S\partial_\m\S^\dagger(\S+\S^\dagger)\right)\\
&&-\ha W_6\;\tr(\hM\S^\dagger+\S\hM)\;\tr (\S+\S^\dagger)
-\ha W_7\;\tr(\hM\S^\dagger-\S\hM)\;\tr(\S-\S^\dagger)\nonumber\\
&&-\ha W_8\;\tr(\hM\S^\dagger\S^\dagger+\S\S\hM)\nonumber\\
&&-\ha^2 W'_6\left(\tr(\S+\S^\dagger)\right)^2-\ha^2 W'_7\left(\tr(\S-\S^\dagger)\right)^2
-\ha^2 W'_8\;\tr(\S\S+\S^\dagger\S^\dagger)\ ,\nonumber
\end{eqnarray}
in which
\begin{equation}
\label{hats}
\hM\equiv 2B_0M\ ,\ \ \ \ \ \ha\equiv 2W_0a\ .
\end{equation}
If one considers (nonperturbatively) $O(a)$ improved QCD with Wilson
fermions, one can use the same chiral lagrangian.  It simply means that
$W_0=0$ in Eq.~(\ref{l2a}).  Note that this does {\em not} mean that we should
set all terms of order $a^2$ equal to zero, despite our convention to include
a factor of $W_0$ in the definition of $\ha$!  The reason is the following.  If the
underlying theory is $O(a)$ improved, all terms of order $a$ in the SET
(\seef\ Eq.~(\ref{cl5})) vanish,
and there is no need to introduce the spurion $A$ at all at this order.  However,
this does not mean that there are no $O(a^2)$ terms in the SET, so we do need
to introduce $O(a^2)$ spurions as in Eq.~(\ref{spurioncl6}).  It just so happens that we can
``recycle'' the same spurion $A$ at order $a^2$ by taking tensor products as in
Eq.~(\ref{cspurions}).  If underlying lattice theory is not improved, there are also
contributions at order $a^2$ proportional to the $O(a)$ coefficients of Eq.~(\ref{cl5}),
in addition to those proportional to appearing in the SET at order $a^2$.  At the level
of $\D\cl^{(4)}$ both types of contribution at order $a^2$ take the same form,
given in Eq.~(\ref{newp4}).

As an application, let us consider the modifications to the $O(p^4)$ result for the pion
mass, \seef\ Eq.~(\ref{oneloopresults}) \cite{Baeretal}.  First, the overall factor $\hm_\ell$
gets replaced by $\hm_\ell+\ha=2B_0 m'_\ell$, with
$m'_\ell$ the shifted light quark masses of Eq.~(\ref{shift}), and likewise, the
logarithms get replaced by
\begin{equation}
\label{newlog}
L(m_\p^2)\to L(2B_0 m_\ell+2W_0a)=L(2B_0m'_\ell)\ ,
\end{equation}
\etc\  Both of these modifications are a consequence of the fact that the $O(a)$
term~(\ref{l2a}) can be absorbed into the shifted quark masses of Eq.~(\ref{shift}).
In this case, as we already saw above, the quark masses that one determines in
a numerical computation are the shifted $m'$ masses, so the
$O(a)$ shift in Eq.~(\ref{newlog}) is not observable.

Second, the terms proportional to the $L_i$'s in Eq.~(\ref{oneloopresults}) also change.
Not surprisingly, instead of only terms proportional to $L_i m_{quark}^2$, one now also
finds terms of the form $W_i am_{quark}$ and $W'_i a^2$.    At order $a$, the
chiral lagrangian only ``knows" about the shifted quark masses $m'_i$ of Eq.~(\ref{shift}),
as the $O(a)$ operator~(\ref{l2a}) is absorbed into the quark masses before tuning them
to be near the critical value.  The same mechanism does not work at order $a^2$.
One may thus imagine working with quark masses so small that the shifted light
quark mass $m'_\ell$ becomes of
order $a^2\L_{QCD}^2$.  This would imply that the power counting has to be
modified; for instance, the $W'_{6,7,8}$ terms in Eq.~(\ref{newp4}) now become
leading order.  This is the so-called LCE (large cutoff effects) regime; the
regime in which $m_\ell\sim m'_\ell\sim a\L_{QCD}^2$ is often referred to as the
GSM (generic small (quark) mass) regime.\footnote{The LCE regime is also commonly
referred to as the Aoki regime.}  So far, our discussion has been in the
GSM regime.  Before we discuss some of the ChPT results in the LCE regime,
however, we consider the phase diagram of the theory, as it turns out that the
competition between the $O(m'_\ell)$ terms and $O(a^2)$ terms in the chiral
lagrangian can lead to various nontrivial phase structures, depending on the sign of
an $O(a^2)$ LEC.

%%####%%
\subsection{\label{phase} Phase diagram}
%%####%%
So far, we have been expanding $\S$ around the trivial vacuum in order
to calculate various physical quantities, such as in Eq.~(\ref{oneloopresults}).
Indeed, with $\chi=2B_0M$ with $M$ as in Eq.~(\ref{M}) with all quark masses positive,
$\S={\bf 1}$ corresponds to the minimum of the classical potential
energy, given by the second term in Eq.~(\ref{l2again}).  For $a=0$, higher order terms
in the chiral expansion of the potential are smaller, and thus do not change this
observation.

When we include scaling violations, there are two small parameters,
$m_{quark}$ and $a\L_{QCD}^2$.  In the previous section, we took these
two parameters to be of the same order, and again the leading order
nonderivative terms in Eqs.~(\ref{l2again}) and~(\ref{l2a}) constitute the leading-order
classical potential.  As long as the shifted quark mass of Eq.~(\ref{shift}) stays
positive and in the GSM regime, the vacuum remains trivial.  However, if we now
make the shifted quark mass smaller, so that at some point $m'_{quark}\sim
a^2\L_{QCD}^3$, the terms with LECs $W'_{6,7,8}$ in Eq.~(\ref{newp4})
become comparable in size, and should be taken as part of the leading-order
classical potential that determines the vacuum structure of the theory.
This is the regime that we denoted as the LCE regime at the end of
Sec.~\ref{WChPT}
\cite{ShSi} (see also Ref.~\cite{MC}).

The strange quark mass is much larger than the up or down quark masses,
so it is interesting to consider  the case that $m_\ell=m_u=m_d\ll m_s$,
with $m_\ell\sim a^2\L_{QCD}^2$, while $m_s$ stays in the GSM regime.\footnote{I will
drop the primes on the shifted quark masses in the rest of this section.}
We then expect the vacuum value of
$\S$ to take the form
\begin{equation}
\label{Svac}
\S_{vacuum}=\pmatrix{\S_2&0\cr 0&1}\ ,
\end{equation}
in which $\S_2$ denotes an $SU(2)$ valued matrix.  The interesting phase structure
is thus essentially that of a two-flavor theory, and in the rest of this subsection we
will analyze the phase structure for $N_f=2$.
For $N_f=2$,
the classical potential, which consists of the second term in Eq.~(\ref{l2again}),
Eq.~(\ref{l2a}),
and the $W'_{6,7,8}$ terms in Eq.~(\ref{newp4}), becomes
\begin{equation}
\label{potential}
V=-\frac{1}{4}f^2\hm_\ell\;\tr(\S_2+\S_2^\dagger)
-\ha^2(W'_6+\frac{1}{2}W'_8)\left(\tr(\S_2+\S_2^\dagger)\right)^2
+{\rm constant}\ ,
\end{equation}
where we have used the $SU(2)$ relation
\begin{equation}
\label{su2}
\tr(\S_2\S_2+\S_2^\dagger\S_2^\dagger)=\frac{1}{2}\left(\tr(\S_2+\S_2^\dagger)\right)^2
-4\ .
\end{equation}
All other terms in the chiral potential are of higher order.

This potential exhibits an interesting phase structure.  First, consider the case
that $W'\equiv 32(W'_6+\frac{1}{2}W'_8)(2W_0)^2/f^2>0$.  In that case, the potential is
minimized for
\begin{equation}
\label{fo}
\S_2=\left\{
\begin{array}{ll}+1\ , & \mbox{$\hm_\ell>0$}
\\
-1\ , & \mbox{$\hm_\ell<0$}
\end{array}\ \ \ \ (W'>0)\ .\right.
\end{equation}
The pion mass is given (to leading order) by
\begin{equation}
\label{fomesonmass}
m_{\p,LO}^2=2|\hm_\ell|+2a^2W'
\end{equation}
(where $LO$ remind us that this is the pion mass to leading order),
and attains a nonzero minimum value at $\hm_\ell=0$.  The phase transition is
first order, with the vacuum expectation value  $\S_2$ exhibiting a
discontinuous jump across $\hm_\ell=0$.

Equation~(\ref{fomesonmass}) suggests what will happen for $W'<0$.%
\footnote{For $W'=0$ higher order terms in the potential would have to be
considered \cite{SharpeObs}.}  The pion mass-squared turns negative when $|\hm_\ell|<
a^2|W'|$, which signals the spontaneous breakdown of a symmetry.
Indeed, writing
\begin{equation}
\label{s2decomp}
\S_2=\s+i{\vec\t}\cdot{\vec\p}\ ,\ \ \ \ \ \s^2+{\vec\p}^2=1
\end{equation}
(with $\vec\t$ the Pauli matrices), we find that the potential is minimized for
\begin{equation}
\label{so}
\s=\left\{\begin{array}{ll} +1\ , & \mbox{$\hm_\ell\ge a^2|W'|$}
\\ \frac{\hm_\ell}{a^2|W'|}\ , & \mbox{$|\hm_\ell|<a^2|W'|$}
\\ -1\ , & \mbox{$\hm_\ell\le -a^2|W'|$}
\end{array}\ \ \ \ (W'<0)\ .\right.
\end{equation}
A phase of width $a^3$ in lattice units opens up, in which isospin and
parity are spontaneously broken, because of the formation of a pion
condensate \cite{ShSi}.  We can take the condensate in the $\p_3=\p_0$
direction in isospin space, in which case the charged pions become massless,
becoming the exact Goldstone bosons for the breakdown of isospin.
Since the condensate now varies continuously across the phase transition,
this is a second order transition.  The possibility of such a phase was first
noticed by Aoki \cite{Aokiphase}, and provides a possible mechanism for
pions to become exactly massless with Wilson fermions at nonzero lattice
spacing by tuning the quark mass.  At the phase transition the charged pions
are still massless, and because of isospin symmetry at that point, the neutral
pion also has to be massless at the transition.
Note however, that if $W'>0$, pions will not be massless at the phase
transition for nonzero $a$.

There is a lot more than can be said about the phase structure.
We will not do so in these lectures, but
end this section with a few comments:
\begin{itemize}
\item[1.]
An important observation is that the chiral lagrangian does not only provide
us with a systematic expansion in the small infrared scales of the theory
(the quark masses and $a\L_{QCD}^2$), but it also provides nonperturbative
information about the phase structure of the theory.
\item[2.]
In two-flavor QCD with Wilson fermions, one may consider a more general
quark-mass matrix
\begin{equation}
\label{tmM}
M=m_\ell{\bf 1}+i\m\t_3\ .
\end{equation}
This leads to tmQCD, Wilson-fermion QCD with a twisted mass.  In the
continuum, one can perform a chiral rotation to make the mass matrix
equal to $M=\sqrt{m^2+\m^2}{\bf 1}$, but with Wilson fermions there is
no chiral symmetry, and the theory for $\m\ne 0$ differs from the one with
$\m=0$.  We note that for $\m\ne 0$ this generalized mass matrix breaks
isospin symmetry explicitly, and $\m$ thus serves as a ``magnetic field''
pointing the condensate in the $\t_3$ direction in the limit $\m\to 0$
(after the thermodynamic limit has been taken).  It is straightforward to
develop ChPT for this case; one simply substitutes this mass matrix into
the (two-flavor) chiral lagrangian.  For detailed discussions of tmQCD,
see for instance the reviews in Refs.~\cite{SRSNara,Sint}.
\end{itemize}

%%####%%
\subsection{\label{LCE} The LCE regime}
%%####%%
We have seen that $O(a)$ effects, while present for (unimproved) Wilson
fermions, can be completely absorbed into a shifted quark mass in the pion
chiral lagrangian.  While we already considered $O(a^2)$ effects in the pion
masses in the GSM regime toward the end of Sec.~\ref{WChPT}, it is interesting to consider effects from $O(a^2)$
scaling violations
in the LCE regime, where $m'_\ell\sim a^2\L_{QCD}^3$.  We will always work with
$m'_\ell$, \seef\ Eq.~(\ref{shift}), and in this section we will again
restrict ourselves to $N_f=2$,
thinking of the strange quark as heavy.  This makes sense, as we may expect
that numerical computations with very small light quark masses, which may
then indeed turn out to be of order $a^2\L_{QCD}^3$, will become more prominent in the
near future.  The results we consider in this section have been obtained in
Ref.~\cite{ABB}.\footnote{For earlier work, see Ref.~\cite{AokiWChPT}.}
They are valid in the phase with unbroken isospin and parity.
In other words, we will choose $m'_\ell$ such that $m_{\p,LO}^2$
of Eq.~(\ref{fomesonmass}) is always nonnegative.

The first thing to note is that, since our power counting changes relative to the
GSM regime, new terms need to be added to the chiral lagrangian if one
wishes to work to order $p^4$:
\begin{eqnarray}
\label{LCEeq}
\mbox{LCE regime:}&O(p^2): &p^2, m'_\ell, a^2\ ,\\
&O(p^3): &ap^2, am'_\ell, a^3\ ,\nonumber\\
&O(p^4): &p^4, p^2m'_\ell, (m'_\ell)^2, a^2p^2, a^2m'_\ell, a^4\ .\nonumber
\end{eqnarray}
Since we will be interested in the nonanalytic terms at one loop, which follow
from the lowest-order chiral lagrangian, I just refer to Ref.~\cite{ABB} for the
explicit form for all new terms not present in Eq.~(\ref{newp4}), which does already
include all $O(p^2)$ terms.

One finds for the pion mass to order $p^4$ \cite{ABB}:
\begin{eqnarray}
\label{LCEpionmass}
m_\p^2&=\tm_{\p,LO}^2\Biggl\{1&+\ \frac{1}{16\p^2 f^2}\left(
(\tm_{\p,LO}^2-10a^2W')\log\left(\frac{\tm_{\p,LO}^2}{\L^2}\right)\right)\\
&&+\ \mbox{analytic terms proportional to $\tm_{\p,LO}^2$ and $a$}\Biggr\}\ .
\nonumber
\end{eqnarray}
Here $\tm_{\p,LO}^2$ is equal to $m_{\p,LO}^2$ (which is given in
Eq.~(\ref{fomesonmass})) up to $O(a^3)$ and $O(a^4)$
terms that have been absorbed into the definition of the quark mass.  This
corresponds to defining the critical quark mass as the
value where the pion mass vanishes, but one should keep in mind that these
scaling violations are not universal.

This result is interesting because it shows that the continuum value of the
coefficient of the chiral logarithm, which is a prediction of ChPT, can be
significantly modified by scaling violations in the LCE regime.  It also shows that at nonzero
lattice spacing the chiral limit is unphysical: for $\tm_{\p,LO}^2\to 0$ the
chiral logarithm diverges, unless we take the lattice spacing to zero as
well, at a rate not slower than the square root of the quark mass.  Note
however, that if $W'>0$, there is a nonvanishing minimum value of the
pion mass which is of order $a^2$, \seef\ Eq.~(\ref{fomesonmass}), so that
this divergence cannot occur.  In the case $W'<0$ the divergence can happen,
and the only way to make mathematical sense of the result~(\ref{LCEpionmass})
is to resum higher powers of $a^2\log(\tm_{\p,LO}^2)$.\footnote{
It is not clear to me that this would lead to a less divergent result after
resummation for all physical quantities in which such enhanced logarithms may occur,
as it turns out to be the case for
the pion mass \cite{AokiWChPT}.}  Similar effects also show up in
pion scattering lengths \cite{ABB}.\footnote{This reference contains an
interesting idea on how to determine the value of $W'$ from the $I=2$ pion
scattering length.}
Because of the absence of chiral symmetry at nonzero $a$, pion interactions do not vanish for
vanishing external momenta; the leading term is of order $a^2$, and divergent
chiral logarithms occur at order $a^4$.
Note that the results of Ref.~\cite{ABB} are obtained approaching the phase transition
from the phase in which isospin is unbroken.  If, for $W'<0$, one calculates scattering
lengths approaching the phase transition from the other side, one would expect them
to vanish.  The reason is that on that side isospin is broken, and two of the three
pions are thus genuine Goldstone bosons at nonzero $a$, with vanishing
scattering lengths.  Since isospin is restored at the phase transition, all pion
scattering lengths should thus vanish approaching the phase transition from the
phase with broken isospin.  This is not a contradiction: scattering lengths are
discontinuous across the phase transition between a symmetric and a broken
phase.

An important general lesson of this section
is that Wilson ChPT allows us to choose quark
masses so small that $m_\p^2\sim a^2\L_{QCD}^4$, \ie, in a regime where
physical and unphysical effects compete.  In contrast, if we only had
continuum ChPT as a tool, we would have to first extrapolate to the
continuum before doing any chiral fits.  However, the examples we have discussed
make it clear that, in general, we would not want to take the chiral and
continuum limits ``in the wrong order'': if one first takes the physical infrared scale
$m_\p$ to zero at fixed nonzero value of the unphysical infrared scale $a\L_{QCD}^2$,
ChPT tells us that infrared divergences may occur, as we have seen above.

We end this section by observing that the result for the
GSM regime to $O(p^4)$ can be recovered from Eq.~(\ref{LCEpionmass}) by expanding
in $a^2$ and dropping all terms of order $a^3$, $a^2m'_\ell$ and $a^4$.

%%####%%
\subsection{\label{axial} Axial current}
%%####%%
Before leaving Wilson fermions for staggered fermions, let us briefly look at
currents in the presence of scaling violations; as an example I will consider the
nonsinglet axial current, and we will work to order $a$, \ie, we are back in the
GSM regime.
I will highlight some of the arguments presented in
Ref.~\cite{ABS}, which contains references to earlier work, and which
also discusses the
vector current.

The issue is that for $a\ne 0$, the axial current for QCD with Wilson fermions is
not conserved.  Since there is no Noether current at the lattice QCD level, there isn't
one at the SET level, nor in ChPT.    Since there is no conserved current, the
lattice current that is usually employed is the local current\footnote{Or an
improved version of the local current.}
\begin{equation}
\label{axialc}
A^a_\m(x)=\bq(x)\g_\m\g_5T^a q(x)\ ,
\end{equation}
in which the $T^a$ are $SU(3)$ or $SU(2)$ flavor generators.  First, this current
needs a finite renormalization,
in order to match to the continuum-regularized (partially) conserved axial current
\cite{KSBetal},
\begin{equation}
\label{matchaxial}
A^a_{\m,ren}=Z_AA^a_\m\ ,
\end{equation}
with $Z_A$ a renormalization constant determined nonperturbatively by enforcing Ward identities.
Since $Z_A$ is determined nonperturbatively, it includes scaling violations.
Then, since $A^a_\m$ is not a Noether current following from the lattice action, the
effective current that represents the axial current in the SET also does not follow from
the SET action; instead, it is just an ``external'' operator for which we have to
find the corresponding expression in the SET.  To order $a$, the ``Symanzik''
current that represents the lattice current $A^a_\m$ in the SET
is\footnote{See the lectures by Peter Weisz at this school.}
\begin{eqnarray}
\label{axialSym}
A^a_{\m,SET}&=&\frac{1}{Z^0_A}(1+\bb_A am)\left(A^a_{\m,cont}+a\bc_A\partial_\m P^a_{cont}\right)\ ,\\
A^a_{\m,cont}&=&\bq\g_\m T^a q\ ,\ \ \ \ \ P^a_{cont}=\bq\g_5 T^a q\ ,\nonumber
\end{eqnarray}
in which $\bb_A$ and $\bc_A$ are Symanzik coefficients that depend on the
underlying lattice action; {\it ``cont''} indicates continuum operators.  (In a fully $O(a)$ improved theory both $\bb_A$ and $\bc_A$ vanish.)
By combining Eqs.~(\ref{matchaxial}) and~(\ref{axialSym}), we see that the overall
multiplicative renormalization factor is $(1+\bb_A am)Z_A/Z^0_A$.  $Z^0_A$ is the
all-orders perturbative matching factor needed to convert a lattice current
into the properly normalized (partially conserved) continuum current; we have
that $Z_A=Z^0_A+O(a)$.\footnote{Only the ratio $Z_A/Z^0_A$ appears when we
combine Eqs.~(\ref{matchaxial}) and~(\ref{axialSym}).}

At this stage, two steps need to be carried out to translate this current to
ChPT.  First, the operator $A^a_{\m,cont}+a\bc_A\partial_\m P^a_{cont}$ needs to be
mapped into ChPT.  Then, the
renormalization factor following from Eqs.~(\ref{matchaxial}) and~(\ref{axialSym})
has to be determined in exactly the same way as the lattice current
is matched to the renormalized continuum current in an actual lattice computation.
This matching has to be done to order $a$ as well, if one consistently wants to include
all $O(a)$ corrections in the chiral theory.  Since this involves the matching of
scaling violations, one does not expect the outcome of this second step to be
universal, and indeed, Ref.~\cite{ABS} finds that already at order $a$ the matching depends
on the precise correlation functions used for the matching, with a nonuniversal result for
the ratio $Z_A/Z^0_A$; I refer to Ref.~\cite{ABS} for further discussion of this second step.
Here, I will address the first step: the mapping to operators in the chiral theory, to order $a$.%
\footnote{This step was already carried out in Ref.~\cite{SW2} using the source method;
see also Refs.~\cite{SRSNara,ABS}.}

In order to find the ChPT expression for the axial current, Ref.~\cite{ABS} first observes that,
in the case that $\bc_A=0$, $\bb_A=0$ and $Z_A^0=1$, the Symanzik current is just the continuum axial current,
which can be rotated into the vector continuum current by a chiral transformation. This
fixes the $\bc_A=0$ part of the axial current in ChPT in terms of the vector
current.  It turns out that the vector current is given by the Noether
current of the chiral theory, because the $O(a)$ terms in the vector current are of the
form $a\partial_\nu(\bq\s_{\m\n}T^a q)$, which is automatically conserved.\footnote{There
is a vector Noether current on the lattice \cite{KSBetal}, but often the local vector current is
used instead.}  The Noether
current following from $\cl^{(2)}+\D\cl^{(4)}$ (\seef\ Eqs.~(\ref{l2again}) and~(\ref{newp4}))
is
\begin{equation}
\label{vectorChPT}
V_\m=( J^R_\m+J^L_\m)\left(1+\frac{8W_4}{f^2}\;\tr(\hA^\dagger\S+\S^\dagger \hA)\right)
+\frac{4W_5}{f^2}\left\{J^R_\m+J^L_\m,\hA^\dagger\S+\S^\dagger \hA\right\}\ ,
\end{equation}
with $J^{R,L}_\m$ defined to lowest order in Eq.~(\ref{conscurrents}); we omitted
terms coming from $\cl^{(4)}$ of Eq.~(\ref{l4}).   Here $\hA\equiv 2W_0A$ with $A$ the spurion
of Eq.~(\ref{Aspurion}).   We should set $\hA=\ha$ of course, but only {\em after}
doing the chiral rotation to find the
$\bc_A=0$ part of the axial current.  We thus find
\begin{equation}
\label{axialChPT}
A_\m=( J^R_\m-J^L_\m)\left(1+\frac{8\ha W_4}{f^2}\;\tr(\S+\S^\dagger)\right)
+\frac{4\ha W_5}{f^2}\left\{J^R_\m-J^L_\m,\S+\S^\dagger\right\}\ \ \ (\bc_A=0)\ ,
\end{equation}
where now we have set $\hA=\ha$.

Finally, we have to worry about the $\bc_A$ term in Eq.~(\ref{axialSym}).  This is
rather simple, because for this we need the lowest-order ChPT operator for the
pseudoscalar density, which is $-i(\S-\S^\dagger)$.   We need to introduce a new
LEC $W_A$ for this contribution, since, unlike the contribution in Eq.~(\ref{axialChPT}),
this term does not relate in any way to the Symanzik effective action, and the same
is thus true in ChPT.   We thus find for
the complete expression for the axial current to order $a$:
\begin{eqnarray}
\label{axialChPTagain}
A_\m&=&( J^R_\m-J^L_\m)\left(1+\frac{8\ha W_4}{f^2}\;\tr(\S+\S^\dagger)\right)
+\frac{4\ha W_5}{f^2}\left\{J^R_\m-J^L_\m,\S+\S^\dagger\right\}\\
&&\hspace{6cm}
-4i\ha W_A\bc_A\partial_\m(\S-\S^\dagger)\ \ \ (\bc_A\ne 0)\ ,\nonumber
\end{eqnarray}
where, following Ref.~\cite{ABS}, I kept $\bc_A$ outside $W_A$.\footnote{This result
can now be multiplied again by the factor $(1+\bb_A a)/Z^0_A$.}   In closing this
section, I should emphasize that the derivation given here is the ``quick and dirty''
derivation --- for a careful analysis of both vector and axial currents, I refer to
Ref.~\cite{ABS}.

%%####%%
\subsection{\label{staggered} Staggered fermions}
%%####%%
Before developing ChPT for staggered QCD, I give a brief review of the
definition and symmetry properties of lattice QCD with staggered fermions,
to make these lectures more self-contained.\footnote{
For more introduction, see the lectures by Pilar Hern\'andez at this school,
as well as the recent review by the MILC collaboration \cite{MILCreview}.}

%%####%%
\subsubsection{\label{review} Brief review}
%%####%%

\vspace*{0mm}
The staggered action is (for the time being I will set the lattice spacing
$a=1$, but I will restore it when we get to the SET and ChPT)
\begin{eqnarray}
\label{stagac}
&&\hspace{-4mm}S=\sum_{x,\m,i}\frac{1}{2}\;\eta_\m(x)\bchi_i(x)\left[U_\m(x)\c_i(x+\m)-
U^\dagger_\m(x-\m)\c_i(x-\m)\right]+\sum_{x,i}m_i\bchi_i(x)\c_i(x)\ ,\nonumber\\
&&\hspace{-4mm}\eta_\m(x)=(-1)^{x_1+\dots+x_{\m-1}}\ .
\end{eqnarray}
Here $\c_i$ is a lattice fermion field with an explicit flavor index $i
=1,\dots,N_f$,
an implicit color index, and {\em no} Dirac index; $U_\m(x)$ are the
gauge-field links, with values in $SU(3)_{color}$.

The action~(\ref{stagac}) has species doubling, and because of the hypercubic
structure of the theory, this doubling is sixteen-fold.
The sixteen components which emerge in the continuum limit can be accounted
for by an index pair $\a,a$, with each index running over the values $1,\dots,4$.
It turns out that the first index can be interpreted as a Dirac index, while
the second index constitutes an additional flavor label; the continuum
limit looks like \cite{STW,GS,Saclay}
\begin{equation}
\label{stagcl}
S_{cont}=\int d^4x\left(\bq_{i\a a}\g_{\m,\a\b}D_\m q_{i\b a}
+m_i\bq_{i\a a}q_{i\a a}\right)\ ,
\end{equation}
in which repeated indices are summed.  In brief, this structure emerges
as follows.  Species doublers live near the corners
\begin{equation}
\label{corners}
\p_A\in\left\{(0,0,0,0),(\p,0,0,0),\dots,(\p,\p,\p,\p)\right\}\ ,
\ \ \ A=1,\dots,16\ ,
\end{equation}
in the Brillouin zone.
It thus makes sense to split the lattice momenta $p$ into
\begin{equation}
\label{split}
p=\p_A+k\ ,\ \ \ \ \ -\frac{\p}{2}\le k_\m\le\frac{\p}{2}\ .
\end{equation}
We thus identify sixteen different fermion fields in momentum space,
$\c(p)=\c(\p_A+k)\equiv\c_A(k)$, which represent the sixteen doublers.
Since the phases $\eta_\m(x)$ insert momenta with values in the set~(\ref{corners}),
it follows that the operation
\begin{equation}
\label{T}
T_{\pm\m}:\ \c_i(x)\to\eta_\m(x)\c_i(x\pm\m)
\end{equation}
mixes the sixteen doublers.  In momentum space, after a basis transformation,
the operations $T_\m$ can be represented by
\begin{equation}
\label{Tmom}
T_{\pm\m}:\ q_{i\a a}(k)\to \g_{\m,\a\b}e^{\pm ik_\m}q_{i\b a}(k)\ ,
\end{equation}
with the $\g_\m$ a set of Dirac matrices.  That these four matrices satisfy
a Dirac algebra follows from the fact that the $T_\m$ anticommute:
\begin{equation}
\label{TT}
T_\m T_\n = -T_\n T_\m\ ,\ \ \ \m\ne\n\ ,
\end{equation}
and that
\begin{equation}
\label{Tsq}
T^2_{\pm\m}:\ \c_i(x)\to\c_i(x\pm 2\m) .
\end{equation}
Using these ingredients, we can see how Eq.~(\ref{stagcl}) follows from Eq.~(\ref{stagac}).
Using Eqs.~(\ref{T}) and~(\ref{Tmom}), the free kinetic term in Eq.~(\ref{stagac}) can be
written as (for one flavor)
\begin{equation}
\label{deriv}
\sum_{x,\m}\frac{1}{2}\bchi(x)\left(T_\m-T_{-\m}\right)\c(x)
=\int\frac{d^4k}{(2\p)^4}\;\sum_\m\bq(k)\g_\m\;i\sin(k_\m) q(k)\ .
\end{equation}
In the continuum limit, $\sin(k_\m)\to k_\m$; note that $k$ is restricted to the
reduced Brillouin zone (\seef\ Eq.~(\ref{split})), on which no species doublers reside.
Finally, since Eq.~(\ref{stagac}) is gauge invariant, the continuum limit~(\ref{stagcl})
has to be gauge invariant as well.  It follows that, at least classically, Eq.~(\ref{stagcl})
is the continuum limit of Eq.~(\ref{stagac}).

The observation that the continuum limit of staggered QCD is given by Eq.~(\ref{stagcl})
has been proven to one loop in perturbation theory, and this result has been
understood in terms of the exact lattice symmetries, making it likely that
this proof can be extended to all orders \cite{GS}.  There is also extensive numerical
evidence \cite{MILCreview}.  We conclude that the lattice theory defined by Eq.~(\ref{stagac})
constitutes a regularization of QCD with $4N_f$ flavors.  In modern
parlance, ``flavor'' is now usually used to denote the explicit multiplicity
associated with the flavor index $i$, while ``taste'' is used to denote
the implicit multiplicity associated with species doubling.  If indeed the
continuum limit is given by Eq.~(\ref{stagcl}), this implies that a full
$U(4)_{taste}$ symmetry emerges in the continuum limit, for each staggered flavor.
This taste symmetry is not
present in the lattice theory --- we will return to this shortly.

We note that the taste degeneracy is somewhat of an embarrassment, if one
wishes to use the flavor index $i$ as physical flavor, as is usually
done in practice.  While this is not a problem for valence quarks
(one simply picks out, say, taste $a=1$ in order to construct operators,
for example), this means that there are too many sea quarks --- too many
quarks on the internal fermion loops.  This problem is solved in practice
by taking the (positive) fourth root of the staggered determinant for
each physical flavor.  While this makes sense intuitively, it is a serious
modification of the theory, which entails a
violation of locality and unitarity at nonzero lattice spacing \cite{BGSobs}.
Much work has recently gone into
attempts to show that the procedure is nevertheless correct in the sense
that it yields the desired continuum limit, in which EFT techniques in
fact play an important role.  There is no space here to discuss this
very interesting (and important!) topic; I refer to the recent reviews
of Refs.~\cite{MGreview,MILCreview}.  Here I will focus on the construction
of staggered ChPT for unrooted staggered fermions.

To construct the SET for staggered QCD it is important to
use all exact symmetries of the lattice theory, in order to restrict the
form of the operators that will appear in the Symanzik expansion, Eq.~(\ref{Sym}).
The action~(\ref{stagac}) has an exact vectorlike $SU(N_f)$ symmetry,
broken only if the quark masses $m_i$ are nondegenerate.
Furthermore, there are spacetime symmetries.  In addition to hypercubic
rotations and parity, the action is invariant under shift symmetry \cite{GS,vdDS}:
\begin{eqnarray}
\label{shiftsym}
&S_{\pm\m}:\ \c_i(x)&\to\z_\m(x)\c_i(x\pm\m)\ ,\ \ \ \ \
\z_\m(x)=(-1)^{x_{\m+1}+\dots+x_4}\ ,\\
&S_{\pm\m}:\ U_\n(x)&\to U_\n(x\pm\m)\ ,\nonumber
\end{eqnarray}
because of the fact that
\begin{equation}
\label{ST}
\z_\n(x)\eta_\m(x)\z_\n(x+\m)=\eta_\m(x+\n)\ .
\end{equation}
On the basis $q_{i\a a}$ shift symmetry is represented by
\begin{equation}
\label{Smom}
S_{\pm\m}:\ q_{i\a a}(k)\to \x_{\m,ab}e^{\pm ik_\m}q_{i\a b}(k)\ ,
\end{equation}
with $\x_\m$ a different set of Dirac matrices (like Eq.~(\ref{TT}) we have that
$S_\m S_\n=-S_\n S_\m$ for $\m\ne\n$) acting on the taste
index, commuting with the $\g_\n$ of Eq.~(\ref{Tmom}), because of Eq.~(\ref{ST}).

Finally,
if $m_i=0$ for a certain flavor, there is a chiral symmetry (``$U(1)_\e$
symmetry'') \cite{KaSm}
\begin{eqnarray}
\label{u1eps}
\c_i(x)&\to& e^{i\theta_i\e(x)}\c_i(x)\ ,\ \ \ \ \
\bchi_i(x)\to e^{i\theta_i\e(x)}\bchi_i(x)\ ,\\
\e(x)&=&(-1)^{x_1+x_2+x_3+x_4}\ .\nonumber
\end{eqnarray}
Since this symmetry is broken by a nonzero $m_i$, it should be interpreted
as an axial symmetry of the theory.  In addition, since it is an exact symmetry of the lattice
theory (for $m_i=0$) it has to correspond to a nonsinglet axial symmetry in the
continuum limit, because the continuum singlet axial symmetry is anomalous.
If all quarks are massless,
it is not difficult to show that this enlarges to an exact $U(N_f)_L\times
U(N_f)_R$ symmetry.\footnote{$U(N_f)$ instead of $SU(N_f)$, because,
as we just noted,
the axial symmetries in this group are all nonsinglet!}  The axial
symmetries in this group are nonsinglet symmetries on the basis $q_{i\a a}$,
because the phase $\e(x)$ inserts momentum, and thus acts nontrivially
on the index pair $\a,a$:
Since
\begin{equation}
\label{epsetazeta}
\e(x)\c_i(x)=\prod_\m T_{-\m}S_\m\c_i(x)\ ,
\end{equation}
it follows that $U(1)_\e$ symmetry is represented on the $q_{\a a}$ basis
as
\begin{equation}
\label{epsmom}
q_{i\a a}\to \left(e^{i\theta_i\g_5\x_5}\right)_{\a a,\b b}q_{i\b b}\ .
\end{equation}

In staggered QCD there is extensive evidence that these axial symmetries
are spontaneously broken, and for each such broken symmetry there is thus
an exact Nambu--Goldstone boson.  When $m_i$ is turned on, this NG boson picks up a
mass-squared proportional to $m_i$, as we will see below.  Contrary
to the case of Wilson fermions, we thus know the critical values of the
quark masses $m_i$ --- the massless limit is at $m_i=0$.  Interpolating fields
for these Nambu--Goldstone bosons are $\e(x)\bchi_i(x)\c_j(x)$ in position
space, or $\bq_i\g_5\x_5 q_j$ in momentum space.

As can be seen from Eq.~(\ref{stagcl}), these lattice symmetries enlarge to
$SU(4N_f)_L\times SU(4N_f)_R$ (and the usual euclidean spacetime symmetries)
in the continuum limit.  In the continuum limit, one thus expects
$(4N_f)^2-1$ pions, associated with the spontaneous breakdown
$SU(4N_f)_L\times SU(4N_f)_R\to SU(4N_f)_V$.  At nonzero $a$, there are
only $N_f^2$ exact Nambu--Goldstone bosons; the rest will pick up additional
contributions to their mass-squared of order $a^2$, as we will see in
more detail below.  These lattice-artifact mass splittings are usually
referred to as ``taste splittings,'' since they occur because of the
breakdown of $U(4)_{taste}$ on the lattice.  Indeed, this taste breaking is
seen on the lattice very clearly in the meson spectrum, which is shown
in Fig.~\ref{fig:MILC}.
Clearly, we would like to understand these results with help of a ChPT
framework.

%%%%%%%%%%%%%%%%%%%
%%% MILC plots
\begin{figure}[t!]
%\vspace*{-5ex}
%\hspace*{5ex}
%
\begin{picture}(75,55)(10,0)
  \put(53,-16){$(a)$}
  \put(20,-10){\includegraphics*[height=6.5cm]{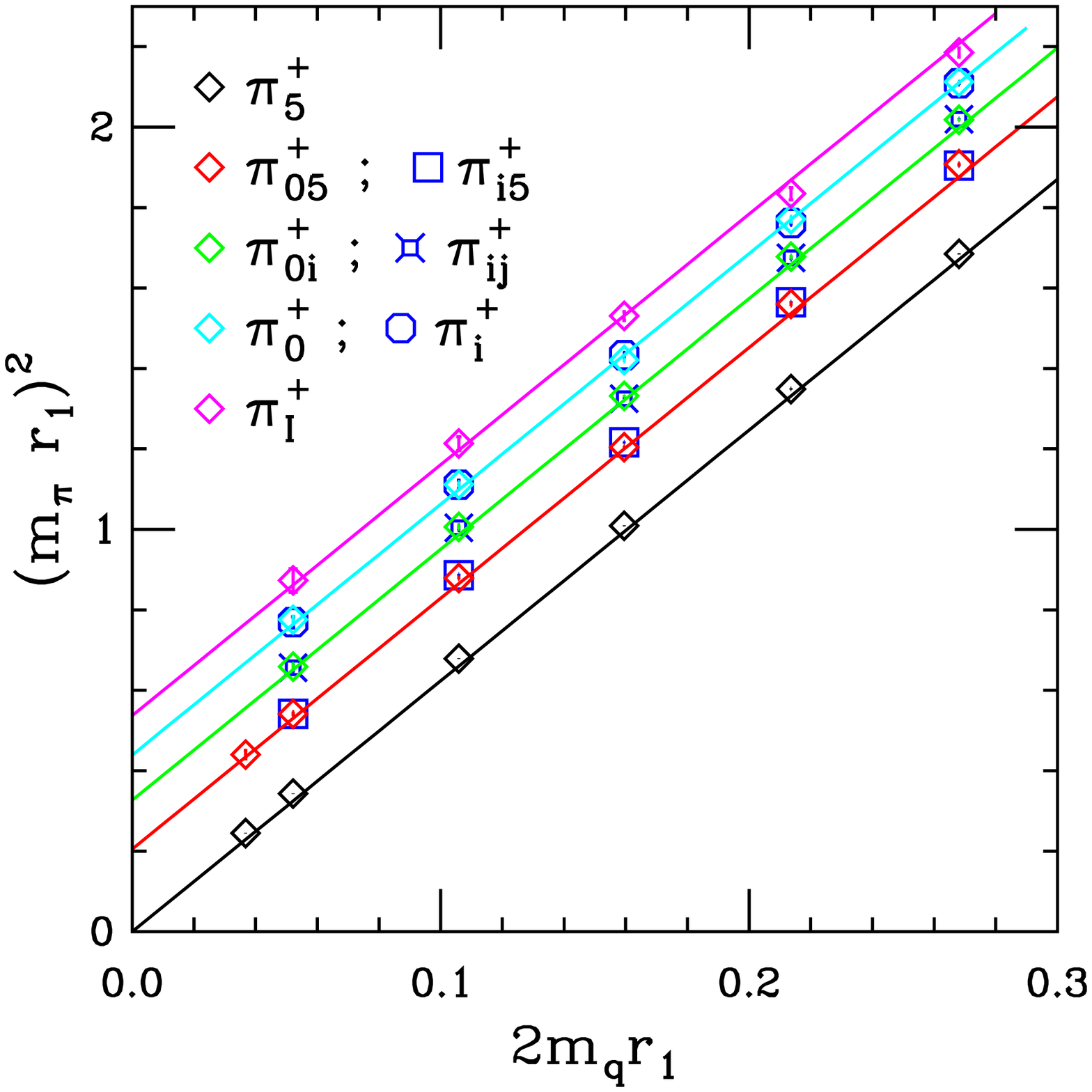}}
\end{picture}
%\hspace*{3ex}
%
\begin{picture}(85,55)(15,0)
  \put(53,-16){$(b)$}
  \put(20,-9){\includegraphics*[height=6.5cm]{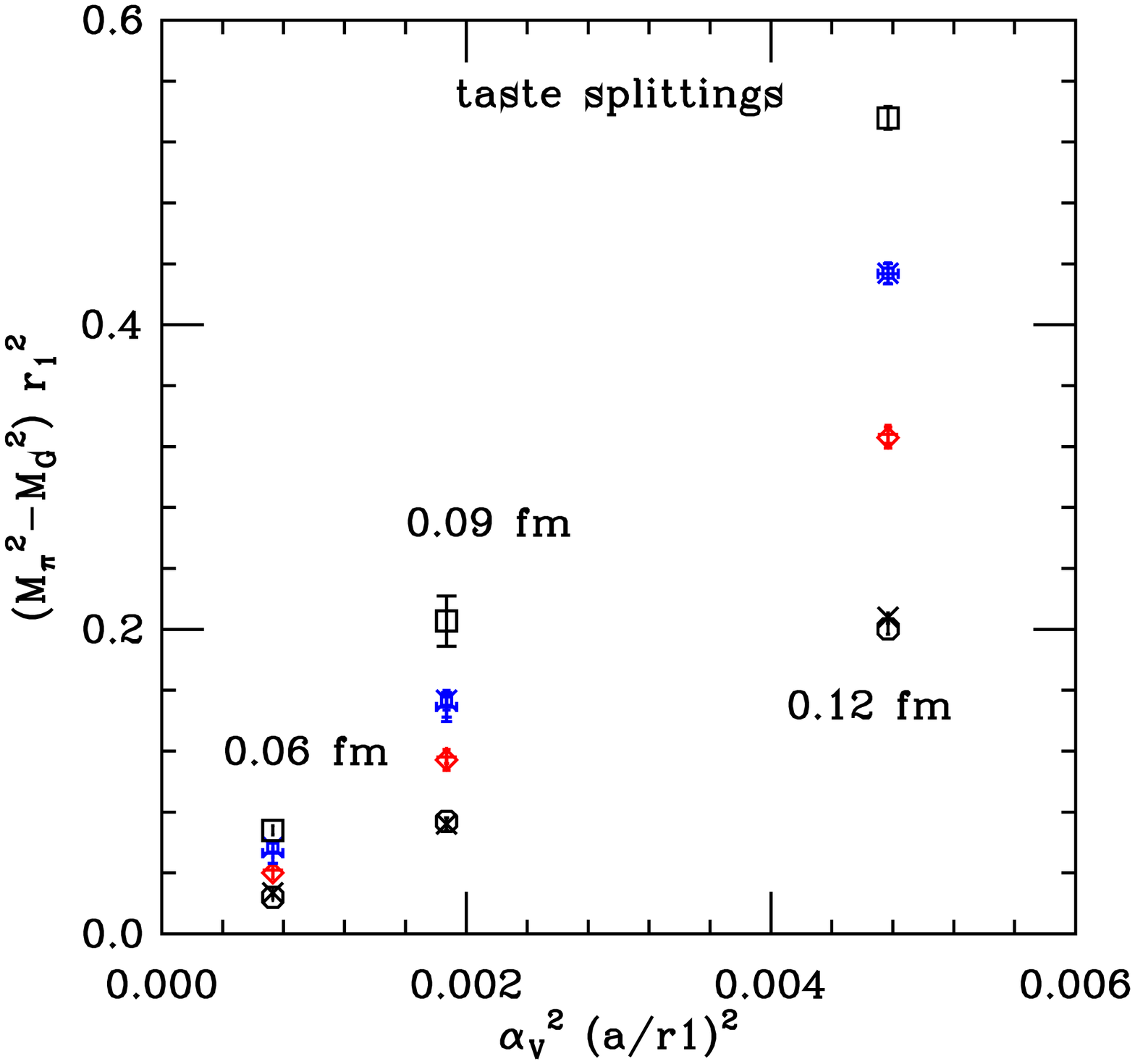}}
\end{picture}
\vspace*{9ex}
\begin{quotation}
\floatcaption{fig:MILC}{({\it a}) Taste splittings between pions made out of an
up and (anti)down staggered quarks, in different representations of the
staggered symmetry group, at fixed lattice spacing, $a=0.125$~fm.  The
black points correspond to the exact Nambu--Goldstone boson, lines are chiral
fits using staggered ChPT.
({\it b}) Taste splittings between the various nonexact Nambu--Goldstone bosons
and the exact Namubu--Goldstone boson as a function of lattice spacing.
These figures show the behavior predicted by Eq.~(\ref{stagmesmasses}).
{}From Ref.~\cite{MILC2004}, to which I refer for a detailed explanation.}
\end{quotation}
\vspace*{-1ex}
\end{figure}
%%%%%%%%%%%%%%%%%%%

%%####%%
\subsubsection{\label{ChPT} Staggered ChPT}
%%####%%

\vspace*{0mm}
The SET is a continuum theory.  This means that it is invariant under
translations over an arbirary distance; in particular, it is invariant
under translations over a distance $a$ in any direction.\footnote{In contrast, staggered
QCD is only invariant under ordinary translations over a distance $2a$ in
any of the principal lattice directions.}  It is also invariant under
any of the exact lattice symmetries.  We may thus combine a shift $S_\m$
in the positive $\m$ direction by a translation over a distance $a$ in the
negative $\m$ direction.  Combining these two, we see that the SET is invariant
under \cite{BGSset,LS}
\begin{equation}
\label{st}
q_i(k)\to\x_\m q_i(k)\ .
\end{equation}
At the level of the SET, the ``translation'' and ``taste'' parts of shift
symmetry thus decouple, and we find that the SET thus has to be invariant
under the $32$-element group $\G_4$ of discrete taste transformations generated by
$\x_\m$, which is a subgroup
of $U(4)_{taste}$.\footnote{Note that this group does not become larger due to the presence of
$N_f$ flavors, because the gluon field also transforms nontrivially
under shifts, \seef\  Eq.~(\ref{shiftsym}).  Thus all quark flavors must be shifted
together.  But because $U_\mu$ shifts by a pure translation, the
analogue of Eq.~(\ref{st}) for gluons is trivial:  $U_\m(p) \to U_\m(p)$.}  We can thus use this group
in order to restrict operators at each order in the Symanzik expansion,
along with the other exact lattice symmetries.

The first term in the Symanzik expansion, $\cl^{(4)}_S$, is given by (the
integrand of) Eq.~(\ref{stagcl}).
There are no dimension-five operators that can be constructed from the
quark fields $q$ and $\bq$ that respect the lattice symmetries \cite{SRSdim5}.
 Ignoring
$U(1)_\e$ symmetry, the possible dimension five operators are those of
Eq.~(\ref{cl5}). The first two
operators in this list are immediately excluded by Eq.~(\ref{epsmom}).  The three
other terms involve powers of the quark mass, and in order to make use
of $U(1)_\e$ symmetry, we thus have make the quark mass a spurion,
as in Eqs.~(\ref{QCD}) and~(\ref{spurion}),
but we now do it with $L$ and $R$ projectors on the quark fields
defined with $\g_5\x_5$, instead of $\g_5$.  It is then not difficult to see that also
the three last operators in Eq.~(\ref{cl5}) are excluded.

This implies that scaling violations for staggered QCD start at order $a^2$.
In fact, before we start the discussion of $\cl^{(6)}$, we should rethink
power counting.  If we use  GSM power counting,
all terms of order $a^2$ would be $O(p^4)$.  However, it turns out that
scaling violations with staggered fermions, even though formally $O(a^2)$,
are numerically rather large, even with improved staggered actions, at
current values of the lattice spacing.\footnote{This state of affairs may
change in the relatively near future, with more highly improved lattice
actions, and even smaller lattice spacings.}  From the taste splittings
in pseudoscalar mesons, it seems that the contribution of the quark
masses and $O(a^2)$ scaling violations to their masses is roughly equal,
and we will therefore use a power counting in which $p^2\sim m_{quark}
\L_{QCD}\sim a^2\L_{QCD}^4$.  This means that $O(a^2)$ effects are
of leading order in ChPT, and we would like to construct at least the
$O(a^2)$ part of the chiral lagrangian, because from the $O(p^2)$
lagrangian we can obtain the $O(p^2)$
and nonanalytic $O(p^4)$ parts of physical quantities in terms of
quark masses and the lattice spacing.  We thus set out to find the
$O(a^2)$ part of $\cl^{(2)}$, \seef\ Eq.~(\ref{l2again}).  This new part
contains no quark masses, and no derivatives, each of which would make
such terms higher order in the chiral expansion.

The $O(a^2)$ part of the SET, $\cl^{(6)}$,  contains a large number of terms
\cite{LS,AB}, and I will only give some examples here.  First, there are
the purely gluonic terms which we already discussed in the context of
Wilson ChPT, and thus do not have to revisit again.  Similarly, fermion
bilinears in $\cl^{(6)}_S$ do not lead to any new terms at the desired
order in ChPT \cite{LS}.  The remaining terms are all four-fermion
operators.

%%%%%%%%%%%%%%%%%%%
%%% four-fermion vertex
\begin{figure}
%\vspace*{-5ex}
\hspace*{5ex}
\begin{picture}(75,55)(0,0)
\put(10,0){\includegraphics*[height=5cm]{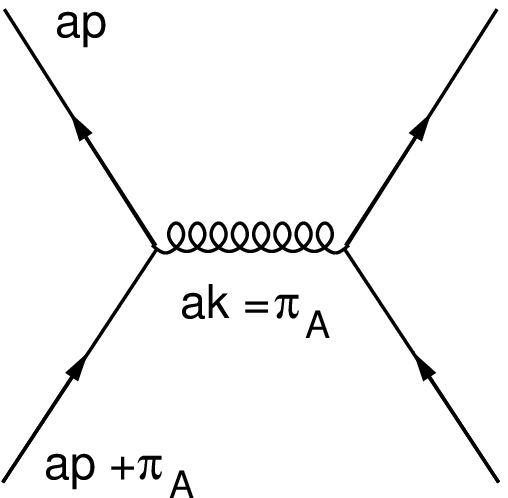}}
\end{picture}
\vspace*{0ex}
\begin{quotation}
\floatcaption{fig:4point}{Generation of taste-breaking four-fermion vertices through
short-distance gluon exchange.}
\end{quotation}
\vspace*{-4ex}
\end{figure}
%%%%%%%%%%%%%%%%%%%

Taste-breaking, short distance four-fermion operators occur in the theory
because of diagrams as shown in Fig.~\ref{fig:4point}.
The gluon carries a momentum near one of the values in
Eq.~(\ref{corners}), with $n>0$ components near $\p/a$, thus changing the taste
of the staggered quarks at both vertices.  For such gluon momenta,
we have that the gluon propagator
\begin{equation}
\label{gluonprop}
\frac{1}{\sum_\m\frac{4}{a^2}\sin^2{(\frac{1}{2}ap_\m)}}\approx \frac{a^2}{4n}\ .
\end{equation}
Effectively, this generates a four-fermion operator in the SET (in which
gluons with momenta of order $1/a$ have been integrated out), for instance
of the form\footnote{Four-fermion operators can be color unmixed or mixed.
For instance, Eq.~(\ref{exff}) can take the form $(\bq_{ia}\x_{5\n} q_{ia})(\bq_{jb}\x_{\n 5} q_{jb})$
or $(\bq_{ia}\x_{5\n} q_{ib})(\bq_{jb}\x_{\n 5} q_{ja})$, where $a$ and $b$ are color
indices.   We see that color indices can be contracted in two different ways.
They will be omitted here, because it makes no difference for our analysis.}
\begin{equation}
\label{exff}
\co_1=(\bq_{i}\x_{5\n} q_{i})(\bq_{j}\x_{\n 5} q_{j})\ ,
\end{equation}
in which $-\x_{\n 5}=\x_{5\n}=\x_5\x_\n$, and
all repeated indices are summed.  This operator is invariant under
the group $\G_4$, but not under the full continuum taste group $U(4)_{taste}$
(it is, in fact, invariant under $SO(4)_{taste}\subset U(4)_{taste}$,
accidentally).  Since the gluon carries no flavor, the flavor index structure
has to be as indicated.  Note that the spin~$\otimes$~taste matrix appearing
in each quark bilinear has to correspond to an odd number of applications
of $T_\m$ or $S_\m$, because gluons couple only to the kinetic term in
Eq.~(\ref{stagac}), which is invariant under $U(1)_\e$.

Note that the appearance of four-fermion operators due to the exchange of
``heavy'' gluons is analogous to the appearance of four-fermion terms with
coefficients of order $g^2_{weak}/M_W^2$ (the
Fermi EFT) in the Standard Model at low energy because of the exchange of
the heavy $W$ and $Z$ bosons.  Likewise, in the case of hand, the four-fermion
operators in the SET are of order $\a_s$, the strong ``fine-structure'' constant.
Light gluons, with momenta $p\ll 1/a$ are still present in the SET, \ie, they have
not been integrated out.

The spin matrices $\g_\m$ can also appear in these four-fermion operators.
This can happen in two ways, here is an example of each:
%\begin{subequations}
%\label{anex}
\begin{eqnarray}
\label{anexa}
\co_2&=&\sum_{\m}(\bq_{i}\g_\m\x_5 q_{i})(\bq_{j}\g_\m\x_5 q_{j})\ ,\\
\label{anexb}
\co_3&=&\sum_{\m<\n}(\bq_{i}\g_\n\x_{\m\n}q_{i})(\bq_{j}\g_\n\x_{\n\m} q_{j})\ ,
\end{eqnarray}
%\end{subequations}
in which $\x_{\m\n}=\x_\m\x_\n$ ($\m\ne\n$), and
where we now showed the sums over Lorentz indices explicitly.  As before,
neither of these operators is invariant under $U(4)_{taste}$, and both
are invariant under $\G_4$ (as well as all other lattice symmetries,
in particular hypercubic rotations).  The first operator is separately
invariant under $SO(4)_{euclidean}\times SO(4)_{taste}$, the second is not.
Operators of the second type can only be represented
in ChPT with at least two derivatives (and one factor $a^2$) \cite{SvdWstag},
which are of order $p^4$.\footnote{Now both derivatives and $\x_\m$ matrices
can provide Lorentz indices, so the argument following Eq.~(\ref{gluonl6}) does
not apply.}  Only
operators of the first type have $O(p^2)$ representations in ChPT.  An
interesting observation is thus that at lowest order staggered ChPT
therefore has $SO(4)_{taste}$ symmetry, and not just the required
$\G_4$ symmetry \cite{LS}.

Finally, we need to make the transition from the SET to ChPT, for
operators such as~(\ref{exff}) and~(\ref{anexa}).  For $N_f=3$, our theory
is now an $SU(12)$ theory, because of the three flavors times four tastes.
The nonlinear field
\begin{equation}
\label{stagS}
\S_{ia,jb}\sim q_{iaL}\bq_{jbR}\ ,
\end{equation}
is thus an $SU(12)$ valued field, with an index structure as shown.
In order to streamline our
notation, we introduce matrices \cite{AB}
\begin{equation}
\label{genxi}
\x_\m^{(3)}=\pmatrix{\x_\m &0&0\cr 0&\x_\m&0\cr 0&0&\x_\m}\ ,
\end{equation}
so that we can write for example Eq.~(\ref{exff}) as
$(\bq\x^{(3)}_{5\n} q)(\bq\x^{(3)}_{\n 5} q)$.   We now rewrite
\begin{eqnarray}
\label{stagspur}
-(\bq\x^{(3)}_{5\n} q)(\bq\x^{(3)}_{\n 5} q)&=&
(\bq_R\x^{(3)}_{5\n} q_L+\bq_L\x^{(3)}_{5\n} q_R)^2\\
&\to&(\bq_R X_R q_L+\bq_L X_L q_R)^2\ ,\nonumber
\end{eqnarray}
where in the last line we introduced two spurions transforming as
\begin{equation}
\label{Xtrans}
X_L\to U_LX_LU_R^\dagger\ ,\ \ \ \ \ X_R\to U_RX_RU_L^\dagger\ ,
\end{equation}
while they are interchanged by parity.  With the field $\S$ we can construct
three types of terms in ChPT with these spurion fields that do not contain
the quark mass matrix or any derivatives:
\begin{equation}
\label{ChPTo1X}
(\bq\x^{(3)}_{5\n} q)(\bq\x^{(3)}_{\n 5} q)\to\left\{
\begin{array}{ll}
&\tr(X_R\S)\;\tr(X_L\S^\dagger)\ ,\nonumber\\
&\Bigl(\tr(X_R\S)\Bigr)^2+\Bigl(\tr(X_L\S^\dagger)\Bigr)^2\ ,\\
&\tr(X_R\S X_R\S)+\tr(X_L\S^\dagger X_L\S^\dagger)\ ,\nonumber
\end{array}\right.
\end{equation}
which, setting the spurions equal to the values that reproduce $\co_1$, gives the
three ChPT operators
\begin{equation}
\label{ChPTo1}
\tr(\x^{(3)}_{5\n}\S)\;\tr(\x^{(3)}_{5\n}\S^\dagger)\ ,\ \ \ \
\left(\tr(\x^{(3)}_{5\n}\S)\right)^2+\left(\tr(\x^{(3)}_{5\n}\S^\dagger)\right)^2\ ,\ \ \ \
\tr(\x^{(3)}_{5\n}\S \x^{(3)}_{5\n}\S)+\tr(\x^{(3)}_{5\n}\S^\dagger \x^{(3)}_{5\n}\S^\dagger)\ ,
\end{equation}
in which sums over the index $\n$ are understood.

Introducing spurions $Y_{L,R}$ for the taste matrices $\x^{(3)}_5$ in
$\co_2$, which transform as $Y_{L,R}\to U_{L,R}Y_{L,R}U_{L,R}^\dagger$,
and working through a similar exercise, one finds the (unique) ChPT
operator representing $\co_2$:
\begin{equation}
\label{ChPT02}
\sum_\m(\bq_{i}\g_\m\x^{(3)}_5 q_{i})(\bq_{j}\g_\m\x^{(3)}_5 q_{j})
\to \tr(\x^{(3)}_5\S\x^{(3)}_5\S^\dagger)\ .
\end{equation}

Referring to Ref.~\cite{AB} for further details, we quote the $O(a^2)$
contribution to the chiral lagrangian, to be added to Eq.~(\ref{l2again}).
Writing this part as $a^2\cv=a^2(\cu+\cu')$, one finds that
\begin{eqnarray}
\label{stagpots}
-\cu&=&C_1\;\tr\left(\x^{(3)}_5\S\x^{(3)}_5\S^\dagger\right)
+\frac{1}{2}C_3
\sum_\n\left[\tr\left(\x^{(3)}_\n\S\x^{(3)}_\n\S\right)+\mbox{h.c.}\right]\\
&&+\frac{1}{2}C_4
\sum_\n\left[\tr\left(\x^{(3)}_{5\n}\S\x^{(3)}_{\n 5}\S\right)+\mbox{h.c.}
\right]
+C_6\sum_{\m<\n}\tr\left(\x^{(3)}_{\m\n}\S\x^{(3)}_{\n\m}\S^\dagger\right)
\ ,\nonumber\\
-\cu'&=&\frac{1}{4}C_{2V}\sum_\n\left[\tr(\x^{(3)}_\n\S)\;\tr(\x^{(3)}_\n\S)
+\mbox{h.c.}\right]
+\frac{1}{4}C_{2A}\sum_\n\left[\tr(\x^{(3)}_{5\n}\S)\;\tr(\x^{(3)}_{\n 5}\S)
+\mbox{h.c.}\right]\nonumber\\
&&+\frac{1}{2}C_{5V}\sum_\n\tr(\x^{(3)}_\n\S)\;\tr(\x^{(3)}_\n\S^\dagger)
+\frac{1}{2}C_{5A}\sum_\n\tr(\x^{(3)}_{5\n}\S)\;\tr(\x^{(3)}_{\n 5}\S^\dagger)
\ .\nonumber
\end{eqnarray}

Let us briefly consider some of the physics results that can be obtained from
the $O(p^2)$ chiral lagrangian, which is the sum of Eqs.~(\ref{l2again}) (for a
twelve-flavor continuum theory)
and~(\ref{stagpots}).  First, consider meson masses at tree level.  Here I will
only quote results for flavored (\ie, off-diagonal in Eq.~(\ref{phi})) masses.
The flavor-neutral sector is more complicated, and I refer to Ref.~\cite{AB}
for details.

Each of the entries in Eq.~(\ref{phi}) is now a four-by-four taste matrix, and we
can thus expand each entry on a set of $U(4)_{taste}$ generators:
\begin{eqnarray}
\label{tastegen}
\phi_{ij}&=&\frac{1}{2}\sum_{F=1}^{16}\X_F\phi^F_{ij}\ ,\ \ \ i,j=u,d,s\ ,\\
\X_F&\in&\{\x_5,i\x_{5\m},i\x_{\m\n},\x_\m,{\bf 1}\}\ .\nonumber
\end{eqnarray}
By expanding $\S$ to quadratic order, we can then read off the meson
masses from Eqs.~(\ref{l2again}) and~(\ref{stagpots}).  In the flavored sector,
only $\cu$ contributes, and we find
\begin{equation}
\label{stagmesmasses}
m_{Fij}^2=B_0(m_i+m_j)+a^2\D(\X_F)\ ,\ \ \ i,j=u,d,s,\ \ \ i\ne j\ ,
\end{equation}
in which
\begin{eqnarray}
\label{stagshifts}
\D(\x_5)&=&0\ ,\\
\D(i\x_{5\m})&=&\frac{16}{f^2}\left(C_1+3C_3+C_4+3C_6\right)\ ,\nonumber\\
\D(i\x_{\m\n})&=&\frac{16}{f^2}\left(2C_3+2C_4+4C_6\right)\ ,\nonumber\\
\D(\x_\m)&=&\frac{16}{f^2}\left(C_1+C_3+3C_4+3C_6\right)\ ,\nonumber\\
\D(\x_{\bf 1})&=&\frac{16}{f^2}\left(4C_3+4C_4\right)\ .\nonumber
\end{eqnarray}
As we expect, there is one exact flavor multiplet of Nambu--Goldstone bosons,
because of $U(1)_\e$ symmetry, with a mass-squared proportional to
the quark masses, as in the continuum.  All other meson masses are shifted
by terms of order $a^2$, and Fig.~\ref{fig:MILC} is in good agreement with the behavior
predicted by Eq.~(\ref{stagmesmasses}).
The numerical results show that $C_4$
must be dominant among $C_{1,3,4,6}$.  This is consistent with the fact that, as previously mentioned,
$U(4)_{taste}$ is broken at this order in ChPT to $SO(4)$, not all the
way down to $\G_4$. Thus there are only five nondegenerate meson
masses \cite{LS}, whereas the irreducible representations
of the staggered symmetry group would allow eight \cite{MGmesons}.

We end this section by quoting a one-loop result that can
be obtained from our lowest-order staggered chiral lagrangian.  But before
we do this, we have to address the fact that the $SU(12)$ theory has too
many sea quarks, which on the lattice side is fixed by taking the fourth root
of each staggered fermion determinant.  The question is how to carry over
this procedure to staggered ChPT.  The way this can be done is through the
``replica trick."  One starts with {\em adding more} fermions to the theory
by using $n_r$ instead of one staggered fermions for each physical flavor.
For each of the $n_r$ up quarks, the quark mass is kept degenerate ---
all up quarks have mass $m_u$, and likewise for down and strange quarks.
In the lattice QCD path integral, this results in raising the staggered determinant
for each flavor to the $n_r$-th power.  The idea is now to do all calculations
keeping the $n_r$ dependence, and then at the end set $n_r=1/4$, which
corresponds precisely to taking the fourth root of the determinants.

As we already mentioned, it is by no means obvious that the fourth-root procedure
is field-theoretically legitimate.  However, if it is, it is possible to show that the
replica trick is the correct way of incorporating the effects of the fourth root
in staggered ChPT (and EFTs for staggered QCD in general) \cite{BGSset,CBrs}.
Assuming that this all works, we quote the $O(p^4)$ result for the two-flavor
$\x_5$-pion decay constant (which is the real Nambu--Goldstone boson for the breakdown of
$U(1)_\e$ symmetry),
with equal light quark masses, $m_u=m_d=m_\ell$: \footnote{This result can be
derived from Eq.~(27) of Ref.~\cite{ABdecay}, by making the strange quark heavy (a
situation
we will discuss more in Sec. 5).  For nondegenerate results, see Ref.~\cite{ABdecay}, for
$O(p^4)$ meson masses, see Ref.~\cite{AB}.}
\begin{eqnarray}
\label{oneloopstag}
\frac{f^{\x_5}_\p}{f}&=1&-\ \frac{1}{8}\sum_B L(2B_0m_\ell+a^2\D(\X_B))\\
&&-\ 4\left(L(2B_0m+a^2\d'_V)+L(2B_0m+a^2\d'_A)-2L(2B_0 m)\right)\nonumber\\
&&+\ \frac{16B_0 m}{f^2}\left(2L_4+L_5\right)+a^2F\ ,\nonumber
\end{eqnarray}
in which
\begin{eqnarray}
\label{hatdeltas}
\d'_V&=&\frac{16}{f^2}\left(C_{2V}-C_{5V}\right)\ ,\\
\d'_A&=&\frac{16}{f^2}\left(C_{2A}-C_{5A}\right)\ ,\nonumber
\end{eqnarray}
and $F$ is a linear combination of $O(p^4)$ LECs.\footnote{The complete
$O(p^4)$ staggered chiral lagrangian has been worked out in Ref.~\cite{SvdWstag}.}
Note that now the LECs are
low-energy constants in the $N_f=2$ theory. The LECs
$C_{2V,A}$ and $C_{5V,A}$ only appear in flavor-neutral
meson masses at tree level (as can be seen by expanding out $\cu'$), and their
appearance reflects the fact that flavor-neutral mesons appear
in the one-loop corrections.  Flavored mesons also appear in the loops; producing the
logarithms in the sum over $B$.

%%%%%%%%%%%%%%%%%%%
%%% MILC meson masses
\begin{figure}[t!]
\vspace*{20ex}
\hspace*{5ex}
\begin{picture}(75,55)(0,0)
\put(-9,0){\includegraphics*[height=9cm]{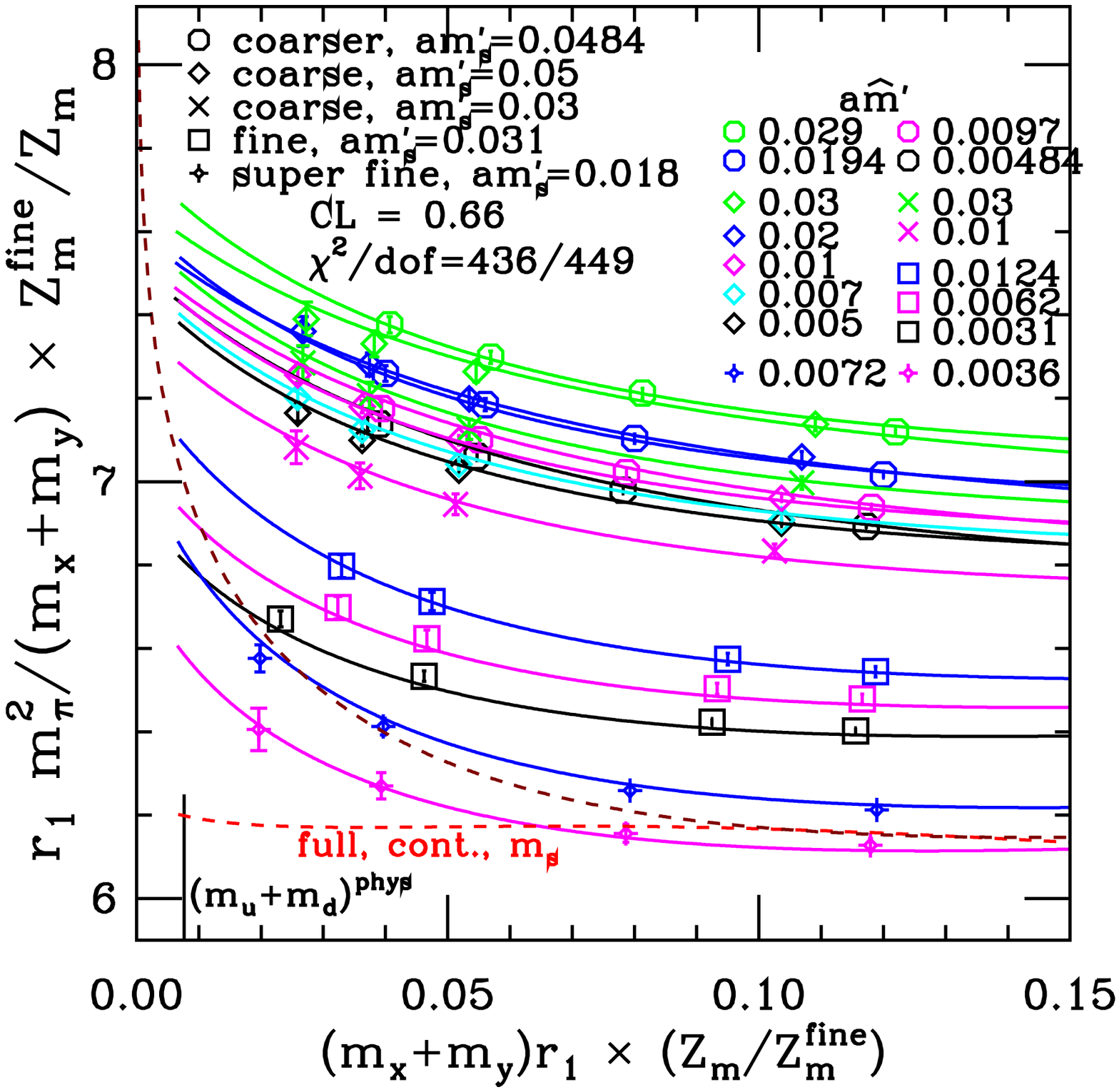}}
\end{picture}
\vspace*{0ex}
\begin{quotation}
\floatcaption{fig:MILCmasses}{$m_\p^2/(m_x+m_y)$ as a function of $m_x+m_y$.
The continuous curves are (partially quenched) chiral fits at fixed lattice spacing; the dashed line
is the (partially quenched) continuum line (with sea quark masses corresponding to those of the
``superfine'' 0.018/0.0072 ensemble), which has a much stronger curvature.
Note how the curvature is reduced by nonvanishing lattice spacing.
The red dashed curve is the fully unquenched continuum curve.
$m'_s$ is the strange sea quark mass and $\hm'$ is the light seaquark mass.
For more explanation of the meaning of all symbols, see Ref.~\cite{MILC2004}.
Figure courtesy C. Bernard.}
\end{quotation}
\vspace*{-1ex}
\end{figure}
%%%%%%%%%%%%%%%%%%%

We note that the masses as they appear in chiral logarithms get
$O(a^2)$ corrections.  This is of course natural, because all lattice mesons can in
principle appear on loops, and most of these masses get $O(a^2)$ corrections,
already at lowest order in ChPT.
This change in the logarithms points at a reason as to why it is important
to incorporate scaling violations in ChPT systematically:
as long as $m_{quark}\sim a^2\L_{QCD}^3$, one cannot expand the logarithm
in Eq.~(\ref{oneloopstag}) in powers of $a^2$.  If we would have naively assumed that
all scaling violations can be understood in terms of simple powers of $a$,
we would have missed this nonanalytic dependence on the lattice spacing.
The $O(a^2)$ term inside the logarithm reduces the curvature for small quark
masses.  Knowing the field-theoretical form of this nonanalytic behavior
makes it thus feasible to explore smaller quark masses at a given lattice
spacing than otherwise would be possible.  In Fig.~\ref{fig:MILCmasses}, I show a plot of the
staggered ChPT fits of the meson masses from the MILC collaboration
(which, of course, use the
full nondegenerate $2+1$-flavor and partially quenched formulas of Ref.~\cite{ABdecay}).
The indices $x$ and $y$ denote the flavors of the valence quarks,
which for most of the data points in this figure have masses different
from the sea quarks, a generalization we will consider in the next section.
In this figure, we see that indeed the curvature on the lattice is significantly
different from that of the continuum curve.

Finally, we observe that the form of the tree-level masses, Eq.~(\ref{stagmesmasses}),
is very similar to Eq.~(\ref{fomesonmass}).  That raises the question whether,
since we are working in the LCE regime in which $m_{quark}\sim a^2\L_{QCD}^3$,
Aoki-like phases could occur for small enough quark mass.
Using techniques similar to those used in QCD inequalities\footnote{For a review,
see Ref.~\cite{SN}.} it can be argued that all $\D(\X_B)\ge 0$, excluding this
route to a possible Aoki-like phase.  However, this is not true for the $O(a^2)$
corrections in the flavor-neutral sector, which are governed by the LECs
$\d'_{V,A}$.  While no sign of an Aoki-like phase has been observed in
numerical computations, this remains an interesting theoretical possibility \cite{AB,AW}.

%%####%%
\section{\label{PQmixed} Choosing valence and sea quarks to be different}
%%####%%
Consider the two-point function for a charged pion in euclidean QCD,
\begin{eqnarray}
\label{pion2pt}
&&\langle\p^+(x)\p^-(y)\rangle=\langle(\bu(x)\g_5 d(x))(\bd(y)\g_5 u(y))\rangle\\&&=
\frac{1}{Z}\int \prod_\m[dU_\m]\prod_i [d\bq_i][dq_i]\;\exp{(-S_{QCD})}\;
\bu(x)\g_5 d(x)\bd(y)\g_5 u(y)\nonumber\\
&&=-\frac{1}{Z}\int \prod_\m[dU_\m]\;\Det(\Sl{D}+M)\;
\tr\left[\g_5\left(\Sl{D}+m_u\right)^{-1}(y,x)\g_5\left(\Sl{D}+m_d\right)^{-1}(x,y)
\right]\ .\nonumber
\end{eqnarray}
In an actual lattice computation, one usually does not use point sources and
sinks; instead one sums with some weight over the timeslices at
$t=x_4$ and $t'=y_4$, but this will not affect the observation we want to make.

In the third line of Eq.~(\ref{pion2pt}), the quarks enter in two clearly distinct ways:
through the determinant, and through the quark propagators that follow from
carrying out the Wick contractions.  We refer to quarks of the first type as
sea quarks, and those of the latter type as valence quarks.  Of course, in
nature these two types of quarks are the same, and they have to be in order
to preserve unitarity of the theory.  But nothing stops us on the lattice from
taking the quark masses in the propagators unequal to those in the
determinant, or, even more drastically, from choosing a completely different
discretization for $\Sl{D}$ in the propagators and in the determinant!
Let us first consider whether this seemingly sacrilegious idea might have any
advantages, and if so, whether we can make any field-theoretical sense out
of it.

First, consider only taking valence quark masses unequal to sea quark masses.
This is a situation commonly referred to as partial quenching.  This name has
its origin in the fact that as we send the sea quark masses to infinity, we
effectively remove the determinant from the path integral, thus reverting to
the quenched approximation.  Unlike the quenched approximation, however,
as long as we keep the sea quark masses in the theory, we can always
recover the physical theory by choosing the valence and sea quark masses
equal (for each flavor) \cite{BGPQ}.  In other words, real or ``full'' QCD is a ``special case''
of partially quenched QCD (PQQCD).\footnote{If one chooses the number of
light sea quarks to be equal to three!  In that case, the name ``partially quenched''
is slightly unfortunate.  However, plenty of lattice computations are still
really partially quenched, in that they include only two sea quarks.}

This observation should then carry over to any EFT for PQQCD, and thus,
in particular, to  partially quenched ChPT (PQChPT).   (Of course, we will have
to reconsider the arguments for the existence of ChPT, a point to which we will
return below.)
What we will see is that, if both
sea and valence quark masses are light enough for the chiral expansion
to apply, being able to vary the valence and sea quark masses
independently is very useful: It gives us an extra tool for
matching lattice QCD computations to ChPT, and thus determine the numerical
values of the LECs at some scale
--- which pin down the EFT quantitatively.  The key observation here is that
those LECs are the {\em same} in the full and partially quenched theories,
because they are independent of the quark masses,
and thus also of the distinction that we make between valence and sea
quark masses in the partially
quenched generalization of QCD \cite{ShSh1}.  In addition, a
very practical consideration is that in most cases it is
much less expensive in terms of computational cost to vary the valence quark
masses (which only show up in the propagators coming from the external
operators) than the sea quark masses, which are part of the effective gauge action used to generate an
ensemble of gauge-field configurations.  We will consider partial quenching in detail in
Secs.~\ref{PQ} and \ref{quark flow}.

The motivation for choosing even the lattice Dirac operators different for
valence and sea quarks is different.   Of course, the basic assumption will be
that in the continuum limit the difference in discretization disappears if
the bare valence and sea quark masses are tuned to yield the same renormalized
quark mass --- an extension
of the notion of universality.   But, the symmetry properties
of a correlation function such as Eq.~(\ref{pion2pt}) are determined by the
symmetry properties of the valence Dirac operator, and in many cases it
is a great advantage to be able to use a discretization with symmetries (in
particular chiral symmetry) as close as possible to those of the continuum.
Often, because of the smaller amount of symmetry on the lattice, operator
mixing is more severe on the lattice, making computations of, for instance,
weak matrix elements much harder, if not impossible.  More symmetry, in
particular more chiral and flavor symmetry, means
less mixing.  However, lattice Dirac operators with good symmetry properties,
such as domain-wall and overlap operators, are numerically very expensive.
Since a large part of the computational effort goes into evaluating the
determinant, it makes sense
to choose a less expensive lattice version for the sea quark Dirac operator,
such as Wilson or staggered operators.  This ``mixed-action'' option will be discussed in
Sec.~\ref{mixed}.

Having extolled the possible virtues, we should address the question whether
these generalizations of QCD make any field-theoretical sense.

%%####%%
\subsection{\label{FT} Path integral}
%%####%%
Of course, operationally, the procedure of choosing different Dirac operators for
the valence and sea quarks outlined in the previous section is
well defined.  But, does it make field-theoretical sense?  For instance, can
we hope to develop EFTs for these generalized, but --- at this stage ---
suspect versions of
QCD?  Clearly, in order to address these questions, it would help
if we can write down a path integral expression that reflects the distinction
between valence and sea quarks.  This can be done using a trick proposed
by Morel \cite{Morel}.  Obviously, separate fermionic fields are needed for
sea and valence quarks, but doing just this would lead to fermion
determinants for both.  The trick is to cancel the valence determinant by
introducing yet another set of quark fields, with Dirac operator and mass
matrix identical to those of the valence quarks, but with opposite (bosonic)
statistics.  We will refer to these fields as ghosts.\footnote{These ghosts
have nothing to do with Faddeev--Popov ghosts!  They are called ghosts, because
they share the similarly ``incorrect'' spin-statistics properties.}  The fermionic part of the
QCD lagrangian generalizes to\footnote{I write the Dirac operators $D_s$ and $D_v$
without a slash, because not all lattice Dirac operators are of the form $\Sl{D}=\g_\m D_\m$.}
\begin{equation}
\label{genQCD}
\cl=\bq_s(D_s+M_s) q_s+\bq_v(D_v+M_v) q_v
+\tq^\dagger(D_v+M_v)\tq\\ ,
\end{equation}
in which $\tq$ denotes the bosonic ghost-quark fields;  $D_s$ is the lattice
Dirac operator for sea quarks, and $D_v$ that for valence quarks.  In general,
we will label sea quantities with subscript $s$, valence quantities with
subscript $v$, and put tildes on entities referring to ghosts.
Note that the
ghost and antighost fields, $\tq$ and $\tq^\dagger$, are not independent.  The path integral
for this theory is
\begin{eqnarray}
\label{pigenQCD}
Z&=&\int \prod_\m[dU_\m]\prod_i [d\bq_{s,i}][dq_{s,i}]
\prod_j [d\bq_{v,j}][dq_{v,j}][d\tq^\dagger_j][d\tq_j]\\
&&\hspace{10mm}
\exp\left(-S_{gauge}-\sum_x[\bq_s(D_s+M_s) q_s+\bq_v(D_v+M_v) q_v
+\tq^\dagger(D_v+M_v)\tq]\right)\nonumber\\
&=&\int \prod_\m[dU_\m]\;\exp(-S_{gauge})\;
\Det\left(D_s+M_s\right)\ ,\nonumber
\end{eqnarray}
which follows because of the exact cancellation of the
valence and ghost determinants.  Intuitively, at the level of diagrams,
valence and ghost loops cancel, because they always come in pairs
with opposite signs between them.  Clearly, it is possible to couple the path integral
to sources for all quark fields, and we can thus generate correlation
functions for operators made out of any of these fields.  In numerical
computations, of course the ghost fields are never introduced in the
first place.  But one still has access to correlation functions that include
ghost quarks, because after doing the Wick contractions one simply
finds that each ghost-antighost contraction is replaced by a valence propagator.
For example, if one considers the pion two-point function of Eq.~(\ref{pion2pt})
for pions made entirely out of ghost quarks, one finds exactly the same
expression as on the last line of Eq.~(\ref{pion2pt}), except without the minus
sign, which would be absent because ghost fields commute, rather than
anticommute.

In order for definition~(\ref{pigenQCD}) to make sense, we have to make
sure that the integral over ghost fields converges.  This is the case if
$D_v$ is antihermitian, as it is in the continuum and for staggered
fermions, for instance.  In addition, we have to require that the
mass matrix, $M_v$, has strictly positive eigenvalues.   For Wilson or
domain-wall
fermions the Dirac operator is not antihermitian (or hermitian), and the
real part of its eigenvalues can be of both signs.  For a method of defining
PQ or mixed-action QCD with valence Wilson fermions, I refer to Ref.~\cite{GSS}.%
\footnote{A similar method as described in Ref.~\cite{GSS} exists for domain-wall
fermions.}
For overlap fermions, the eigenvalues are also complex, but one can prove
that if the overlap operator satisfies $D^\dagger=\g_5 D\g_5$, all
eigenvalues have a nonnegative real part.  Thus, the ghost integral
converges again as long as $M_v$ has only positive eigenvalues.

It is clear that, while a euclidean path-integral definition can be given,
this theory is sick.  The ghost quarks have the wrong statistics for a
spin-$1/2$ field.  Remember that if you try to quantize a spin-$1/2$ field
with the wrong statistics, one of the problems that occurs is that the
hamiltonian of the theory is unbounded from below \cite{PSQFT}!
This and other \cite{BGPS} aspects indicate that it may be problematic
to continue the theory to a theory in Minkowski space with a positive
hamiltonian.
But even if we could continue to Minkowski space, there is a
problem with unitarity if one considers correlation functions with valence
quarks on the external lines, while {\em different} sea quarks run on the
loops.\footnote{It is clear that this sickness already occurs in weak-coupling
perturbation theory.}

Fortunately, lattice QCD, as applied in modern numerical computations,
does not want to be in Minkowski space.  It is perfectly fine to match
euclidean lattice computations to a euclidean EFT.  Once the EFT has been
completely defined by obtaining the values of all its LECs, we can
take the continuum limit, set valence quark masses equal to sea quark
masses, and continue to Minkowski space.  The fact that PQQCD and mixed-action
QCD can only be defined in euclidean space is not a problem.

However, for all this to work, we do need to develop ChPT for
partially quenched and mixed-action theories, and it is not obvious that this can be done.
The generalized theory does not satisfy some of the cherished properties of a
healthy quantum field theory, in particular unitarity,\footnote{And
causality, if these generalized theories only live in euclidean space.  However,
since PQQCD and mixed-action QCD are local theories, we expect that the
cluster property is realized on correlation functions.}
 which are usually invoked to argue that a
local, unitary EFT such as ChPT has to exist, and, is in fact the
most general possibility \cite{SWphysica}.
We will say more about this in Sec.~\ref{PQ} for PQQCD, and in Sec.~\ref{mixed}
for mixed-action QCD.

Before we delve into the specifics, let us discuss some generalities
considering the symmetries of Eq.~(\ref{genQCD}).
If we have $N_s$ sea quarks (we will always take $N_s=3$ or $2$, as in
previous sections) and $N_v$ valence quarks (we can always pick as
many valence quarks as we need, so $N_v$ is arbitrary), it looks like the
chiral symmetry group of Eq.~(\ref{genQCD}) (for $M_s=0$ and $M_v=0$) is
$[U(N_s)_L\times U(N_s)_R]\times[U(N_v|N_v)_L\times
U(N_v|N_v)_R]$.\footnote{In the partially quenched case, for which $D_s=D_v$
in Eq.~(\ref{genQCD}),
the symmetry group is larger, as we will discuss in Sec.~\ref{PQ}.}
Here  $U(N|N)$ is a ``graded'' group, which we will
discuss in more detail in the next section;\footnote{See the text starting around
Eq.~(\ref{gradedm}).} for now, all we need is that the
symmetry group of the ghost part of the lagrangian Eq.~(\ref{genQCD}) alone
would be the subgroup $U(N_v)_L\times U(N_v)_R$.  It turns out that
this is not quite correct \cite{DOTV,MZ,GSS}, as I will now explain.

Since $\tq^\dagger$ is not independent of $\tq$, it follows that the ghost
part of Eq.~(\ref{genQCD}) is invariant under
\begin{equation}
\label{ghsym}
\tq_L\to V\tq_L\ ,\ \ \ \ \ \tq_R\to V^{\dagger -1}\tq_R\ ,
\end{equation}
with $V\in GL(N_v)$.  Here
\begin{equation}
\label{ghproj}
\tq_L=\frac{1}{2}(1-\g_5)\tq\ ,\ \ \ \ \ \tq_R=\frac{1}{2}(1+\g_5)\tq\ ,
\end{equation}
from which it follows that (unlike for quark fields $\bq$ and $q$, for which the chiral
projections can be chosen independently, \seef\ Eq.~(\ref{chproj}))
\begin{equation}
\label{ghproj2}
\tq_L^\dagger=\tq^\dagger\frac{1}{2}(1-\g_5)\ ,\ \ \ \ \ \tq_R=\tq^\dagger\frac{1}{2}(1+\g_5)\ .
\end{equation}
Note that the definition of the projected $q^\dagger$'s is opposite to that of the
projected $\bq$'s.
Equation~(\ref{ghsym}) then follows from the fact that
\begin{equation}
\label{ghkin}
\tq^\dagger D_v\tq=\tq_L^\dagger D_v\tq_R+\tq_R^\dagger D_v\tq_L\ .
\end{equation}
Here I assumed that $\{D_v,\g_5\}=0$, which holds in the continuum.
The lattice equivalent of the decomposition~(\ref{ghkin})
will depend on the type of lattice fermion employed,
as already noted above.\footnote{For instance, for staggered fermions, one can
define $\tq_R$ and $\tq_L$ using the generator $\e(x)$ of $U(1)_\e$ transformations
instead of $\g_5$.}  But in all cases, the fact that $\tq^\dagger$ is not
independent of $\tq$ determines the precise form of the chiral symmetry
group, and in all cases it has to coincide with $GL(N_v)$ in the continuum
limit.

The vector subgroup, defined as the group which leaves
the ghost condensate $\tq^\dagger_L\tq_L+\tq^\dagger_R\tq_R$ invariant,
is the unitary subgroup, $U(N_v)\subset GL(N_v)$.    If we restrict ourselves
momentarily to the ghost sector only, this implies that $\S$ has to be an
element of the coset $GL(N_v)/U(N_v)$.  If we want to write $\S$ in terms of
a meson field $\tphi$ analogous to $\phi$ in Eq.~(\ref{phi}), but with its meson
component fields ``made out of ghost quarks and antiquarks,'' we should
write $\S=\exp{(2\tphi/f)}$, with $\tphi$ a {\em hermitian} matrix.  The number
of mesonic degrees of freedom is the same as it would be for the coset
$[U(N_v)_L\times U(N_v)_R]/U(N)_V$, but the expression for $\S$ in terms of the
meson fields is different.\footnote{I am skipping a number of further
technical details about the precise symmetry group of Eq.~(\ref{genQCD}).
Also, when all is said and done, the anomaly will have to be taken into
account, as usual.  For the latter, see Sec.~\ref{PQ}.}

For nonperturbative issues,  it is crucial to work with the correct symmetry group.  This can
already be seen from Eq.~(\ref{genQCD}): If we introduce spurion fields
$M$ and $\bM$ to write the ghost mass term as
\begin{equation}
\label{ghmass}
\tq^\dagger M_v\tq\to\tq^\dagger_L M\tq_L+\tq^\dagger_R\bM\tq_R\ ,
\end{equation}
transforming as
\begin{equation}
\label{ghmasstr}
M\to V^{\dagger -1}MV^{-1}\ ,\ \ \ \ \ \bM\to V\bM V^\dagger\ ,
\end{equation}
we see that both $M$ and $\bM$ remain strictly positive under $GL(N_v)$
transformations, consistent with the convergence of the ghost path integral.
This would not be true if we replace $GL(N_v)$ by $U(N_v)_L\times U(N_v)_R$.

One may want to use ChPT to find the phase structure of
a quenched or partially quenched theory, as we did for $N_f=2$ Wilson fermions in Sec.~\ref{phase}
\cite{GSS}.  In that case, one will have to be careful to start from the correct
symmetries in constructing the chiral lagrangian.  However, if one is in a phase
in which the correct vacuum has been determined (and usually in applications,
is given by $\S={\bf 1}$), and one is only interested
in calculating the chiral expansion of physical quantities, it can be shown that
instead the symmetry group $U(N_v|N_v)_L\times U(N_v|N_v)_R$ can be used
for constructing the chiral lagrangian \cite{GSS,ShSh2}.  This is what we will do
in the next section.  The reason is that {\em in perturbation theory}, which is
defined by expanding $\S$ in terms of $\tphi$, we can
perform a field redefinition $\tphi\to i\tphi$ that brings $\S$ into
the ``standard form,''  $\S=\exp(2i\tphi/f)$,
\ie, the form that would have followed from starting with the ghost-sector symmetry group $U(N_v)_L\times U(N_v)_R$.

As an aside for the interested reader,
there is another way to define PQQCD and mixed-action QCD, through the replica
trick \cite{replica}.
Instead of introducing ghost quarks, one only introduces separate
valence quarks, $n_i$ for each flavor.  One then takes the limit $n_i\to 0$ at the end
of a calculation.\footnote{The idea is similar to the usage of the replica
trick in defining the fourth root with staggered fermions, hence the same name.
Note however, that the aim is different: here we take $n_i\to 0$, rather than
continuing it to $1/4$.}
Formally, this sends the valence determinant to a constant, thus removing it
from the QCD partition function.    For all positive integer values of the $n_i$,
this defines a physically sensible theory, in which we just happen to discretize
different quarks in different ways.  The disadvantage is that not much is known
about the properties of the limit $n_i\to 0$.  In this case one bases the construction
of the chiral lagrangian on the symmetry group $[U(N_s)_L\times U(N_s)_R]\times[U(N_v)_L\times U(N_v)_R]$ (with $N_v=\sum_i n_i$), and
considers ChPT in the $n_i\to 0$ limit.   In all cases where explicit calculations
have been done, the results agree from those obtained with Morel's trick,
using ghost quarks.

%%####%%
\subsection{\label{PQ} Partially quenched ChPT}
%%####%%
In this section, we will consider ChPT for
PQQCD, the generalization of QCD in which we only
choose the masses of valence and sea quarks to be independent, while
using the same discretization of the Dirac operator.
The first question is whether indeed EFTs such as ChPT can be constructed
for PQQCD.  We will assume that it can, as a euclidean EFT for the underlying
euclidean theory.  There exists no proof  of this assertion.  Some
observations in support can be made, however.
\begin{itemize}
\item[1.]   One expects the SET to exist, with almost the same level of rigor
as in the physical case.  This expectation is based on two observations:
First, I expect that the perturbative construction of the SET goes through
just the same for the partially quenched case.
Then, the fact that the SET also is valid
beyond perturbation theory amounts basically to the assumption that the
coefficients in the Symanzik expansion have a well-defined meaning beyond
perturbation theory,\footnote{An assumption that is used in nonperturbative
improvement.} and it does not look like a drastic step to expect that this
assumption is also valid in the partially quenched case.  While this observation
does not directly address the validity of partially quenched ChPT (PQChPT),
the SET is a ``stepping stone,'' and its validity is important for the construction
of the appropriate chiral theory.
\item[2.]   PQQCD contains full QCD as a subset: setting a valence quark mass
equal to a sea quark mass makes that valence quark the same as a sea quark,
and correlation functions with only sea quarks on the external lines are exactly
those of full QCD.   The collection of all correlation functions of PQQCD thus contains
all correlation functions of full QCD as a subset.
This allows us to make two observations:  First, chiral symmetry breaking in the sea sector
(\ie,\ through a sea-quark condensate) takes place as usual.  In PQQCD, one can
use vectorlike symmetries to rotate sea quarks into valence quarks,
thus producing also a valence condensate (and, by extension, a ghost
condensate equal to the valence condensate).\footnote{The argument
that vector symmetries are not spontaneously broken \cite{VW} extends
to continuum PQQCD.  However, not all discretizations obey the conditions
for this theorem, as the possible existence of the Aoki phase in QCD
with Wilson fermions shows \cite{ShSi}.}
Second, for full QCD the SET and ChPT exist, and it seems not
unreasonable to assume that the match between PQQCD and PQChPT can
be extended away from $m_v=m_s$, at least as long as the difference
$m_v-m_s$ does not become too large, relative to $m_s$ and $m_v$.
Below, we will have more to say about what ``too large'' means.
\item[3.]  The assumption that EFTs such as SET and ChPT exist for
purely euclidean theories appears rather natural if one would imagine
constructing the EFTs through the renormalization group.  In this framework
one would expect the EFT to be local below the scale above which
one has integrated out all modes through some type of renormalization
group blocking procedure.  To be sure, there is no proof that the EFT
obtained if one actually knew how to carry out the renormalization group
blocking is local.  But it is a ``folklore'' that seems equally applicable to
PQQCD as to the standard case of full QCD.
\end{itemize}
Based on these observations, we will assume that, as before, the lagrangian
of the EFT of interest should be taken to be the most general local
function of the fields consistent with the symmetries of the theory.  In fact,
it is interesting to follow this route, and see where it will lead us.  Hopefully,
PQChPT will exhibit, in a more concrete form, what the diseases of the
theory are, and, hopefully, these will be the only diseases!  If this is true,
this makes it possible to use PQChPT in the interpretation of lattice results
(read: use the predictions of PQChPT to fit data).  If this works to a high degree
of precision, that constitutes a nontrivial and important test of PQChPT, and
thus of our understanding of PQQCD.  As we will see, the tests become
nontrivial once numerical computations ``can see'' the nonanalytic terms
predicted by PQChPT.   Numerical evidence for such nonanalytic behavior
exists, and I expect that it will fairly rapidly become more extensive, at least
in the Nambu--Goldstone-boson sector.

PQQCD has a much larger symmetry than Eq.~(\ref{genQCD}), because now
sea, valence, and ghost quarks can all be rotated into each other.  (Only the
mass matrices cause a soft breaking of this symmetry.)  Following our earlier
discussion, we will now think of $\tq^\dagger$ as independent of $\tq$, just
as $\bq$ is independent of $q$.  The full chiral symmetry group (ignoring the
mass matrix) is then $SU(N_s+N_v|N_v)_L\times SU(N_s+N_v|N_v)_R$, where we
now omit an anomalous axial $U(1)$, as well as the $U(1)$ for quark number,
which is trivially represented on mesons.  The nonlinear field $\S=\exp{(2i\Phi/f)}$
now lives in the coset $[SU(N_s+N_v|N_v)_L\times SU(N_s+N_v|N_v)_R]/
SU(N_s+N_v|N_v)_V$, and correspondingly, the field $\Phi$ describes
$(N_s+2N_v)^2-1$ mesons, counting precisely the number of pseudoscalar
meson fields one can construct from the sea, valence, and ghost quarks
and antiquarks when we leave out the singlet field that corresponds to the axial
$U(1)$ (see below).

Before we continue, let us have a brief look at the graded group $SU(N_s+N_v|N_v)$
\cite{deWittFreund}.  An element $U\in SU(N_s+N_v|N_v)$ can be written
in block form as
\begin{equation}
\label{gradedm}
U=\pmatrix{A&B\cr C&D}\ ,
\end{equation}
in which $A$ is an $(N_s+N_v)\times (N_s+N_v)$ matrix of commuting numbers,
$D$ an $N_v\times N_v$ matrix of commuting numbers, while $B$ and $C$
are $(N_s+N_v)\times N_v$ and $N_v\times(N_s+N_v)$ matrices of
anticommuting numbers, respectively.  It is straightforward to check that if $\S$ has this
structure, also $\Phi$ has to have this structure.
The quark fields
\begin{equation}
\label{Qs}
Q=\pmatrix{q_s\cr q_v \cr \tq}\ ,\ \ \ \ \ \bQ=\pmatrix{\bq_s &\bq_v&\tq^\dagger}
\end{equation}
transform in the fundamental and antifundamental representations of
$SU(N_s+N_v|N_v)$,\footnote{Not all products of irreducible representations of graded
groups are fully  reducible.  This plays a (minor) role
for weak matrix elements in the partially-quenched theory with
$N_s=2$, see Ref.~\cite{MGEP},
and references therein.} and in terms of these ``super'' quark fields, the
PQQCD lagrangian takes the simple form
\begin{equation}
\label{PQQCDlag}
\cl_{PQQCD}=\cl_{gauge}+\bQ(D+\cm)Q\ ,
\end{equation}
in which $\cm$ is the mass matrix
\begin{equation}
\label{cm}
\cm=\pmatrix{M_s& 0& 0\cr 0& M_v &0\cr 0&0&M_v}\ .
\end{equation}

In the chiral theory,
invariants are constructed using traces and determinants, and we need the
generalization of these to graded groups.  The ``supertrace'' of a matrix $U$
as in Eq.~(\ref{gradedm}) is defined as
\begin{equation}
\label{strace}
\str(U)=\tr(A)-\tr(D)\ .
\end{equation}
The minus sign maintains the cyclic property,
$\str(U_1U_2)=\str(U_2U_1)$.  The ``superdeterminant'' is then
defined through
\begin{equation}
\label{superdet}
\sdet(U)=\exp(\str\log(U))=\det(A-BD^{-1}C)/\det(D)\ ,
\end{equation}
for $U$ of the form Eq.~(\ref{gradedm}); this definition
implies that $\sdet(U_1U_2)=\sdet(U_1)\sdet(U_2)$.
To understand this expression, first decompose $U$ as
\begin{equation}
\label{Udecomp}
U=\pmatrix{A&B\cr C&D}=\pmatrix{1&BD^{-1}\cr 0&1}\pmatrix{A-BD^{-1}C&0\cr 0&D}
\pmatrix{1&0\cr D^{-1}C&1}\ .
\end{equation}
Now, using the definition $\sdet(U)=\exp(\str\log(U))$, the result follows,
with the superdeterminant of the first and last factors on the right-hand side
being equal to one.

Hermitian conjugation is defined as usual, with the proviso that
complex conjugation reorders: $(ab)^*=b^*a^*$,\footnote{Of course, this
only makes a difference when both $a$ and $b$ are Grassmann variables.}
from which it follows that $(U_1 U_2)^\dagger=U_2^\dagger U_1^\dagger$.
The group $SU(N_s+N_v|N_v)$ is now defined as the group of all unitary
graded $(N_s+2N_v)\times(N_s+2N_v)$ matrices, with grading as in
Eq.~(\ref{gradedm}), and superdeterminant equal to one.  Clearly, if
$\S=\exp(2i\Phi/f)$ has $\sdet(\S)=1$, this is equivalent with
$\str(\Phi)=0$.

The matrix $\Phi$ can be written in block form
\begin{equation}
\label{block}
\Phi=\pmatrix{\phi&\eta\cr\be&\tphi}\ ,
\end{equation}
in which $\phi$ contains all meson fields made out of fermionic quarks,
$\tphi$ contains all fields made out of ghost quarks, $\eta$ contains
those made out of a fermionic quark and a ghost antiquark, whereas
$\be$ contains meson fields made out of a ghost quark and a fermionic
antiquark.  Note that $\phi$ is not the same as Eq.~(\ref{phi}), because it
also contains meson fields made out of valence and sea quarks (``mixed''
pions) and meson fields made out of valence quarks only (``valence''
pions).  The upper lefthand $N_s\times N_s$ block inside $\phi$ can
be identified with Eq.~(\ref{phi}), for $N_s=3$.

In practical applications, it turns out to be  useful to relax the restriction that $\str(\Phi)=0$, which
makes $\S\in U(N_s+N_v|N_v)$ instead of $SU(N_s+N_v|N_v)$.
In conjunction, it is  also useful to parametrize the diagonal of $\Phi$
not in terms of physical fields (\ie, those found by diagonalizing
the mass matrix), but in terms of ``single-flavor'' fields:
\begin{equation}
\label{diag}
\mbox{diag}(\Phi)=(U,D,S,X,Y,\dots,\tX,\tY,\dots)\ .
\end{equation}
Here we have introduced a now commonly used notation for the valence
quarks $x,y,\dots$, with $U\sim u\bu$, $X\sim x\bx$, \etc, instead of $u_v,d_v,\dots$, to save writing indices.
One can have as many valence quarks as one needs --- since they do
not contribute to the dynamics, their number need not be fixed.
This explains the dots in Eq.~(\ref{diag}); of course, the number of
ghost quarks needs to be equal to the number of valence quarks.\footnote{
Within these lectures, I'll need only two valence quarks, $x$ and $y$.
But for instance for nonleptonic kaon decays, one needs three.}
The ``super''-singlet field (the ``super-$\eta'$''),
\begin{equation}
\label{superetap}
\Phi_0\equiv \str(\Phi)=U+D+S+X+Y+\dots-(\tX+\tY+\dots)\ ,
\end{equation}
is not a Goldstone meson, because of the axial anomaly.  This can
easily be seen from the contribution of all quarks to the triangle
diagram: in order for the ghost-quark contributions to {\em add} to
the anomaly, one should include their contributions with an
explicit minus sign, because they do not get a sign from the loop.
This means that we should really remove the field $\Phi_0$ from the
chiral lagrangian (as we have been doing thus far).  We will do this
by giving it a large mass, that we will eventually send to infinity,
thus decoupling the super-$\eta'$.\footnote{For a proof that this
is a correct procedure, see Ref.~\cite{ShSh2}.}

It is time to get to the partially quenched chiral lagrangian!  The
lowest order form is
\begin{equation}
\label{pql2}
\cl^{(2)}=\frac{1}{8}f^2\;\str(\partial_\m\S^\dagger\partial_\m\S)-
\frac{1}{8}f^2\;\str(\c^\dagger\S+\S^\dagger\c)+\frac{1}{6}m_0^2
\left(\str(\Phi)\right)^2\ ,
\end{equation}
in which $\c$ is a spurion for the quark masses, that should be
set equal to $2B_0\cm$.  We note that the field $\Phi_0=\str(\Phi)$
is not constrained by symmetry (except parity), and thus we should
really multiply every term in Eq.~(\ref{pql2}) by an arbitrary (even)
function of $\Phi_0$.\footnote{And add more terms, see Refs.~\cite{GL,BGQ}.}
However, we will be decoupling the super-$\eta'$, and that removes the
dependence on all parameters contained in these potentials \cite{ShSh2}.

{}From this lagrangian, we can read off the lowest-order
expression for all meson masses in the theory.  This is straightforward
for the off-diagonal fields in $\Phi$.  By expanding Eq.~(\ref{pql2}) to quadratic
order, one find that the mass of a meson with flavors $i$ and $j$ is
given by
\begin{equation}
\label{tlmass}
m_{ij}^2=B_0(m_i+m_j)\ .
\end{equation}
The only thing ``different'' is that the ghost-meson propagators
get an extra minus sign because of the supertrace in Eq.~(\ref{pql2}),
\begin{equation}
\label{ghostprop}
\langle\tphi_{ij}(p)\tphi_{ji}(q)\rangle=\frac{-1}{p^2+m_{ij}^2}\;\d(p-q)\ .
\end{equation}
This is a first example of how PQChPT makes the ``diseases'' of PQQCD
visible.  For fermionic mesons, one has to remember that the ordering
of $\eta$ and $\be$ matters.  The fermionic nature of these fields
will provide minus signs when calculating loops, thus providing a
mechanism through which the cancellation between valence quark and
ghost quark loops is built into PQChPT.

The flavor-neutral sector is a little more complicated, because of the
$m_0^2$ term (or, equivalently, because of mixing between the $U$, $D$,
\etc\ fields).  Let us calculate the propagator $\langle\Phi_{ii}\Phi_{jj}\rangle$.
From the first two terms in Eq.~(\ref{pql2}), there is a contribution
\begin{eqnarray}
\label{conn}
G_{ij}\equiv\langle\Phi_{ii}\Phi_{jj}\rangle&=&
\frac{\e_i\d_{ij}}{p^2+m_{ii}^2}\ ,\\
\e_i&=&\left\{
\begin{array}{ll}
+1\ ,& \mbox{$i$ sea or valence}\\
-1\ ,&\mbox{$i$ ghost}
\end{array}\right.
\ \ \ \ ,\nonumber
\end{eqnarray}
from the first two terms in Eq.~(\ref{pql2}),
with a minus sign as in Eq.~(\ref{ghostprop}) if $i$ refers to a ghost.
There are additional contributions coming from the $m_0^2$ term, that
can be found by treating this term as a two-point vertex, and doing the
geometric sums.  In detail, we can rewrite
\begin{equation}
\label{2ptvertex}
\frac{1}{6}m_0^2
\left(\str(\Phi)\right)^2=\frac{1}{6}m_0^2\;\Phi_{ii}K_{ij}\Phi_{jj}\ ,
\end{equation}
with
\begin{equation}
\label{K}
K_{ij}=\e_i\e_j\ .
\end{equation}
One finds, using that
\begin{equation}
\label{KGK}
(KGK)_{ij}=K_{ij}\sum_k G_{kk}\ ,
\end{equation}
for the contribution proportional
to $m_0^2$ \cite{BGPQ,ShSh1}:
\begin{eqnarray}
\label{dppart}
&&-m_0^2/3\;(GKG)_{ij}+
\left(-m_0^2/3\right)^2 (GKGKG)_{ij}+\dots\\
&&\hspace{2cm}=\frac{-m_0^2/3}{(p^2+m_{ii}^2)(p^2+m_{jj}^2)}\left(
\frac{1}{1+\sum_{k=u,d,s}\frac{m_0^2/3}{p^2+m_{kk}^2}}\right)\nonumber\\
&&\hspace{2cm}=\frac{-m_0^2/3}{(p^2+m_{ii}^2)(p^2+m_{jj}^2)}\;
\frac{(p^2+m_U^2)(p^2+m_D^2)(p^2+m_S^2)}
{(p^2+m_{\p^0}^2)(p^2+m_\eta^2)(p^2+m_{\eta'}^2)}\nonumber\\
&&\hspace{2cm}\to \frac{-1/3}{(p^2+m_{ii}^2)(p^2+m_{jj}^2)}\;
\frac{(p^2+m_U^2)(p^2+m_D^2)(p^2+m_S^2)}
{(p^2+m_{\p^0}^2)(p^2+m_\eta^2)}\ .\nonumber
\end{eqnarray}
Let us step through this equation line by line.  In the first expression on the
right-hand side,
the sum in the denominator really extends over all flavors, but the
ghost terms cancel the valence terms, reducing the sum to run over
the sea flavors $u$, $d$ and $s$ only.  The second expression is just a
rewriting of the first line.  It is clear that the denominator of the
second factor is a third-order polynomial in $p^2$, and that it has to
have poles at the physical masses (this factor only refers to the sea
sector).  These are of course the masses of the $\p^0$, $\eta$ and
$\eta'$.  The $\eta'$ mass-squared contains a term proportional to $m_0^2$,
\begin{equation}
\label{etapmass}
m_{\eta'}^2=\frac{N_s}{3}\;m_0^2+\mbox{terms proportional to sea quark masses}\ ,
\end{equation}
whereas
the $\p_0$ and $\eta$ become independent of $m_0^2$ when we take it
to infinity.  This observation explains the last line in Eq.~(\ref{dppart}).
In Eq.~(\ref{etapmass}) I made the dependence on the number of sea quarks,
$N_s$, explicit --- of course, in our explicit example above $N_s=3$.

This result for the neutral propagators exhibits a new property of the
partially quenched theory.  The easiest way to see this is to note
that, if we take $i=j$ to be a valence flavor, there is a double
pole in Eq.~(\ref{dppart}).  This is a clear sickness of the partially
quenched theory, reflecting the fact that it does not correspond to
a standard field theory in Minkowski space.  We should expect, however,
that if we take both $i$ and $j$ to correspond to sea flavors, that
everything works as in the full theory: by design the sea sector does not know
about valence and ghost quarks!  Indeed, if we pick both $i$ and $j$ one
of $u,d,s$, we see that at least one of these poles cancel against a factor
in the numerator, leaving us with an expression that can be written as
a sum over single poles.  (This also works if there is a degeneracy in
the sea sector.)  It is a nice exercise to work out the propagators for
the sea $\p^0$ and $\eta$, and find that indeed they agree with the
expressions one finds in the full theory.  A very simple example is
the case of degenerate sea masses, $m_U=m_D=m_S\equiv m_{sea}$, for which we
obtain for the neutral sea propagator the expression
\begin{equation}
\label{seaprop}
\langle\Phi_{sea}\Phi_{sea}\rangle=\left(1-\frac{1}{3}\right)
\frac{1}{p^2+m_{sea}^2}\ ,
\end{equation}
where the ``1'' term comes from the first two terms in Eq.~(\ref{pql2}), and the
``$1/3$'' term comes from Eq.~(\ref{dppart}), and projects out the $\eta'$.

So, while the sea sector is healthy, as it should, the neutral valence
sector is sick.  So what?  Maybe this just tells us that we should
not consider flavor-neutral valence pions.\footnote{They
are numerically hard, because of disconnected diagrams, anyway.}
However, even if we make this choice, the sickness permeates to the
flavored sector as well, as I will now argue.

We should be able to calculate chiral logarithms in meson masses,
decay constants, \etc, from Eq.~(\ref{pql2}) as usual.  The four-point vertices
following from $\cl^{(2)}$ have a flavor structure
$\Phi_{ij}\Phi_{jk}\Phi_{kl}\Phi_{li}$, dressed up with derivatives
or quark mass factors.  Consider a one-loop contribution to
$\langle\Phi_{ij}\Phi_{ji}\rangle$, with $i\ne j$ so as to stay away from the
dangerous neutral sector.  Because we only have four-point vertices,
all one-loop diagrams are tadpoles, with $\Phi_{ij}$ and $\Phi_{ji}$
connected to the same vertex.  One of the Wick contractions is the one
in which the index structure on the vertex is $\Phi_{ij}\Phi_{jj}\Phi_{ji}
\Phi_{ii}$, where $\Phi_{ii}$ and $\Phi_{jj}$ have to contract and form
the loop!  We see that the neutral propagator terms of Eq.~(\ref{dppart})
unavoidably show up, if we want to make use of partial quenching at all.
For simplicity, let us work out the degenerate case
$m_U=m_D=m_S=m_{sea}$, $m_{ii}=m_{jj}\equiv m_{val}$.
Simplifying Eq.~(\ref{dppart}) accordingly, and
integrating over $p$, as one would in the one-loop diagram, we obtain
\begin{eqnarray}
\label{echlog}
-\frac{1}{3}\int\frac{d^4p}{(2\p)^4}\;\frac{p^2+m_{sea}^2}{(p^2+m_{val}^2)^2}
&=&\frac{1}{3}\;\frac{d\phantom{m_{val}^2}}{dm_{val}^2}
\int\frac{d^4p}{(2\p)^4}\;\frac{p^2+m_{sea}^2}{p^2+m_{val}^2}\\
&\to&\frac{1}{48\p^2}(m_{sea}^2-2m_{val}^2)
\log\left(\frac{m_{val}^2}{\L^2}\right)\ .\nonumber
\end{eqnarray}
The arrow in the second line indicates that I only kept the chiral
logarithm.  This chiral logarithm is not of the structure we encounter
in full QCD, where no infrared divergences can occur.  Here there is
an infrared divergence: Eq.~(\ref{echlog}) diverges when we take the valence
mass to zero at fixed sea mass.  Only when we take these masses in
a fixed ratio can we define the limit.  This means that the chiral
expansion in the partially quenched theory will not converge
if we take the sea and valence masses too different.\footnote{For a
heuristic argument that such infrared divergences are a property of
QCD, and not an artifact of ChPT, in the quenched case, see Ref.~\cite{BGHS}.}  We should not
only keep $m_{val}^2$ and $m_{sea}^2$ small in the sense of the chiral
expansion, but we should also keep the unphysical infrared scale
$m_{sea}^2-m_{val}^2$ small compared to either of those,
$|m_{sea}^2-m_{val}^2|\,\ltap\,{\rm min}(m_{sea}^2,m_{val}^2)$.  For
current numerical simulations, these conditions are well satisfied,
so that the infrared divergences of the partially quenched theory do
not constitute a problem in practice.

There are many examples of the unphysical infrared behavior of the
partially quenched theory --- the phenomenon is generic.  Here I briefly
mention two of them; for detailed explanations, see the original papers
quoted below.

The first example occurs in
scalar-isoscalar and scalar-isovector propagators \cite{scalarpapers}.
In that case, there are  one-loop
contributions with $\p-\p$ or $\p-\eta$ intermediate states.  In the isopin limit,
Eq.~(\ref{dppart}) contributes to the $\eta$ internal line.  With the
additional pion internal line, this leads to an integral like Eq.~(\ref{echlog}),
but with three powers of $p^2+m_{val}^2$ in the denominator.\footnote{Using the
quark-flow picture of the next section, it is straightforward to see that all
propagators in this contribution are valence propagators.}  This leads to
an infrared divergence (at fixed $m_{sea}^2$) that goes like $(m_{sea}^2-m_{val}^2)/m_{val}^2$,
worse than the logarithmic divergence of Eq.~(\ref{echlog}).

A second example where the double pole has a dramatic effect is
in the contribution from one-pion exchange to the nucleon-nucleon potential.
In full QCD one-pion exchange leads to a Yukawa potential, $\exp(-m_{val} r)/r$,
whereas double-pole terms in the exchange lead, instead, to a potential
of the form $(m_{sea}^2-m_{val}^2)\exp(-m_{val}r)/m_{val}$, as can be seen
by differentiating with respect to $m_{val}^2$, like we did in Eq.~(\ref{echlog}) \cite{SBMS}.
The double pole thus leads to an unphysical interaction that dominates the
physical term at large distance.

In all cases, the unphysical infrared effects are, of course, proportional
to $m_{sea}^2-m_{val}^2$.  In general, the message is that partial quenching
can be a useful tool, but in order to make use of PQChPT, we do not only need
$m_{sea}^2$ and $m_{val}^2$ to be small enough, but also their difference,
$m_{sea}^2-m_{val}^2$.

Having done all the preparatory work, let us consider the $O(p^4)$ result for the
valence pion mass, in the case of degenerate sea quarks, and for
$m_x=m_y$ \cite{pqsh}:\footnote{See also Ref.~\cite{pqgl}, which also
discusses the case in which $\chi_{sea}$ is not very small compared to $m_0^2$;
this can lead to a reduction of the coefficients of chiral logarithms.}
\begin{eqnarray}
\label{pqpionmass}
m_X^2&=&2B_0 m_x\Biggl\{1+\frac{2B_0}{24\p^2 f^2}\left((2m_x-m_{sea})
\log\left(\frac{2B_0 m_x}{\L^2}\right)+m_x-m_{sea}\right)\nonumber\\
&&\hspace{11mm}+\frac{32B_0}{f^2}\left((2L_8-L_5)m_x+3(2L_6-L_4)m_{sea}\right)
\Biggr\}\ .
\end{eqnarray}
The chiral logarithms in this expression precisely come from the
``double-pole'' part of the neutral propagator, Eq.~(\ref{dppart}), \ie, they
are of the type Eq.~(\ref{echlog}).\footnote{All contributions to the self energy
with mesons containing sea quarks on the loop get absorbed by the
wavefunction renormalization.}

This result, finally, allows us to return to the point as to why
partial quenching is useful.  As we have seen, full QCD is contained
in PQQCD --- the only difference is that we choose different
masses for the valence and the sea quarks.   Since the LECs are, by
construction, independent of the quark masses, the LECs in PQChPT
and full ChPT are identical.  In other words, partially quenched
lattice computations give us access to the real-world values of the
LECs of ChPT \cite{ShSh1}; the fact that we can vary the valence
quark masses independently simply gives us another ``knob to turn.''
This can be very useful: it is clear that by varying the valence
quark masses independently, one can extract $2L_8-L_5$ from the
pion mass, without varying the sea quark masses.  This stands in
contrast to the full theory: as can be seen from Eq.~(\ref{oneloopresults}),
in order to separate $2L_8-L_5$ from $2L_6-L_4$, one would have to
vary the sea quark mass.  It is important to note that the LECs {\em do}
depend on the number of ``dynamical'' flavors, $N_s$ --- one still has to
do the numerical lattice computation with the correct number of sea
quarks.

While it is true that the {\em values}
of the LECs in PQChPT and full ChPT are the same, there can be {\em more}
LECs in the partially quenched case.  This already shows up at order $p^4$:
the operator $\tr(L_\m L_\n L_\m L_\n)$ is not independent  from
those in Eq.~(\ref{l4}) for $SU(3)$
\seef\ Sec.~\ref{lagr}, but $\str(L_\m L_\n L_\m L_\n)$ is for
$SU(3+N_v|N_v)$ \cite{pqshvdw}.\footnote{For another example
relevant for weak matrix elements, see Ref.~\cite{LaSo}.  For a discussion of the
implications of graded group representation theory for ``external operators''
such as electroweak operators, see Ref.~\cite{MGEP}.}
One thus obtains the most general
$O(p^4)$ lagrangian for PQChPT by replacing traces with supertraces in
Eq.~(\ref{l4total}), and by adding
\begin{equation}
\label{l4pqop}
L_{PQ}\left(\str(L_\m L_\n L_\m L_\n)+2\;\str(L_\m L_\m L_\n L_\n)
+\frac{1}{2}\left(\str(L_\m L_\m)\right)^2-
\str(L_\m L_\n)\;\str(L_\m L_\n)\right)\ ,
\end{equation}
where the new operator is written in this way because this combination
vanishes for $SU(3)$.  This new operator does not contribute at tree level,
because at tree level there is no distinction with full QCD, for which $SU(3)$ is the
relevant symmetry group.  It can
contribute to one-loop diagrams, and does so in the case of pion
scattering, for instance \cite{pqshvdw}, where, of course, this only
happens if valence masses are chosen different from sea masses.

Finally, we remark that quenched QCD, and thus quenched ChPT corresponds
to the special case that $N_s=0$ \cite{BGQ,SQ}.  While we are not pursuing quenched
QCD in these lectures, it is worth noting that this special case has,
in fact, some special properties.  For example, one cannot take the limit
$m_0^2\to\infty$, as inspection of Eq.~(\ref{dppart}) reveals (the factor in
parentheses on the first line gets replaced by one).  Of course, quenched
QCD does not have the correct number of light sea quarks, and is thus
not only sick, but terminally ill.  No physical result with controlled
errors can ever be obtained from the quenched approximation.

%%####%%
\subsection{\label{quark flow} Quark flow}
%%####%%
In the previous section, we have seen that PQChPT can be systematically
developed for PQQCD.  Once we assume that this EFT exists just as in the
full-QCD case, the construction is quite straightforward.  But, one may like to
have a more pictorial understanding how it all works, and this is provided by
the quark-flow picture (which, in a sense, is formalized through the replica trick,
but here I will not attempt to present it that way).

We have two types of quarks: valence and sea quarks.  Valence quarks are
represented by propagators, as in Eq.~(\ref{pion2pt}), and in a diagrammatic
language, can be represented by valence quark lines.  Each valence quark line
has to have a beginning and an end at some external source or sink.  Valence
quark loops can be formed for instance if more than one valence line connects the same
source and sink, as in the example of Eq.~(\ref{pion2pt}).  The ``valence quark
part'' of a contribution to a correlation function can thus be precisely defined,
and the topology of valence lines represents the various possible Wick contractions
which contribute to a particular correlation function.

%%%%%%%%%%%%%%%%%%%
\begin{figure}
\setlength{\unitlength}{1mm}
\vspace*{4ex}
\begin{center}
\begin{picture}(60,40)(0,0)
\put(30,3){\makebox(0,0){($a$)}}
\put(10,10){\includegraphics*[width=4cm]{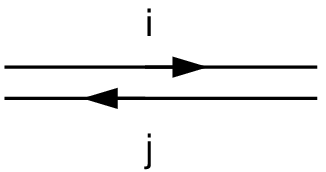}}
\end{picture}
\begin{picture}(60,40)(0,0)
\put(20,3){\makebox(0,0){($b$)}}
\put(10,19){\includegraphics*[width=4cm]{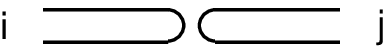}}
\end{picture}
\begin{picture}(60,50)(0,0)
\put(30,-5){\makebox(0,0){($c$)}}
\put(10,1){\includegraphics*[width=4cm]{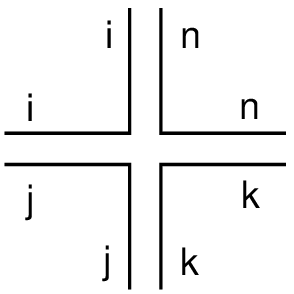}}
\end{picture}
\begin{picture}(60,50)(0,0)
\put(30,-5){\makebox(0,0){($d$)}}
\put(10,1){\includegraphics*[width=4cm]{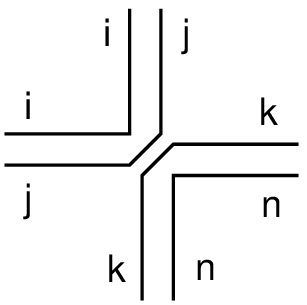}}
\end{picture}
\end{center}
\vspace*{3ex}
\begin{quotation}
\caption{Quark flow diagrams for ChPT Feynman rules.  ($a$): flavored
propagator, ($b$): $m_0^2$ part of the flavor-neutral propagator, \seef\ Eq.~(\ref{dppart}),
($c$): single-trace four-point vertex, ($d$): double-trace four-point vertex.
Arrows on quark lines are shown in ($a$), and understood in the other figures.  See text.
\label{fig:quarkflow}}
\end{quotation}
\vspace*{0ex}
\end{figure}
%%%%%%%%%%%%%%%%%%%

%%%%%%%%%%%%%%%%%%%
\begin{figure}
\setlength{\unitlength}{1mm}
\vspace*{4ex}
\begin{center}
\begin{picture}(45,40)(0,0)
\put(22,3){\makebox(0,0){($a$)}}
\put(2,10){\includegraphics*[width=4cm]{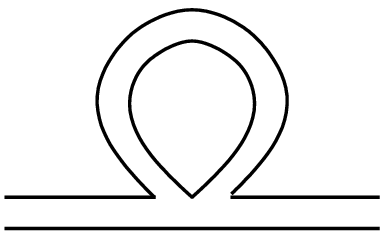}}
\end{picture}
\begin{picture}(45,40)(0,0)
\put(22,3){\makebox(0,0){($b$)}}
\put(2,10){\includegraphics*[width=4cm]{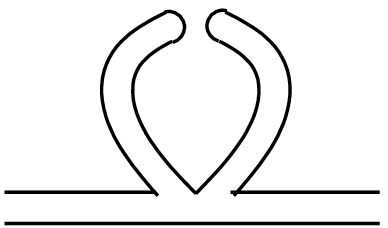}}
\end{picture}
\begin{picture}(45,40)(0,0)
\put(22,3){\makebox(0,0){($c$)}}
\put(2,10){\includegraphics*[width=4cm]{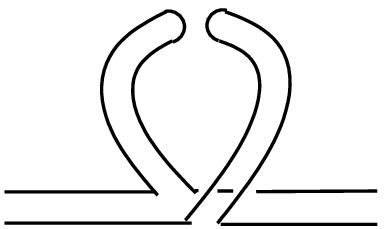}}
\end{picture}
\end{center}
%\vspace*{3ex}
\begin{quotation}
\caption{Quark-flow diagrams for various one-loop contributions to the pion self energy.
\label{fig:selfenergyqf}}
\end{quotation}
\vspace*{-4ex}
\end{figure}
%%%%%%%%%%%%%%%%%%%

In perturbation theory, the (logarithm of the) fermion determinant, which comes only from sea quarks, can be represented by an infinite sum over one-loop diagrams with any number
of gluon lines attached to the sea-quark loop.  One thus
imagines the diagram coming from a certain Wick contraction in the valence sector
to be dressed up with gluons and sea quark loops in all possible ways.  In this
way one can also represent sea quark contributions through loops, but clearly
this picture is not rigorous outside perturbation theory.
In this picture, however, the reason that only
internal (\mbox{\ie,} not connected to a source or sink) sea quark loops occur is because
valence and ghost loops, which in principle also are present,\footnote{For instance,
they would be present
if one takes the ghost masses different from the valence masses.} cancel each
other.  It is instructive to consider PQChPT in terms of quark loops.

Figure~\ref{fig:quarkflow}
shows quark-flow representations of the meson propagators at tree level,
as well as the four-point meson vertex.  The flavored propagators that follow from
Eq.~(\ref{pql2}) are represented in ``double-line'' notation, with each line denoting a
quark of a certain flavor, Fig.~\ref{fig:quarkflow}a.
Note that we only draw valence lines --- one is to
imagine these diagrams arbitrarily dressed up with gluons and sea quark loops.
In the flavor-neutral sector, we have seen that there
are additional contributions, which follow from iterating the two-point vertices
proportional to $m_0^2$.   Flavor-neutral mesons get a contribution as in Fig.~\ref{fig:quarkflow}a
(just set $i=j$), but also get a contribution from Fig.~\ref{fig:quarkflow}b, where the ``double hairpin''
represents the $m_0^2$ two-point vertices.  This is what we usually call a
``disconnected'' diagram in lattice QCD, because it represents a disconnected
valence Wick contraction, but again, it is not really disconnected in terms of
gluons.  Note that the double hairpin stands for the full geometric sum~(\ref{dppart}) of all
insertions of the $m_0^2$ vertices, where each additional insertion of a double-hairpin
vertex corresponds to another sea-quark loop inserted between the two hairpins.

Figure~\ref{fig:quarkflow}c represents a vertex with flavor structure $\Phi_{ij}\Phi_{jk}\Phi_{kn}
\Phi_{ni}$, which follows from the single-trace terms in $\cl^{(2)}$, whereas
Fig.~\ref{fig:quarkflow}d represents the flavor structure $\Phi_{ij}\Phi_{ji}\Phi_{kn}\Phi_{nk}$,
which is a double-trace term.  The latter do not occur in $\cl^{(2)}$, but they
do occur in $\cl^{(4)}$.    Various one-loop contributions to a pion propagator
are shown in Fig.~\ref{fig:selfenergyqf}.  Each internal quark loop has to be summed over only the
sea quarks, because valence and ghost internal loops cancel each other.
It so happens that only diagrams of type~\ref{fig:selfenergyqf}c contribute at order $p^4$ to meson masses; more types
contribute to decay constants.

\begin{figure}
\setlength{\unitlength}{1mm}
\vspace*{4ex}
\begin{center}
\begin{picture}(55,40)(0,0)
\put(28,3){\makebox(0,0){($a$)}}
\put(3,10){\includegraphics*[width=5cm]{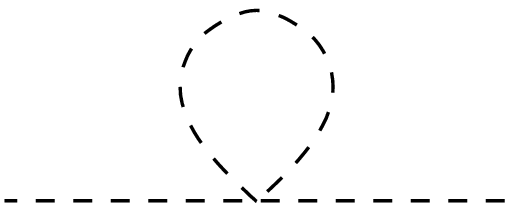}}
\end{picture}
\begin{picture}(55,40)(0,0)
\put(28,3){\makebox(0,0){($b$)}}
\put(3,10){\includegraphics*[width=5cm]{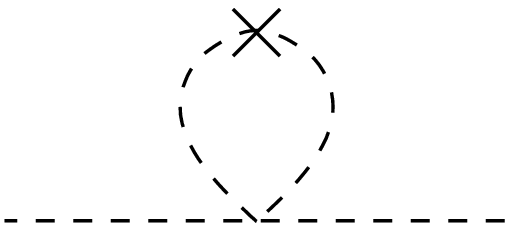}}
\end{picture}
\end{center}
\begin{quotation}
\caption{ChPT diagrams for the pion one-loop self energy.  A dashed line stands for
one of the meson fields in $\Phi$.  The cross stand for an insertion of Eq.~(\ref{dppart}).
\label{fig:selfenergy}}
\end{quotation}
\vspace*{-4ex}
\end{figure}

{}From these examples, it is clear that there is
no one-to-one correspondence between PQChPT diagrams and quark-flow
diagrams.  The PQChPT diagram representing all diagrams of Fig.~\ref{fig:selfenergyqf} is just the
tadpole shown in Fig.~\ref{fig:selfenergy}a.  One can refine this a little by representing the hairpin
vertex by a cross, as in Fig.~\ref{fig:selfenergy}b.  Of course, one can also use the quark-flow
picture for correlation functions in full ChPT, and one can use those as a starting
point to develop PQChPT intuitively, as was originally done for quenched ChPT
\cite{SQ}.  It can be very helpful in seeing how ChPT should work for some ``modified'' version of QCD.  For an example of applying this to staggered ChPT for
staggered QCD with the fourth-root procedure, see Ref.~\cite{AB}.  However, it is important
to have a path-integral representation for each modified version of QCD,\footnote{This
is the problem with staggered QCD with the fourth-root procedure, see Ref.~\cite{MGreview}.}
 as this gives us
the opportunity to use many of the standard tools of field theory.  Examples of these
tools are the precise definition and role of the symmetries of the modified theory,
and the availability of field redefinitions \cite{BGPQ,BGQ}.

%%%%%%%%%%%%%%%%%%%
%%%%%%%%%%%%%%%%%%%

Double-hairpin diagrams as in Fig.~~\ref{fig:quarkflow}b contribute only to flavor-neutral
propagators (at tree level).  The quarks on each of the hairpin lines are valence quarks,
and the quark-antiquark pair that each of these hairpins represents gets
bound into a valence meson --- we thus expect each of these hairpins to
represent a single meson pole with the mass of a valence pion.
Any other quark loops in this diagram are not explicitly shown, and
can only come from the fermion determinant.  In other words, any
such quark-loop contributions can only involve the sea quarks, and,
if sea quark masses do not equal valence quark masses, nothing in the
diagram can remove the two valence-pion poles.  We conclude, at this
heuristic level, that double pole terms have to occur in neutral pion
propagators in PQQCD, and this is, of course, confirmed in PQChPT.\footnote{For
a much more detailed argument along these lines, see Section 6 of Ref.~\cite{ShSh2}.}
Because double poles cannot occur in full QCD, we also conclude that
the residue of these double poles has to be proportional to $m_{sea}-m_{val}$.

Another observation that is easily made in the quark flow picture is
how PQQCD gives access to Wick contractions not available in full QCD.
Suppose we do take all valence masses equal to the sea masses, so that
the theory is not really partially quenched.  The only effect of
partial quenching is that we have more flavors available to put on the
external lines than run around on the internal loops which come from the
fermion determinant.  We can use this, for example, to separate the
disconnected part, Fig.~\ref{fig:quarkflow}b, from the connected part,
Fig.~\ref{fig:quarkflow}a, of any
flavor-neutral meson propagator.   In other words, partial quenching
separates the contribution of different Wick contractions to the same
correlation function.
At the level of quark flow diagrams,
this is obvious.  But since all that happens at the quark flow level is
encoded in PQChPT, we can use this to obtain systematic chiral
expansions for these separate Wick contractions.
A simple example follows from Eqs.~(\ref{conn}) and~(\ref{dppart}).  In QCD
with one flavor, these two contributions cannot be separated.  But, one can
add valence quarks to the theory (keeping the number of sea quarks at just
one), and use this to separate the connected and disconnected parts~(\ref{conn})
and~(\ref{dppart}) even in QCD with one flavor.  This can for instance be used
to define what it means to set the quark mass to zero, by requiring that the
pion mass obtained from the connected part vanishes \cite{Fetal}.

Another example of this
is given in Ref.~\cite{ShSh2}, in which it is shown how this can be used
for a determination of $L_7$.  For other work  using the
``extra'' quarks available in the partially quenched theory to separate
connected and disconnected Wick contractions, see Ref.~\cite{JCMS} for
an application to electromagnetic properties of baryons, Ref.~\cite{QKD}
for an application to quenched and partially quenched nonleptonic kaon
decays, and Ref.~\cite{JDM} in the context of the hadronic contribution 
to $g-2$.

%%####%%
\subsection{\label{mixed} ChPT for mixed actions}
%%####%%
QCD with a mixed action is a generalization of PQQCD, because not only
do we treat the valence and sea quark masses as independent, but also
their lattice Dirac operators.   An important consequence is that
mixed-action QCD has fewer flavor symmetries than its partially quenched
cousin \cite{Baeretal2}.  Revisiting the general observations in support of the
validity of PQQCD in Sec.~\ref{PQ}, we note that observations 1 and 3 still
apply, but that observation 2 has to be modified.  It is still true that full QCD
is a subset of mixed-action QCD, but it is no longer true that one can rotate
sea quarks into valence quarks.

If we would consider a theory with two
sets of sea quarks, with  each set  discretized with a different lattice Dirac
operator, there would be little doubt that this constitutes a valid discretization
of QCD, because of universality.  One would expect that the framework of
Sec.~\ref{lattice spacing} applies to such a theory as well, with the necessary
technical modifications.  In mixed-action QCD, however, one of these
sets is quenched, making it less obvious that a universal continuum limit
exists.  But if it exists, it will clearly have to be the corresponding partially
quenched version of QCD.
Hence, if one combines the assumption that PQChPT is the correct
EFT for PQQCD with the notion of universality of different discretizations,
it appears reasonable that we can combine the techniques of Secs.~\ref{lattice spacing}
and \ref{PQ} to construct a chiral EFT for mixed-action QCD.

The relevant symmetry group clearly depends on what discretization we
use for the valence and sea quarks.  For definiteness, let us consider the case
that we choose Wilson fermions for the sea quarks, and overlap fermions\footnote{
Or domain-wall fermions with a very small residual mass.} for
the valence quarks.\footnote{For the case of overlap fermions on a staggered sea,
see Ref.~\cite{BBRS}.}  The lattice symmetry group (ignoring the soft breaking
by quark masses, and removing ``overall'' $U(1)$ factors as usual) is then
$SU(N_s)_V\times[SU(N_v|N_v)_L\times SU(N_v|N_v)_R]$.  With overlap fermions,
the symmetry group in the valence sector is the same as in the continuum,
but Wilson  fermions have no chiral symmetry, and the symmetry group is thus
the smaller group of vector-like transformations only.

We can now construct the chiral lagrangian (valid in the GSM regime to
order $p^4$, and in the LCE regime to order $p^2$) by writing down the
SET for the mixed-action theory, introducing spurions, and making the transition
to the chiral lagrangian as we did in Sec.~\ref{lattice spacing}, and this has been
done systematically in Ref.~\cite{Baeretal}.  However, it is rather simple to understand
the result, so, rather than repeat the whole analysis, I will first quote the
result, and then comment on the various terms.  The result is
\begin{eqnarray}
\label{mixedChPT}
\cl&=&\cl_{cont}^{PQQCD}(2B_0M_s\to 2B_0M'_s\equiv 2B_0M_s+\ha{\bf 1})\\
&&+\D\cl^{(4)}(\ha\to\ha P_s)-\ha^2 W_M\;\str(P_s\S P_s\S^\dagger)\ ,\nonumber
\end{eqnarray}
with $\D\cl^{(4)}$ given by Eq.~(\ref{newp4}), and the subscript $cont$ stands for
``continuum.''  Here $P_s$ is the projector on the sea
sector, defined by $P_s q_s=q_s$ and $P_s q_v=0$, $P_s\tq=0$.  While this
equation contains a lot of information, it is easy
to understand this result.  First, since the valence quarks have exactly the
same chiral symmetries as in the continuum, there are no explicit $O(a^2)$
terms in the valence sector of the chiral lagrangian.  The only such terms
come from the Wilson sea sector, and this explains the appearance of the
projector $P_s$ in Eq.~(\ref{mixedChPT}), and the fact that only the sea quark
masses get shifted by an $O(a)$ term, as in Eq.~(\ref{shift}).  Of course, there
are scaling violations in the valence sector as well, but they can only occur
as $O(a^2)$ corrections to the continuum LECs, $f$, $B_0$ and $L_i$;
but in the GSM regime these are of order $p^6$.\footnote{To the best of my
knowledge, the $O(p^4)$ chiral lagrangian with LCE power counting has
not been constructed.}
Rotational symmetry is always broken on the lattice, but, as we have seen
before, the effects of that only shows up at higher orders in the chiral lagrangian.
Essentially, the form of the lagrangian~(\ref{mixedChPT}) is thus that of the continuum
lagrangian, plus the symmetry-breaking lattice part for the sea sector.
However, one completely new term does appear in Eq.~(\ref{mixedChPT}), and
that is the term with the new LEC $W_M$.   This term is allowed, because there are
no symmetries that connect the valence and sea sectors.   The new term does not
break (continuum) chiral symmetry restricted to the sea sector, so a similar term
also appears if we choose different types of sea quarks, such as staggered
or domain-wall quarks.   Since this new term does not break chiral symmetry,
a similar term also arises in the valence sector.  But since we have that
$P_v={\bf 1}-P_s$ and $\S\S^\dagger={\bf 1}$, we can write
\begin{equation}
\label{vsrel}
\str(P_v\S P_v\S^\dagger)=\str(P_s\S P_s\S^\dagger)+\str({\bf 1})-2\,\str(P_s)\ ,
\end{equation}
so only one of these terms is independent.   If we take the valence quarks to
be the same as the sea quarks, the
projector $P_s$ is replaced by the unit matrix, and the new term is an irrelevant
constant.  But with a mixed action this is not the case, and
the new term should be taken into account.

Let us consider the $O(p^2)$ expressions for meson masses in the LCE regime, beginning with
the flavored ones.  By expanding the lagrangian to quadratic order in $\Phi$,
we find, using now Latin indices for the valence flavors and Greek indices for sea
flavors, in order to distinguish them more clearly:
\begin{eqnarray}
\label{mixedmasses}
m_{ij}^2&=&B_0(m_i+m_j)\ ,\\
m_{\a\b}^2&=&B_0(m'_\a+m'_\b)+\frac{32\ha^2}{f^2}\left( N_sW'_6+W'_8\right)
\ ,\nonumber\\
m_{i\a}^2&=&B_0(m_i+m'_\a)+\frac{4\ha^2}{f^2}\left(4N_sW'_6+2W'_8+W_M\right)
\ .\nonumber
\end{eqnarray}
The factors $N_s$ in these expressions come from factors $\str(P_s)$ which
occur working out the expansion in $\Phi$.
The new LEC $W_M$ only shows up in the ``mixed'' pion mass, \ie, the pion
made of a valence and a sea quark, hence there is no $O(p^2)$ relation
between the pure valence, pure sea, and mixed pion masses.\footnote{In the
GSM regime there is, because the $O(a^2)$ terms are of higher order in that
regime.  In the GSM regime we do thus have that $m_{mixed}^2=(m_{sea}^2+
m_{val}^2)/2$ to lowest order.}  In any application of QCD with a mixed action, the mixed pion
mass gives thus an important indication whether one should use GSM or LCE
power counting.

In the flavor-neutral sector, all the complications of PQQCD of course are
inherited by the mixed-action case, including the appearance of double-pole
terms, but a new issues also arises.  In the sea sector, to quadratic order, there is an additional contribution
to the $m_0^2$ term in Eq.~(\ref{pql2}), coming from the $W'_7$ term in Eq.~(\ref{newp4})
shifting the value of $m_0^2$ {\em only in the sea sector} (because of the
$P_s$ projectors, \seef\ Eq.~(\ref{mixedChPT})) effectively to
\begin{equation}
\label{m0shift}
m_{0,eff}^2=m_0^2+\frac{96\ha^2}{f^2}W'_7\ .
\end{equation}
For a theory with degenerate valence and degenerate sea quarks, one finds
for the flavor-neutral propagator in the valence sector \cite{GIS}
\begin{eqnarray}
\label{mixeddp}
\langle\phi_{ii}\phi_{jj}\rangle&=&
\frac{\d_{ij}}{p^2+m_{val}^2}-\frac{1}{3}\;
\frac{m_0^2(p^2+m_{sea}^2)+(N_s m_0^2/3)(m_{0,eff}^2-m_0^2)}
{(p^2+m_{val}^2)^2(p^2+m_{sea}^2+N_s m_{0,eff}^2/3)}\\
&\to&
\left(\d_{ij}-\frac{1}{N_s}\right)\frac{1}{p^2+m_{val}^2}-
\frac{(m_{sea}^2-m_{val}^2)/N_s+32\ha^2 W'_7/f^2}
{(p^2+m_{val}^2)^2}\ ,\nonumber
\end{eqnarray}
where the second line is obtained by taking $m_0\to\infty$ in the first line
of this equation.

The importance of this result lies in the fact that scaling violations are
not universal.  Naively, one might think that the best way to conduct a mixed-action
lattice computation is to tune the quark masses such that the valence pion
mass is equal to the sea pion mass.  However, comparison between Eqs.~(\ref{mixedmasses}) and~(\ref{mixeddp}) shows that this is not the same as
tuning quark masses such that the residue of the double pole vanishes.
If one tunes quark masses to make sea and valence pion masses equal,
there will still be a double pole in the theory, with a residue of order $a^2$.
Since the double pole can lead to infrared ``enhancement'' effects (such as
the enhanced chiral logarithms and other such examples we encountered in
Sec.~\ref{PQ}), such scaling violations could be larger than one would expect based
on a simple estimate of the size of $(a\L_{QCD})^2$.
This phenomenon does not occur at this order in the case that one uses overlap valence
quarks on a staggered sea; because both valence and staggered quarks have exact
chiral symmetries, in that case the effect can occur at the earliest at order $a^2m_{quark}^2$ \cite{GIS,BBRS}.

%%####%%
\subsubsection{\label{universality} One-loop universality}
%%####%%

\vspace*{0mm}
As soon as the lattice spacing is large enough that we need to include
it as a variable in chiral fits, the version of ChPT we need to use will
depend in some detail on what kind of lattice action is being used.  For instance,
ChPT for Wilson fermions is quite different from ChPT for staggered fermions.
This is in general also true when a mixed action is used.  This is not
a problem of principle, but for each new type of lattice action, the
necessary ChPT calculations need to be repeated.

It would be very nice if some general prescription existed that allowed one
to take a calculation carried out in continuum PQQCD, and modify it using
some simple rules into the appropriate mixed-action ChPT expressions.
While this is clearly not possible in general, it is possible under
some limitations, which however cover many cases of practical interest.

As we noted before, the use of a mixed action is particularly useful
when the valence fermions have exact chiral symmetry (broken only by the physical
quark masses), while the sea sector has much less symmetry.  If we allow
only valence quarks on the external lines of any correlation function of
interest, in order to access what we will refer to as ``valence quantities,'' and we work to leading order (\ie, tree level) in ChPT, one
clearly needs just that part of the $O(p^2)$ chiral lagrangian that
refers only to the valence sector --- one can simply set all sea-meson
and mixed-meson fields equal to zero.  This is nothing else than the
observation that sea quarks (and thus sea and mixed mesons) can only
appear on loops.  If the lattice valence quarks have exact chiral symmetry,
the $O(p^2)$ chiral lagrangian is the same as in the continuum, \ie, it takes
the form of Eq.~(\ref{l2again}).

What is remarkable is that a similar argument holds at order $p^4$,
\ie, to one loop \cite{COW}.  For one-loop contributions we need
only the $O(p^2)$ lagrangian, and, in one-loop diagrams with {\em only
valence quarks on the external legs}, only one sea quark loop can occur, as can easily be
seen in the quark-flow picture.\footnote{For disconnected contributions, \seef\ the
second term in Eq.~(\ref{mixeddp}), see below.}
Let us see how this works in the case that
the sea quarks are Wilson-like, and that we are in the LCE regime.  To order $p^2$,
the chiral lagrangian is then
\begin{eqnarray}
\label{mixedLCE}
\cl_{LO} &=&\frac{1}{8}f^2\,\str\left((D_\m\S)^\dagger D_\m\S\right)-\frac{1}{4}B_0f^2\,\str\left(\cm\S+\S^\dagger\cm\right)\\
&&-\ha^2 W'_6\left(\str(P_s\S+\S^\dagger P_S)\right)^2-\ha^2 W'_7\left(\str(P_s\S-\S^\dagger P_s)\right)^2\nonumber\\
&&-\ha^2 W'_8\;\str(P_s\S P_s\S+\S^\dagger P_s\S^\dagger P_s)
-\ha^2 W_M\;\str(P_s\S P_s\S^\dagger)\ ,\nonumber
\end{eqnarray}
from Eqs.~(\ref{l2again}) and~(\ref{newp4}), where I have inserted the projectors $P_s$, and in which the $O(a)$ term has been absorbed into
the sea-quark mass matrix, as in Eq.~(\ref{mixedChPT}).    As noted,
we can have only one sea quark on the loop.  This implies that if we expand out $\S$ in
terms of $\Phi$, we will only use those terms in the expansion of any of the
$O(a^2)$ operators in which two of the indices on the $\Phi$ fields correspond
to sea quarks (these two indices correspond to the ``beginning'' and the ``end'' of the
sea quark loop), with the rest of the indices referring to valence quarks.  Because of the
$P_s$ projectors in the $O(a^2)$ terms, such terms only arise when we set
one $\S$ (or one $\S^\dagger$) equal to one, in all possible ways in which this
can be done.  Applying this ``1-loop valence'' rule to Eq.~(\ref{mixedLCE}), it simplifies to
\begin{eqnarray}
\label{mixedLCEsimple}
\cl_{LO}\bigg|_\mathrm{1-loop\ valence\ rule}&=&\frac{1}{8}f^2\,\str\left((D_\m\S)^\dagger D_\m\S\right)-\frac{1}{4}B_0f^2\,\str\left(\cm\S+\S^\dagger\cm\right)\nonumber\\
&&-\ha^2\left(4N_s W_6'+2W_8'+W_M\right)\;\str\left(P_s\S+\S^\dagger P_s\right)\ .
\end{eqnarray}
In other words, all dependence (to one loop) on the LECs $W_{6,8}'$ and $W_M$
can be absorbed completely into a shift of the sea-quark mass matrix of the form
\cite{CGOW}
\begin{equation}
\label{2ndshift}
\mbox{1-loop\  valence\ rule:}\quad
B_0M_s\to B_0M_s+\frac{4\ha^2}{f^2}\left(4N_s W_6'+2W_8'+W_M\right){\bf 1}\ .
\end{equation}
Note, however, that this simplified form {\em cannot} be used to calculate any quantity involving
sea quarks!  For instance, the leading-order meson mass for a meson made only out
of sea quarks is given by the middle equation of Eq.~(\ref{mixedmasses}), which does
{\em not} correspond to shifting the sea quark masses as in Eq.~(\ref{2ndshift});
in particular, the LEC $W_M$ does not occur.
It also does not apply to the disconnected term in Eq.~(\ref{mixeddp}), in which again
the sea meson masses are given by Eq.~(\ref{mixedmasses}).   However, since this is the
only place where the sea meson mass shows up at one loop in valence quantities, the LEC $W_7'$
can be absorbed into the sea meson mass-squared, for the one-loop contributions to
valence quantities.

In summary, the one-loop
expressions of continuum PQChPT for valence quantities, such as meson masses, decay
constants and scattering amplitudes \cite{COWW},
carry over to the mixed case, if one takes the
mixed meson mass to be given by the last line of Eq.~(\ref{mixedmasses}),
\begin{equation}
\label{mmsubst}
m^2_{i\a}=B_0(m_i+m'_\a)+\frac{4\ha^2}{f^2}\left(4N_sW_6'+2W_8'+W_M\right)\ ,
\end{equation}
and the sea meson mass by
\begin{equation}
\label{smsubst}
m^2_{\a\b}=B_0(m'_\a+m'_\b)+\frac{32\ha^2}{f^2}\left(N_sW_6'+W_8'+N_sW_7'\right)\ ,
\end{equation}
by combining Eqs.~(\ref{mixedmasses}) and~(\ref{mixeddp}).  It is only through these
masses that the LECs $W_{6,7,8}'$ and $W_M$ occur.

In a calculation to order $p^4$, there are also tree-level contributions originating from
terms in the chiral lagrangian of order $p^3$ and $p^4$, \seef\ Eq.~(\ref{LCEeq}).
Single-trace terms containing a $P_s$ thus do not contribute.  In double-trace terms
any $\S$ multiplied by a $P_s$ should be set equal to one.  That means that such
terms do contribute, but again they can be absorbed into a redefinition of other
LECs.    For instance, the $W_4$ and $W_6$ terms in Eq.~(\ref{newp4}), reduce to
\begin{eqnarray}
\label{W6red}
&&\ha W_4\;\str\left(\partial_\m\S\partial_\m\S^\dagger\right)\;\str\left(P_s\S+\S^\dagger P_s\right)\\
&&-2\ha B_0W_6\;\str\left(\cm\S^\dagger+\S\cm\right)\;\str\left(P_s\S+\S^\dagger P_s\right)\nonumber\\
\to&&
2N_s\ha W_4\;\str\left(\partial_\m\S\partial_\m\S^\dagger\right)
-4N_s\ha B_0W_6\;\str\left(\cm\S^\dagger+\S\cm\right)\ ,\nonumber
\end{eqnarray}
or
\begin{equation}
\label{B0W6}
f^2\to f^2+\frac{16\ha}{f^2}N_s W_4\ ,\ \ \ \ \
B_0\to B_0\left(1+\frac{16\ha}{f^2}N_s W_6\right)\ .
\end{equation}
{\em All} double-trace terms in the chiral lagrangian need to be treated this way, leading
to more redefinitions of the $O(p^2)$ LECs $B_0$ and $f^2$.  In this way the redefined
LECs $B_0$ and $f$ also pick up dependence on the sea quark masses, from the $L_4$
and $L_6$ terms in Eq.~(\ref{l4total}).  Whether this is convenient or not depends on the
application.   Note that if one trades sea quark masses for sea-meson masses, one has to account
for the leading-order $O(a^2)$ shifts of Eq.~(\ref{mixedmasses}).  For more discussion,
including the example of $I=2$ $\p\p$ scattering to order $p^4$, see Ref.~\cite{COW}.

In summary, what we find is that, to order $p^4$, the formulas of continuum PQChPT can be used
for correlation functions with only valence quarks on the external legs,
with a suitable redefinition of the mixed and sea meson masses and the $O(p^2)$ LECs.
In other words, once these redefinitions have been made, the theory does not
remember the fact that the sea quarks are Wilson like.  That means that we uncovered
something universal \cite{COW}:  The same argument should also apply when the
sea quarks are staggered!  And indeed, it does: For example, it is straightforward to check
from Eq.~(\ref{stagpots}) that
the shift in $M_s$ corresponding to Eq.~(\ref{2ndshift}) is
\begin{equation}
\label{stagshift}
\mbox{1-loop\  valence\ rule:}\quad
B_0M_s\to B_0M_s+\frac{4a^2}{f^2}\left(C_1+4C_3+4C_4+6C_6+C_{mix}\right)\ ,
\end{equation}
where $C_{mix}$ is the staggered equivalent of $W_M$ \cite{BBRS}.

%%####%%
\section{\label{kaon} $SU(2)$ {\it versus} $SU(3)$: physics with heavy kaons}
%%####%%
So far, we have mostly discussed ChPT for QCD with three flavors
to order $p^4$.  An important question is
whether this is sufficiently high order for practical applications to
lattice QCD, and the answer is, unfortunately, that in general it is
not.  It is a fact that the kaon mass is not so small in the
real world, and even if ChPT applies to the strange quark mass as well,
the expansion to order $p^4$ may not suffice.   Indeed, $m_K^2/m_\p^2\approx
13$, suggesting that an expansion that works very well for $m_\p^2$ may not
work so well when $m_K^2$ is the expansion parameter.

One option for dealing with this problem is to go to higher order in the chiral
expansion.  In applications to phenomenology, there has been much work to
extend ChPT calculations to order $p^6$ \cite{phenop6}.  For applications
to the lattice, $O(p^6)$ expressions, which involve two-loop calculations
(as can be seen from Eq.~(\ref{pc})), need to be extended to the partially quenched
case, and ideally, also to include scaling-violation effects.   Meson masses
and decay constants have been calculated in partially quenched, three-flavor
ChPT to order $p^6$ \cite{lat2loop}, and it will be interesting to see what
happens if the two-loop chiral logarithms  are included in fits to lattice data.  Scaling violation effects have
not been calculated in ChPT to the same order.\footnote{In the GSM regime,
$m_{quark}\sim a\L_{QCD}^2$, this would be easier to carry out than in the
LCE regime, $m_{quark}\sim a^2\L_{QCD}^3$.}   While this means that
in principle lattice results would have to be first extrapolated to the continuum,
before chiral fits are performed, it may be interesting to include ``what we know,''
\ie, the continuum part at order $p^6$, because the continuum $O(p^6)$ terms are expected
to be important at larger quark masses, where scaling violations are less
important.  Up to date, something similar has been done by MILC, which included
all the analytic terms to order $p^8$ in their mass and decay constant fits,
but no logarithms beyond order $p^4$
(yet).\footnote{This means that the fitted LECs at order $p^6$
from the lattice cannot be compared to any continuum results.  While this ``hybrid"
fitting method might also affect lower-order LECs, they are expected to be much
less sensitive, because their values are predominantly determined by lattice
results at lower quark masses, where the $O(p^6)$ effects are less important
\cite{MILC2004}.}  For an early attempt to fit $N_f=2+1$ full-QCD overlap results to
$O(p^6)$ ChPT including two-loop nonanalytic terms, see Ref.~\cite{jlqcd}.

It has recently been suggested that kaon loops are not a reliable part of
ChPT \cite{JD}, because the relevant scale is $2m_K\sim 1$~GeV, rather
than $m_K$ itself.   Diagrams can be reconstructed from their cuts and poles,
and, for a meson of mass $m$, these cuts typically start at $4m^2$, and not $m^2$.
For instance,  the contribution of a kaon pair to the imaginary part of the $\p\p$
scattering amplitude starts at $s=4m_K^2$; for other examples ($F_\p$ and
the pion electromagnetic form factor), see Ref.~\cite{JD}.   Of course,
$2m_K\sim 1$~GeV is too large to trust a chiral expansion in that scale.
However, this appears to be a quantity-dependent observation.  It applies
to purely pionic quantities, since (virtual) kaons always have to contribute
in pairs to these quantities.  But if we consider for instance $\p K$ scattering,
the cut starts at $(m_\p+m_K)^2\approx m_K^2$, and the chiral expansion
for this scattering amplitude as a function of $m_K$ may have better
convergence properties.  A similar argument applies to $F_K$.

For lattice computations relevant for phenomenology, there is a different
option.  It is not difficult to do lattice computations with the strange quark
mass adjusted to (close to) its physical value, so that only extrapolations
in terms of the light quarks masses ($m_u$ and $m_d$) are needed.
In this case, we do not need three-flavor ChPT, but, rather, we can work with two-flavor
ChPT, treating the strange quark as ``heavy.''  Note, however, that we do not
integrate out the strange quark, which we cannot do, because $m_s$ is
not large compared to $\L_{QCD}$.  The actual lattice computations still
have to be done with three flavors, up, down and strange.

In the pion sector it is straightforward to develop ChPT again, one simply
starts from the group $SU(2)_L\times SU(2)_R$, instead of $SU(3)_L\times
SU(3)_R$.\footnote{In this section we will use the word ``pion'' to refer to
$\p^\pm$ and $\p^0$, because we will not think of the kaon as an approximate
Nambu--Goldstone boson.}   But the properties of kaons (its mass, decay
constant, the kaon $B$ parameter, \etc) will depend on $m_u$ and $m_d$
through interactions of kaons with pions, and since lattice computations will
still typically use unphysical values for the up and down quark masses, we
will still need to know how to extrapolate kaon properties in those light
quark masses.  We thus need to extend two-flavor ChPT systematically
to include the interactions of kaons with pions.   How to do this will be the topic of
this section.    Studies using this EFT in application to three-flavor lattice data
were recently carried out in Refs.~\cite{RBCkaon,pacscs}.
While the case of kaons is particularly simple, the basic setup
also applies to EFTs for the coupling of baryons and heavy-light mesons
to pions.

While this alternative approach may turn out to be a useful approach,
it is a more limited approach, because the kaons are not treated as
Nambu--Goldstone bosons.  For instance, while the dependence of the kaon mass
on $m_u$ and $m_d$ is predicted by this EFT, it will not go to zero in the
chiral limit.  Instead, the chiral-limit value of the kaon mass is a new
parameter in the EFT.   Another example is that in this approach, one cannot
ask certain questions, such as the value of the three-flavor condensate in the
chiral limit.  This particular question is interesting, because it has been
argued that if this condensate were very small, that would lead to the
need to redefine the power counting of ChPT, and thus rearrange the chiral
lagrangian order by order \cite{genchpt}.   To answer this question, one needs
to work with the three-flavor theory, and vary the strange mass.
 In the setup we will investigate in
this section, one only gets to ask about the two-flavor condensate, at the
physical value of the strange mass.\footnote{Recent results for both
condensates in the three-flavor theory can be found in Ref.~\cite{MILC2007}.
It is found that while the three-flavor condensate is smaller than the
two-flavor condensate, it is not very small.}

%%####%%
\subsection{\label{kaon EFT} Including a kaon in two-flavor ChPT}
%%####%%
We need to develop two-flavor ChPT for the three pions, coupled to the
isospin-$1/2$ kaon.   In the special case of the kaon, it should be possible
to deduce the results from ChPT with three flavors.  For instance, one
may take the result for $f_K/f_\p$ in Eq.~(\ref{oneloopresults}), and expand it
in the light quark mass for fixed strange quark mass.  One then absorbs the
strange quark mass dependence into the LECs, and this parameterizes the
dependence of $f_K/f_\p$ on the light quark mass, while giving up information
about the dependence on the strange quark mass.  However, this method is
not available for other hadrons such as baryons, and it is useful and instructive
to develop a method that does not rely on three-flavor ChPT.   After developing the
method, we will return to the matching with three-flavor ChPT in some examples.

The systematic method for coupling pions to a non-Goldstone hadron was worked
out a
long time ago in Ref.~\cite{CCWZ}, which shows how to couple any field with a given
isospin to pions in the most general way.   In fact, the method generalizes to
other groups as well: The general case is that of some continuous group $G$
that is spontaneously broken to some subgroup $H$.  Since only $H$ is
realized in Wigner mode, all fields need to transform in representations of $H$,
and not $G$, as we will see in more detail below.  Here we will restrict
ourselves to $G=SU(2)_L\times SU(2)_R$, and $H=SU(2)_V$, \ie, isospin.

We will present the construction
as explained in Ref.~\cite{roessl}.  As we have seen in Sec.~\ref{lagr}, the pion
fields parametrize the coset $[SU(2)_L\times SU(2)_R]/SU(2)_V$ ($G/H$ in the
general case), because
$SU(2)_V$ transformations leave the vacuum invariant.   An element of the
coset can be represented by picking an element $(u_L,u_R)$ of the
full symmetry group $SU(2)_L\times SU(2)_R$, and defining an
equivalence class by making any two such elements equivalent when they
differ by a transformation in $SU(2)_V$.  Each equivalence class is an element
of the coset.  We can pick a ``standard'' representative of each equivalence
class by imposing a condition on $u_L$ and $u_R$; we pick the condition that
$u_L=u_R^\dagger\equiv u$.  We can do this: by definition, if we multiply
$(u_L,u_R)$ on the right by an element $(h^\dagger,h^\dagger)\in SU(2)_V$, obtaining $(u_Lh^\dagger,u_Rh^\dagger)$,
this group element is in the same equivalence class, and thus represents the
same coset element.   If we pick $h$ such that it solves $u_R=hu_L^\dagger h$,
this brings the coset element into standard form.
If we now multiply the group element
$(u_L,u_R)$ by another group element, $(v_L,v_R)$:
\begin{equation}
\label{mult}
(u_L,u_R)\to (v_L u_L,v_R u_R)\ ,
\end{equation}
the new group element in general corresponds to a different element of the
coset.  But again we can choose a $(h^\dagger,h^\dagger)\in SU(2)_V$ such that
\begin{equation}
\label{mult2}
(u_L,u_R)\to (v_L u_Lh^\dagger,v_R u_Rh^\dagger)
\end{equation}
keeps the transformed element in standard form, where, of course, $h=h(u,v_{L,R})$ now
depends both
on $u$ and $(v_L,v_R)$.  If we choose $(v_L,v_R)\in SU(2)_V$, \ie, $v_L=v_R=v$, we can choose the ``compensator'' field $h$ equal to $v$, and it
does not depend on $u$.   It follows that, as expected, an $SU(2)_V$ transformation
stays in the same equivalence class.

We now may define $\S$ in terms of $u$ by\footnote{Whether with the symbol ``$u$'' I refer
to the up quark or the nonlinear pion field should always be clear from the context.}
\begin{eqnarray}
\label{Sigmaagain}
\S&=&u_Lu_R^\dagger=u^2\to v_L\S v_R^\dagger\ ,\\
u&=&\exp(i\phi/f)\ ,\nonumber
\end{eqnarray}
where now
\begin{equation}
\label{phisu2}
\phi=\pmatrix{\frac{\p^0}{\sqrt{2}}&\p^+\cr\p^-&-\frac{\p^0}{\sqrt{2}}}\ ,
\end{equation}
because then $\S$ also parametrizes the coset, and has the correct transformation
properties under the chiral group if we identify $v_{L,R}=U_{L,R}$, \seef\ Eq.~(\ref{sigma}).

This construction provides a straightforward way to include the kaon.  Introducing
a field $K$ in the isospin-$1/2$ representation of $SU(2)_V$, we can extend the
nonlinear represention of the chiral group to
\begin{eqnarray}
\label{bb}
u&\to& v_Luh^\dagger(u,v_{L,R})=h(u,v_{L,R})uv_R^\dagger\ ,\\
K&\to& h(u,v_{L,R})K\ .\nonumber
\end{eqnarray}
These two fields, their (covariant, see below)
derivatives and the sources $\ell_\m$, $r_\m$, $\c$ and
$\c^\dagger$ now form the set of building blocks from which to construct the
chiral lagrangian.

Note that we only have to specify the $SU(2)_V$ representation of the kaon field $K$,
and not the representation of the full chiral group $SU(2)_L\times SU(2)_R$.  There is a
simple intuitive reason for that \cite{DBK}:  Suppose, for example, we specify that a kaon field $K'$ transforms
in the $(1/2,0)$ representation of $SU(2)_L\times SU(2)_R$,
$K'\to v_LK'$.  Then, performing the
field redefinition $K''=\S^\dagger K' = \left(1+O(\p)\right)K'$ changes the representation
to $(0,1/2)$, since $K''\to v_RK''$.  We see that the presence of pions (or, in general,
the Goldstone bosons associated with $G/H$) give us the freedom to choose
any representation of $SU(2)_L\times SU(2)_R$ we want!  The field $K$ used in
Eq.~(\ref{bb}) is obtained by defining $K=u^\dagger K'$.\footnote{It can be shown that any other nonlinear representation of the group
$SU(2)_L\times SU(2)_R$ with the same field content can be brought into the
form~(\ref{bb}) by a field redefinition \cite{CCWZ}.}   We note that the field $K'$
transforms under parity to
\begin{equation}
\label{Kparity}
K'\to K'^{(P)}=-\S^\dagger K'\ ,
\end{equation}
because the parity transform of $K'$ should transform as $K'^{(P)}\to v_R K'^{(P)}$
under the chiral group; the minus sign indicates that the kaon is a pseudoscalar.
The field $K$ transforms into $-K$ under parity, consistent with its transformation
rule Eq.~(\ref{bb}) under the chiral group ($h\in SU(2)_V$).

In order to construct the chiral lagrangian, we need to construct invariants out of
$u$ and $K$, which transform as in Eq.~(\ref{bb}).  Since $h$ is local, we will need
a gauge connection for $SU(2)_V$, constructed from $u$.  We thus introduce
\begin{eqnarray}
\label{pionbbungauged}
u_\m^L&=&u^\dagger\partial_\m u\ ,\\
u_\m^R&=&u\partial_\m u^\dagger\ ,\nonumber
\end{eqnarray}
which transform into each other under parity.  Under Eq.~(\ref{bb}), the parity
odd and even combinations transform as
\begin{eqnarray}
\label{chtr}
\D_\m\equiv\frac{1}{2}\left(u_\m^R-u_\m^L\right)
&\to& \frac{1}{2}h\left(u_\m^R-u_\m^L\right)h^\dagger\ ,\\
\frac{1}{2}\left(u_\m^R+u_\m^L\right)&\to&\frac{1}{2}h\left(u_\m^R+u_\m^L\right)h^\dagger
-\partial_\m hh^\dagger\ .\nonumber
\end{eqnarray}
We see that $\D_\m$ transforms homogeneously, while the other combination
transforms as an $SU(2)_V$ gauge connection.  The latter can thus be used
to define a covariant derivative for $K$:
\begin{equation}
\label{kcovder}
D_\m K=\partial_\m K+\frac{1}{2}\left(u^\dagger\partial_\m u
+u\partial_\m u^\dagger\right)K\ .
\end{equation}
If we wish to include the sources $\ell_\m$ and $r_\m$, this can be done by
replacing $\partial_\m\to\partial_\m-i\ell_\m$ in $u_\m^L$ and
$\partial_\m\to\partial_\m-ir_\m$ in $u_\m^R$.

The chiral lagrangian consists of three parts.  First, there is the purely pionic
part of the lagrangian, which is given by $\cl^{(2)}+\cl^{(4)}$ of Eqs.~(\ref{l2again})
and~(\ref{l4total}).  For two flavors, a number of simplifications occur because
of special properties of $SU(2)$.  Then, there is the quadratic term in the kaon
fields that defines the kaon propagator in ChPT,
\begin{equation}
\label{kaonfree}
\cl_K^{(0)}=(D_\m K)^\dagger D_\m K+M^2K^\dagger K\ .
\end{equation}
Here $M$ is a free parameter equal to the kaon mass in the chiral
limit.

As already noted, even with $\ell_\m=r_\m=0$ the covariant derivative~(\ref{kcovder})
is required, because
$h$ in Eq.~(\ref{bb}) is a local, field-dependent transformation.   Interactions between
pions and kaons are therefore necessarily present in our EFT.   In this respect, it is
instructive to see how this same conclusion would have been reached if we started
from the field $K'=uK$ discussed above.  It looks like we can write down a kaon
kinetic term without any interactions in terms of this field: $\partial_\m K'^\dagger
\partial_\m K'$ is invariant under the chiral group.  However, this is not invariant under
parity.  Using  the parity transform $K'^{(P)}$ introduced in Eq.~(\ref{Kparity}),
a lagrangian invariant under the chiral group and parity is
\begin{equation}
\label{Kplag}
\frac{1}{2}\left(\partial_\m K'^\dagger \partial_\m K'
+\partial_\m (K'^\dagger\S) \partial_\m(\S^\dagger K')\right)\ ,
\end{equation}
which, using $K'=uK$, is equal to the first term in Eq.~(\ref{kaonfree}), modulo a term
proportional to $K^\dagger L_\m L_\m K=2\,\tr(\D_\m\D_\m)K^\dagger K$
that can be absorbed into the $A_1$ term in Eq.~(\ref{kaon2}) below.

Many more terms
involving pion-kaon couplings can be constructed.
For a more extensive discussion
of building blocks and possible terms in the chiral lagrangian, I refer to Ref.~\cite{roessl}.%
\footnote{Note that conventions in Ref.~\cite{roessl} differ from those used here.}
Here I give only one more building block:
\begin{equation}
\label{massbb}
\c_+=u^\dagger\c u^\dagger+u\c^\dagger u\ ,
\end{equation}
which is even under parity (the odd combination, with a minus sign instead of a plus
sign, appears in the chiral lagrangian at order $p^4$).

With these definitions, noting that $\D_\m=\frac{1}{2}u^\dagger L_\m u$, with
$L_\m=\S\partial_\m\S^\dagger$ (\seef\ Eq.~(\ref{LR})),
Eq.~(\ref{l2again}) can be written as
\begin{equation}
\label{l2again2}
\cl^{(2)}=-\frac{1}{2}f^2\;\tr(\D_\m\D_\m)-\frac{1}{8}f^2\;\tr(\c_+)\ ,
\end{equation}
and more terms involving the kaon field are given by
\begin{eqnarray}
\label{kaon2}
\cl_K^{(2)}&\!\!=\!\!&A_1\;\tr(\D_\m\D_\m)\;K^\dagger K
+A_2\;\tr(\D_\m\D_\n)\;(D_\m K)^\dagger D_\n K\\
&&-A_3 K^\dagger\c_+ K-A_4\;\tr(\c_+)K^\dagger K\nonumber\\
&&+A_2'\;\tr(\D_\n\D_\n)\;(D_\m K)^\dagger D_\m K
+A_3'(D_\m K)^\dagger\c_+D_\m K+A_4'\;\tr(\c_+)(D_\m K)^\dagger D_\m K\ ,
\nonumber
\end{eqnarray}
where $A_{1,2,3,4}$ and $A_{2,3,4}'$ are seven new LECs.\footnote{The primed $A$'s
are not present in $\cl_K^{(2)}$ of Ref.~\cite{roessl}, and can probably be traded for
higher-order terms by field redefinitions.  However, this is not mandatory, and
I choose to keep them in  $\cl_K^{(2)}$.}  Formally, we see that
$\cl_K^{(0)}$ and $\cl_K^{(2)}$ are the first two terms in a
small-derivative/light quark mass
expansion, because only derivatives on the pion field $u$ should be counted
as small, and not those acting on the kaon field $K$.  We note that some of the
terms in $\cl_K^{(2)}$ would be counted as $O(p^4)$ if the kaon is treated as
a Nambu--Goldstone boson, \seef\ Sec.~\ref{chpt}.

%%####%%
\subsection{\label{kaon pc} Power counting}
%%####%%
At this point, we need to revisit power counting.  First, there are
now three scales in the problem, $m_\p$, $m_K$ and $4\p f$, \seef\
Sec.~\ref{power}, and we are now {\em not} assuming that $m_K^2/(4\p f)^2$
is small.  Therefore, it is not obvious that a systematic chiral
expansion in $m_\p^2/(4\p f)^2$ and/or $m_\p^2/m_K^2$ can be set up.
Second, our new lagrangian also contains interactions contributing
for example to the scattering process $\p^+\p^-\to K^+K^-$.  But, clearly,
with the kaon mass not treated as a small parameter, the initial-state
pions in this process also do not have small momenta in the sense of a chiral expansion.
The prediction for this process from our new lagrangian can thus not be
argued to be the leading order in a systematic expansion.

We conclude that  ``kaon ChPT'' cannot be used for any process near or above the two-kaon threshold.  However, it can be used to calculate correlation
functions in which a kaon goes in and comes out, so that the large
energy residing in the kaon mass does not get converted to pions.  This
means that one can use the new lagrangian to calculate the dependence
of kaonic quantities, such as its mass and decay constant, on the
light quark masses $m_u$ and $m_d$.  Note that in the leptonic decay
of a kaon, its mass does get converted into the energy of the outgoing
leptons, but that does not affect the calculation of its dependence on
light quark masses.\footnote{Using kaon ChPT for semi-leptonic decays
only works when the momentum of the outgoing pion is small.}

While this claim turns out to be correct, we still need to deal with the
first problem mentioned above, \ie, the appearance of a new ``large''
scale $m_K$, in order to show its validity.
{}From Eq.~(\ref{kaonfree}) it follows that the parameter $M^2$ only
shows up in the kaon propagator.  It can therefore only get promoted
to the numerator of some contribution to a physical quantity by
appearing in a loop.  The simplest case is that of a closed kaon loop;
for instance, the kaon tadpole diagram contribution to the pion
self-energy that would follow from $\cl_K^{(0)}+\cl_K^{(2)}$.  In
dimensional regularization this would give a contribution of order
$m_K^2\log{(m_K^2/\L^2)}$, which is not small.  However, such
contributions can be absorbed into the low-energy constants of the
pion lagrangian, because they are independent of the pion mass.\footnote{Similar
to the way that the physics of $\r$ mesons and other heavy hadrons appears
through the values of the LECs in the pion lagrangian.}
We conclude from this example that we can thus {\em omit}
closed kaon loops from the calculation of correlation functions
with kaon ChPT.\footnote{The same argument was used for baryon
ChPT already in Ref.~\cite{GSSb}.}  With this rule that we omit
closed kaon loops, we conclude that correlation functions
with only pions on the external legs should be calculated with the
pion chiral lagrangian only.

In effect, because the kaon is heavy
compared to the pion, we have integrated it out, and including closed kaon
loops as discussed above would amount to double-counting.  This is
of course the reason that we can use two-flavor ChPT for pion physics well below
the kaon mass.  The new element is that we can also consider processes with
a kaon going in and coming out --- as long as no kaons get annihilated
or produced, the energy stored in the kaon mass does not play a
dynamical role.  An example is $\pi K\to\p K$ scattering: As long as the energy
of the incoming pion is small, the energy of the outgoing pion also has to
be small, and two-flavor ChPT (in the presence of a ``heavy'' kaon)
should apply.  For a detailed analysis of this scattering process, see
Ref.~\cite{roessl}.

%%%%%%%%%%%%%%%%%%%
\begin{figure}
\vspace*{4ex}
\begin{center}
\includegraphics*[width=8cm]{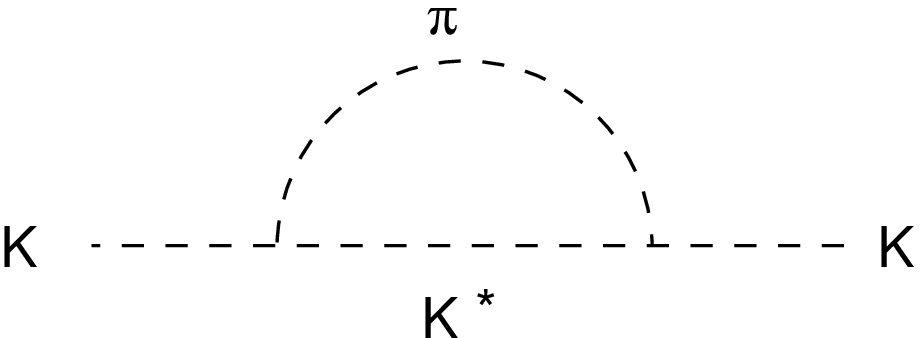}
\end{center}
\begin{quotation}
\caption{One-loop contribution to the kaon self energy with a $K^*\p$ intermediate
state.
\label{fig:kaonself}}
\end{quotation}
\vspace*{-4ex}
\end{figure}
%%%%%%%%%%%%%%%%%%%

There are also loops with internal kaon lines where the
kaon does not form a closed loop, for instance in the just-mentioned
example of kaon-pion scattering.
To one-loop order, they do not occur in loop corrections to pion and kaon
masses and decay constants, because all one-loop diagrams for these
quantities are tadpole diagrams, which follows from the fact that there
are no three-meson vertices in our EFT.  However, such vertices do occur in the kaon theory if we also include the $K^*$,
described by a vector field $K_\m$ also in the isospin-$1/2$ representation
of $SU(2)_V$.  In this extended version of kaon ChPT, a possible invariant is
\begin{equation}
\label{kstar}
-igfK_\m^\dagger(u^\dagger\partial_\m u)K=gK_\m^\dagger K\partial_\m\phi
+\dots\ ,
\end{equation}
\ie, now the theory does contain a $K^*K\p$ three-meson vertex.

Let us consider an example.  The vertex~(\ref{kstar})
leads to a correction to the kaon self energy through the diagram of Fig.~\ref{fig:kaonself}.
The amputated part of this diagram stands for an integral of the form \cite{RBCkaon}
(using dimensional regularization)
\begin{equation}
\label{integral}
I=\int\frac{d^dq}{(2\p)^d}\;\frac{q_\m q_\n\left(\d_{\m\n}+(p-q)_\m (p-q)_\n/
M_*^2\right)}{(q^2+m_\p^2)((p-q)^2+M_*^2)}\ ,
\end{equation}
in which $M_*$ is the mass of the $K^*$.  We need this integral for an onshell kaon,
so we need to set $p=Mv$, with $v$ a four-vector with $v^2=-1$.   The denominator
of the $K^*$ propagator then becomes
\begin{equation}
\label{denom}
q^2-2Mv\cdot q+M_*^2-M^2\ .
\end{equation}
Nonanalytic dependence on $m_\p$
comes from the region $q\sim m_\p$.
The behavior of the integral now depends on the various other scales.  If $M_*$ and $M$ were
the masses of spin-zero and spin-one heavy-light mesons (instead of the $K^*$ and the kaon),
the term linear in $M$ would be the dominant term, because
$M_*^2-M^2\sim\L_{QCD}^2$ and $M_*\sim M\gg\L_{QCD}$.\footnote{Of course, $Mv\cdot q$
can be smaller than $\L_{QCD}^2$, but the $q$ integral gets cut off in the infrared by $m_\p$.
The situation is more complicated if there are also other infrared scales present, such as
momentum transfer in a form factor or scattering amplitude.  For detailed
discussions, see for instance Refs.~\cite{BL,AWL}.}
Instead, for the
$KK^*$ system both masses as well as their difference are of order $\L_{QCD}$,
and the difference $M_*^2-M^2$ is the dominant term.
This leads to a different infrared behavior for each of these cases.    For the heavy-light case, the
denominator~(\ref{denom}) can be of order $m_\p$ (times $M$),\footnote{This leads to the
terms of order $m_\p^3$ which appear in the heavy-light and baryon cases.}
while for the kaonic
case, the denominator does not become small.  In the latter case, we can thus effectively replace
this denominator by $\L_{QCD}^2$, and read off the infrared behavior from the
rest of the integral, which gives rise to a contribution of order $m_\p^4\log(m_\p^2/\L^2)$ \cite{RBCkaon}.   The power $m_\p^4$ has to occur because of dimensional
analysis, and the fact that the kaon propagator, even though it appears inside a
loop integral, can be ``power-counted'' as order $\L_{QCD}^{-2}$.  I would expect this
type of argument to extend to higher loop contributions, thus explaining how higher-loop contributions
lead to nonanalytic terms at higher orders in $m_\p^2$.
For an extensive discussion of all this in the
context of baryon ChPT, I refer to Refs.~\cite{JM,BL}.

It is straightforward to show that there are no $m_\p^2\log(m_\p^2/\L^2)$
terms coming from $\cl_K^{(0)}+\cl_K^{(2)}$, and thus \cite{roessl}
\begin{equation}
\label{kaonmass}
m_K^2=M^2-2(A_3+2A_4)\hm_\ell+O(\hm_\ell^2\log(\hm_\ell))\ .
\end{equation}
This  is consistent with the result for $m_K^2$ shown in Eq.~(\ref{oneloopresults}),
which does not contain a term of the form $m_\p^2\log(m_\p^2/\L^2)$.\footnote{Expanding
$L(m_\eta^2)$ in $m_\pi^2/m_K^2$ gives
rise to analytic terms in $m_\p^2$ only.}  It should be noted that Ref.~\cite{roessl}
obtains this result in the theory without the $K^*$.  Indeed, since the $K^*$ is
significantly heavier than the kaon, we should be able to integrate it out.   Doing
this leads to an effective coupling of the form
\begin{equation}
\label{effkstar}
g^2\frac{f^2}{M_*^2}K^\dagger\D_\m\D_\m K
=g^2\frac{f^2}{2M_*^2}\;\tr(\D_\m\D_\m)K^\dagger K
\end{equation}
(where we used that $\D_\m=-u^\dagger\partial_\m u$ is an element of the Lie algrebra
for $SU(2)$), and we see that this is of the form of the first term in Eq.~(\ref{kaon2}).

%%####%%
\subsection{\label{decay} An application: decay constants}
%%####%%
The pion decay constant follows simply from Eq.~(\ref{oneloopresults}), by expanding
out $L(m_K^2)$ in $\hm_\ell$:
\begin{equation}
\label{fpikaon}
f_\pi=f'\left(1+\frac{4l_4}{f^2}B_0 m_\ell-2L(2B_0 m_\ell)+O(\hm_\ell^2)\right)\ ,
\end{equation}
in which we can match the LECs $f'$ and $l_4$ to their three-flavor counterparts \cite{GL},
\begin{eqnarray}
\label{match}
f'&=&f\left(1-L(B_0 m_s)+\frac{16L_4}{f^2}\,B_0 m_s\right)\ ,\\
l_4&=&4(L_5+2L_4)-\frac{1}{(4\p)^2}\left(\log\left(\frac{B_0 m_s}{\L^2}\right)+1\right)\ .
\nonumber
\end{eqnarray}
The matching formulas~(\ref{match}) only make sense if $m_s$ is small enough
for three-flavor ChPT to apply; otherwise $f'$ and $l_4$ should be treated as
two-flavor LECs that depend in an unknown way on the strange quark mass.
The expressions~(\ref{match}) give the first few orders of the expansion of
$f'$ and $l_4$ in $m_s$.  It is possible that the strange quark mass is so
large that this expansion simply does not converge.  But it is also possible that
this expansion does (asymptotically) converge, but too slowly for the
expressions~(\ref{match}) to be of practical use.
In that case, one should either use kaon ChPT, or three-flavor ChPT to higher
order than $p^4$.  Whether the expansion in the strange quark mass converges
to some order in ChPT or not is quantity dependent.

In order to obtain the expansion of $f_K$ in terms of $m_\ell$, we can follow two
routes.  One is to simply use the three-flavor result for $f_K$ in Eq.~(\ref{oneloopresults}),
and expand in $m_\ell$, as we did above for $f_\p$.  However, it is instructive to
see how one obtains this expansion in kaon ChPT independently of the three-flavor
ChPT calculation \cite{RBCkaon}.  In order to do this, we need to translate the
currents
\begin{eqnarray}
\label{scurrents}
J^{L,R}_\m&=&\bq_{L,R}\g_\m s_{L,R}\ ,\\
\bq&=&\pmatrix{\bu&\bd}\ ,\nonumber
\end{eqnarray}
into the appropriate  operators in kaon ChPT.  These currents now do not correspond to
the Noether currents of (softly broken) symmetries, because we are not assuming
that the strange quark is small.  Therefore, their translations into kaon ChPT do
not follow from the lagrangian $\cl_K^{(0)}+\cl_K^{(2)}$.
We thus introduce spurions $H_{L,R}$
into these currents, with transformation rules such as to make these currents
invariant under $SU(2)_L\times SU(2)_R$:
\begin{equation}
\label{curragain}
J^{L,R}_\m=\bq_{L,R}H_{L,R}\g_\m s_{L,R}\ ,\ \ \ \mbox{with}\ \ \ H_{L,R}\to v_{L,R}H_{L,R}\ ,
\end{equation}
in which $v_{L,R}\in SU(2)_{L,R}$.  Making the transition to kaon ChPT,
the first two terms in the expansion in terms of derivatives of the pion fields are
\begin{eqnarray}
\label{chptcurr}
J^R_\m&=&C_1(D_\m K)^\dagger uH_R+iC_2K^\dagger\D_\m uH_R\ ,\\
J^L_\m&=&-C_1(D_\m K)^\dagger u^\dagger H_L+iC_2K^\dagger\D_\m u^\dagger H_L\ ,\nonumber
\end{eqnarray}
where $C_{1,2}$ are new LECs (not present in the lagrangian).  The left- and right-handed currents are related by parity.  In these ChPT expressions for the currents,
one can now set $H_{L,R}=\pmatrix{1\cr 0}$, or $H_{L,R}=\pmatrix{0\cr 1}$ in order
to pick out the charged or neutral kaon, corresponding to picking out
the up or down quark in Eq.~(\ref{curragain}), respectively.
Taking the matrix element of the
axial current $J^R_\m-J^L_\m$ between a kaon state and the vacuum, one finds that
\begin{equation}
\label{fKkaon}
f_K=2C_1\left(1+\frac{C}{f^2}\hm_\ell-\frac{3}{4}L(2\hm_\ell)+O(\hm_\ell^2)\right)\ ,
\end{equation}
where $C$ is a linear combination of LECs in the kaon EFT lagrangian.\footnote{It turns
out that $C_2$ does not contribute to this order.}   Note that the LEC $C_1$ is
unrelated to $f'$ in Eq.~(\ref{fpikaon}), because there is no $SU(3)$ symmetry in
kaon ChPT.  If we do use $SU(3)$, we may
again match the LECs in Eq.~(\ref{fKkaon}) to the three-flavor ChPT result, from
which we find
\begin{eqnarray}
\label{match2}
C_1&=&\frac{1}{2}f\left(1-3L(\hm_s)+\frac{8(2L_4+L_5)}{f^2}\;\hm_s\right)
\ ,\\
C&=&8(4L_4+L_5)-\frac{5}{2}\;\frac{1}{(4\p)^2}\left(\log\left(\frac{\hm_s}{\L^2}\right)+1\right)
\ .\nonumber
\end{eqnarray}
As it should, the coefficient of $L(m_\p^2)$ in Eq.~(\ref{fKkaon}) matches that following
from Eq.~(\ref{oneloopresults}).

We close this section with another simple example of the translation of a weak
operator into kaon ChPT \cite{RBCkaon}.  The $\D S=2$ operator $(\bs_L\g_\m d_L)(\bs_L\g_\m d_L)$
can be written in $SU(2)_L$-invariant form by again using the spurion $H_L$ as in
\begin{equation}
\label{deltaSeq2}
(\bs_L\g_\m H_L^\dagger q_L)(\bs_L\g_\m H_L^\dagger q_L)
\to B(H_L^\dagger u K)(H_L^\dagger u K)\ ,
\end{equation}
where in the second step I gave the leading-order translation into kaon ChPT,
introducing another weak LEC $B$.  Finally, one sets $H_L^\dagger=\pmatrix{0&1}$,
because for this choice $H_L^\dagger q_L=d_L$.
This technique for finding the ChPT representations of
electro-weak operators is of course not special to kaon ChPT, and the same
method can also be used in three-flavor ChPT \cite{CBtasi}.

It turns out that $SU(2)$ ChPT can also be applied to semi-leptonic
and nonleptonic kaon decays, despite the fact that the final-state pions
can have an energy up to half the kaon mass.  This is not at all obvious,
because in the $SU(2)$ framework the kaon mass is not a small parameter.
For more on this topic, see Refs.~\cite{FS,BC}.

%%####%%
\section{\label{finite volume} Finite volume}
%%####%%

As a final topic, I would like to touch, rather briefly, on the fact that all lattice QCD computations
are, necessarily, performed in a finite spatial volume, and with a finite maximal extent $L_4$ in the
(euclidean) time direction.

Usually,  the time dimension of the four-dimensional euclidean
volume is chosen larger than the spatial directions; large enough so that only the lowest
(or lowest few) intermediate states contribute to the correlation functions of interest at large times.
In order to mimimize, or eliminate, systematic effects from the finite extent in spatial
directions, the conceptually easiest method
would be to choose the spatial volume of a cubic box
$V_3=L^3$
(where $L$ is the linear size of the spatial box, and we will assume periodic
boundary conditions for all fields), sufficiently large that one may ignore the effects of $L$
being finite.
However, the cost of numerical computations grows rapidly with $L$,
and one would thus like to choose the volume just large enough that finite-volume effects
can be neglected, but not much larger.  The question then arises whether we have any
theoretical insight into what ``just large enough'' means.

In fact, as we will see, it is possible to develop systematic expansions which provide
quantitative information on finite-volume effects, without the introduction of any new
parameters in the chiral lagrangian.  This allows us to quantitatively estimate the effects
of working in a finite volume, and even to make use of finite-volume effects as a probe of
hadronic quantities such as the pion decay constant and the chiral condensate.

The reason that again ChPT gives us the tool to develop these systematic expansions
is that pions\footnote{In this
section, I will use the word ``pions'' to refer only to $\p^\pm$ and $\p^0$, and
assume that their masses are equal, \ie, that isospin is conserved.} are the lightest
hadrons.  Thus, if we imagine reducing $L$ from some very large value, the pions
feel the effects of being in a finite volume first.  Therefore, ChPT should give us access
to the dominant finite-volume effects on many hadronic quantities, if we know how to apply it in this setting.

{}It already follows from this qualitative discussion that the ``figure of merit'' for what
``large'' or ``small'' volume means is the linear size of the box in terms of the wavelength
of the pion, $z=m_\p L$.  Clearly, large volume means $z\gg 1$.   But only a more
quantitative analysis can tell us whether this means, say, $z\geqx 3$, or whether maybe
$z\geqx 10$ is needed.

The key theorem in setting up finite-volume ChPT states that the lagrangian to
be used is the infinite-volume lagrangian \cite{GLfv1}.   The argument goes as
follows.  First
assume that the spatial volume is infinite, but that the
euclidean time extent $L_4$ is finite, and that
the pion fields~(\ref{phisu2}) obey periodic boundary conditions in that direction.
The pion effective theory with these boundary conditions describes the theory at a
finite temperature $T=1/L_4$.  The important observation here is that to calculate
pion correlation functions at finite temperature, one uses the {\em same} chiral
lagrangian as at zero temperature (\ie, infinite $L_4$).  The temperature, and thus
the dependence on $L_4$ enters only through the boundary conditions.\footnote{For
field theory at finite temperature, see the lectures by Owe Philipsen at this school.}
But, if this is the case, this then also has to be true for the dependence on the
three spatial dimensions of the box, if we use periodic boundary conditions also
in those directions \cite{GLfv2}.  In other words, the LECs in a finite-volume
calculation should be taken the same as in infinite volume.

This
result makes sense from the EFT point of view.  If we remember that the LECs
originated from integrating out high-energy degrees of freedom, they can pick up
exponentially suppressed finite-volume corrections of order $\exp(-EL)$, where
$E$ is the energy of the integrated mode, relative to their infinite-volume values.
If we take for example $E=m_\r$ and
$L=2$~fm, $\exp(-EL)=4.5\times 10^{-4}$, which is very small in comparison with
the present precision of our knowledge of LECs.   We conclude that we can safely
ignore such effects, and thus take the LECs to be those of the infinite-volume
theory.\footnote{A corollary is that the chiral theory at nonzero temperature only
makes sense for small $T$, so that $\exp(-E/T)$ is small.}
This more intuitive argument also shows that the theorem should hold
for more general boundary conditions, such as anti-periodic ones.

It follows that
the way in which the volume dependence enters the Feynman rules is through the
propagator, which now is periodic, since it solves, for particle with mass $m_\p$,
\begin{equation}
\label{propper}
(-\bo+m_\p^2)G_L(x-y)=\bdelta(x-y)=\prod_\m\sum_{n_\m}\d(x_\m-y_\m+n_\m L_\mu)\ ,
\end{equation}
where $L_\m$ is the linear extent of the volume in the $\m$ direction, and
$\bdelta$ is the periodic delta function, with period $L_\m$ in the $\m$
direction.   The solution
is, of course,
\begin{equation}
\label{propsol}
G_L(x-y)=\sum_{n_\m}G_\infty(x-y+n L)\ ,
\end{equation}
with $G_\infty$ the infinite-volume propagator, and in which $nL$ is the four-vector
with components $n_\m L_\m$.   In words, finite volume effects occur because pions
can ``travel around the world,'' multiple times in each direction, in a world that is a box
with periodic boundary conditions.
Because of the periodicity,
momenta are quantized, $p_\m=2\p n_\m/L_\m$, with $n_\m$ integer, and
the propagator can also be represented by
\begin{equation}
\label{propmom}
G_L(x-y)=\frac{1}{L_1L_2L_3L_4}\sum_p\frac{e^{ip(x-y)}}{p^2+m_\p^2}\ ,
\end{equation}
where the sum is over the discrete momenta.  Feynman rules for vertices are
the same as in the infinite-volume theory.  With these rules, one may extend the
calculations of meson masses, decay constants, and other quantities to include
finite-volume (and/or finite-temperature) corrections.  We will return to this in
Sec.~\ref{p-regime}.

However, before we extend ChPT calculations to finite volume in this way, we need
to address a serious physics issue: it is well known that in the chiral limit, the
chiral condensate vanishes in a finite volume.  In order to obtain a nonvanishing condensate, one
should take the chiral limit after taking the infinite-volume limit, and not before; these two limits do not commute.
But, if there is no chiral symmetry breaking, that brings into question the
whole framework of ChPT!

Let us see what happens in the chiral theory at zero quark mass in
finite volume.  The condensate can be found from the logarithmic derivative of the
partition function with respect to $\chi^\dagger$:
\begin{equation}
\label{logder}
\langle\bq_R q_L\rangle=-\frac{2B_0}{Z}\,\tr\left(\frac{\partial Z}{\partial\chi^\dagger}\right)\bigg|_{\chi=0}
=-\frac{1}{4}f^2B_0\langle\,\tr(\S)\rangle\ ,
\end{equation}
where in the second step we made the transition to the chiral theory.  We can decompose
\begin{equation}
\label{decomp}
\S(x)=U\,\exp(2i\phi(x)/f)\ ,
\end{equation}
with $U$ a constant $SU(N_f)$ matrix, and in which
$\phi$ does not contain any constant mode,
\ie, $\int d^4x\,\phi(x)=0$.\footnote{Here I have arbitrarily placed $U$ on the left of
$\exp(2i\phi(x)/f)$.
More generally, we can write $\S(x)=U_L\exp(2i\phi(x)/f)U_R$,
but if we set $U_L=UU_R^\dagger$, we have that $\S(x)=U\exp(2iU_R^\dagger\phi(x) U_R/f)$,
and $U_R^\dagger\phi U_R$ does not contain a constant mode if $\phi$ does not.}
The matrix $U$ represents an element of the coset $[SU(N_f)_L\times SU(N_f)_R]/SU(N_f)$,
and varying $U$ over the group $SU(N_f)$ covers the vacuum manifold, \seef\ Sec.~\ref{lagr}.
Since $U$ represents an $SU(N_f)_L$ symmetry transformation, it follows that for vanishing
quark masses, the chiral lagrangian does not depend on $U$.  This, in turn, implies that
the condensate~(\ref{logder}) vanishes, because
\begin{equation}
\label{intU}
\int dU\,U=0\ ,
\end{equation}
with $dU$ the Haar measure on the group $SU(N_f)$.

What we see is that ChPT
reproduces the correct behavior in finite volume.  The intuitive insight is that as long as the volume is
large compared to the typical hadronic scale, $L\gg 1/\L_{QCD}$, QCD is nonperturbative,
and hadrons form.  The lightest hadrons are the pions, and all that happens if we
lower the value of $z=m_\p L$ is that the pions get very distorted.  However, if this is
the correct picture, ChPT should still be the correct effective theory.  Only when we
make $\L_{QCD}L$ too small QCD becomes perturbative, and ChPT is no longer
the appropriate effective theory.  As long as $\L_{QCD}L\gg 1$, ChPT is the
appropriate EFT for the low-energy physics of QCD, even if $L$ is small in units
of $m_\p$.

When we turn on quark masses, the chiral condensate no longer vanishes.  For
a degenerate quark mass $m$ and a zero mode
$\S_0\equiv(1/V)\int d^4x\,\tr(\S)$, the size of the leading-order mass term in the chiral
lagrangian is of order
\begin{equation}
\label{sizemterm}
N_f f^2B_0mV\S_0\ ,
\end{equation}
with $V=L^4$ the four-dimensional finite volume, if we take our volume to be a four-dimensional
box with equal size in each (euclidean) direction.  We may now consider different regimes.
If we keep Eq.~(\ref{sizemterm}) fixed, we choose $m\sim 1/L^4$, or, equivalently,
$m_\pi=\sqrt{2B_0m}\sim1/L^2$; we will refer to this regime as the ``$\e$-regime'' \cite{GLfv3}.%
\footnote{In this small four-dimensional volume, this basically
constitutes our definition of the pion mass, as we can clearly not define it as the parameter
characterizing the large-$t$ behavior of the euclidean pion two-point function.}  This is a
different regime from the large-volume regime we considered in Sec.~\ref{chpt}, in which we
found that $m_\p\sim p\sim 1/L$, and thus $m\sim 1/L^2$; this regime is usually referred to
as the ``$p$-regime.''  In the $\e$-regime, with $\e\equiv 1/L$, we have that $p\sim\e$,
while $m_\p\sim\e^2\ll p$, while in the $p$-regime they are of the same order.

The kinetic term of the chiral lagrangian, $\int d^4x\,\frac{1}{2}\,\tr(\partial_\m
\phi\partial_\m\phi)$, suppresses the nonzero modes $\phi$, and limits the fields $\phi$
to be of order $1/L$.  If we expand
\begin{equation}
\label{Sinepsregime}
\S=U\;\exp(2i\phi/f)=U\left(1+\frac{2i}{f}\phi-\frac{2}{f^2}\phi^2+\dots\right)\ ,
\end{equation}
it is easy to see that, with $m_u=m_d\equiv m$, the first term in
\begin{eqnarray}
\label{chl2espregime}
\int d^4x\;\frac{1}{4}f^2mB_0\;\tr(\S+\S^\dagger)&=&
\frac{1}{4}f^2B_0mV\,\tr(U+U^\dagger)\\
&&
-\int d^4x\;\frac{1}{2}\,mB_0\;\left(\tr(U+U^\dagger)\phi^2\right)+O(\phi^4)
\nonumber
\end{eqnarray}
is of order $L^0$ in the $\e$-regime, while the second term is of order
$1/L^2$.  To leading order, $U$
is nonperturbative: The zero modes in $\S$ do not have a gaussian damping, and
can thus not be taken into account through standard perturbation theory.  When we
increase the quark mass from $m\sim 1/L^4$ to $m\sim 1/L^2$, the first term in
Eq.~(\ref{chl2espregime})
grows to be of order $L^2$, still dominating the second term, which is now of order
$L^0$.  This forces $U$ to approach $\bf 1$, making also the zero mode perturbative.  We thus make the transition from the $\e$-regime to
the $p$-regime.   This is also reflected by the behavior of the $p=0$ mode contribution
to the pion propagator~(\ref{propmom}) in finite volume, which, taking all $L_\m=L$, is equal
to  $1/(L^4m_\p^2)$.  For $m_\pi^2=2B_0m\sim 1/L^4$, this is of order one, whereas all other momentum
modes contribute terms of order $1/L^2$.  When $m_\p$ is of order $1/L$, also the
zero-momentum contribution is of order $1/L^2$, and it does not dominate the
propagator.

There is thus a region in which the chiral expansion in the $\e$-regime
coincides with that in the $p$-regime.  Both representations have to agree for
$m\sim 1/L^2$, making it
possible to match correlation functions between the two regimes \cite{GLfv3}.  It is only
in the $p$-regime that the Feynman rules in infinite volume, where only the
propagator~(\ref{propsol}) knows about the finite-volume, apply.

Next, we will consider each of these two regimes separately.  I will only consider
the finite-volume physics of pions,
because in the real world they are much lighter than kaons, and this is now also
the case in state-of-the-art lattice QCD computations.  Therefore, finite-volume effects
for kaons and other hadrons will be dominated by their interactions with pions.   To put it
differently, the kaons are always in the $p$-regime ($m_K\gg 1/L$;
for example, for $L=2$~fm, $m_K L\approx 5$), and in the $p$-regime finite-volume
effects due to kaons traveling around the world are much suppressed compared to
those due to pions.\footnote{For an investigation including both pions in the
$\e$-regime and kaons in the $p$-regime, see Ref.~\cite{BDFH}.}

%%####%%
\subsection{\label{p-regime} $p$-regime}
%%####%%
In the $p$-regime, it is quite straightforward to adapt the infinite-volume expressions
of ChPT to finite volume.  Here, for definiteness, we will take a spatial volume $V_3=L^3$,
with periodic boundary conditions, while, for simplicity, we will assume that the (euclidean) time extent
is large enough to be taken as infinity.  We will also assume that only finite-volume effects
due to pions are significant.

Finite-volume effects start occurring at one loop,
because in the loops the pion propagator gets replaced by its finite-volume version,
Eq.~(\ref{propsol}).  For quantities like meson masses and decay constants, \seef\
Eq.~(\ref{oneloopresults}), all one-loop integrals are tadpole integrals, for instance of
the form
\begin{equation}
\label{tadpole}
G_\infty(0)=\int\frac{d^4p}{(2\p)^4}\;\frac{1}{p^2+m_\p^2}\doteq\frac{m_\p^2}{16\p^2}\left(\frac{2}{d-4}+\log\left(\frac{m_\p^2}{\L^2}\right)
+\mbox{finite}\right)\ ,
\end{equation}
where the second equation holds using dimensional regularization, and, of course,
we have to stay away from the pole at $d=4$.  In finite volume, this gets replaced by
\begin{equation}
\label{fvdiff}
G_L(0)=G_\infty(0)+(G_L(0)-G_\infty(0))\ .
\end{equation}
Using that
\begin{equation}
\label{propeval}
G_\infty(x)=\frac{1}{4\p^2}\frac{m_\p}{\sqrt{x^2}}K_1(m_\p\sqrt{x^2})\ ,
\end{equation}
in which $K_1$ is the modified Bessel function of the second kind of order one,
we find that
\begin{eqnarray}
\label{diffeval}
G_L(0)-G_\infty(0)&=&\sum_{\vec n}G_\infty(x=({\vec n}L,0))-G_\infty(0)\\
&=&\sum_{{\vec n}\ne 0}G_\infty(x=({\vec n}L,0))\nonumber\\
&=&\sum_{n=1}^\infty k(n)\;\frac{1}{4\p^2}\frac{m_\p}{\sqrt{n}L}K_1(\sqrt{n}m_\p
L)\ .\nonumber
\end{eqnarray}
Here $k(n)$ counts the number of vectors $\vec n$ with integer-valued components of
length $\sqrt{n}$.  For instance, $k(1)=6$, $k(2)=12$, $k(3)=8$, $k(4)=6$, \etc\ Note that,
as one would expect, the UV divergence of $G_L(0)$ is the same as in infinite volume,
because finite-volume effects come from pions traveling a ``long'' distance.  Of course,
in the one-loop calculations leading to Eq.~(\ref{oneloopresults}), we also encounter integrals
such as Eq.~(\ref{tadpole}) with an extra $p^2$ in the numerator.  Here we can use that,
for $x\ne 0$,
\begin{equation}
\label{trick}
\int\frac{d^4p}{(2\p)^4}\;\frac{p^2e^{ipx}}{p^2+m_\p^2}=-m_\p^2\int\frac{d^4p}{(2\p)^4}\;\frac{e^{ipx}}{p^2+m_\p^2}
=-\frac{1}{4\p^2}\frac{m_\p^3}{\sqrt{x^2}}K_1(m_\p\sqrt{x^2})\ .
\end{equation}
We thus find, for example, for the finite-volume
pion and kaon masses and decay constants to one loop \cite{GLfv1}
\begin{eqnarray}
\label{ffmassdecay}
m_\p(L)&=&m_\p\left(1+\frac{m_\p^2}{(4\p f_\p)^2}\sum_{n=1}^\infty
\frac{4k(n)}{\sqrt{n}m_\p L}K_1(\sqrt{n}m_\p L)\right)\\
&=&m_\p\left(1+\frac{m_\p^2}{(4\p f_\p)^2}
\frac{24}{(m_\p L)^{3/2}}\sqrt\frac{\p}{2}\;e^{-m_\p L}+O\left(e^{-\sqrt{2}m_\p L}\right)\right)\ ,\nonumber\\
f_\p(L)&=&f_\p\left(1-2\frac{m_\p^2}{(4\p f_\p)^2}\sum_{n=1}^\infty
\frac{4k(n)}{\sqrt{n}m_\p L}K_1(\sqrt{n}m_\p L)\right)\nonumber\\
&=&f_\p\left(1-\frac{m_\p^2}{(4\p f_\p)^2}
\frac{48}{(m_\p L)^{3/2}}\sqrt\frac{\p}{2}\;e^{-m_\p L}+O\left(e^{-\sqrt{2}m_\p L}\right)\right)\ ,\nonumber\\
f_K(L)&=&f_K\left(1-\frac{3}{4}\frac{m_\p^2}{(4\p f_\p)^2}\sum_{n=1}^\infty
\frac{4k(n)}{\sqrt{n}m_\p L}K_1(\sqrt{n}m_\p L)\right)\nonumber\\
&=&f_K\left(1-\frac{m_\p^2}{(4\p f_\p)^2}
\frac{18}{(m_\p L)^{3/2}}\sqrt\frac{\p}{2}\;e^{-m_\p L}+O\left(e^{-\sqrt{2}m_\p L}\right)\right)\ ,\nonumber
\end{eqnarray}
where $m_\p$ and $f_\p$ are the pion mass and decay constant in infinite volume.
The kaon mass does not receive any finite-volume corrections due to pions,
as we can infer from the fact that the infinite-volume one-loop contribution
has no contribution from pion loops (but only from an $\eta'$ loop, \seef\
Eq.~(\ref{oneloopresults})).
In the second of each of these equations, we used the asymptotic expansion
$K_1(z)\sim\sqrt{\p/(2z)}\;\exp(-z)$.  The dominant corrections do not come from higher
orders in the asymptotic expansion of $K_1$, but from the term $n=2$ in the sum in
Eq.~(\ref{ffmassdecay}).  For $m_\p\approx 300$~MeV and $m_\p L\approx 3$, the size
of the corrections in Eq.~(\ref{ffmassdecay})
is of order one percent; making $L$ smaller the corrections
grow rapidly in size.  Detailed studies of finite-volume effects for  the pion and kaon
masses and decay constants as well as the $\eta$ mass have been made in Ref.~\cite{CDHfv} using a partial
resummation of higher loops based on a different approach \cite{Lfv1}, and, for the pion mass
to two loops in ChPT \cite{CHfv}.\footnote{For earlier work, see these references.}

Many investigations of finite-volume effects in the $p$-regime have been carried out,
using both ChPT and other techniques, and it is beyond the scope of these lectures to
cover this topic in any more detail.  However, there is one application in which
ChPT can play an uniquely important role, and that is the effect of partial quenching on the
study of two-particle correlation functions in a finite box.

In a finite volume, the energy spectrum of QCD is discrete, and that also holds for
two-particle (and many-particle) states.   In fact, the energy levels of two-particle
states in finite volume give information on scattering phase shifts
in infinite volume \cite{Lfv2}.  The intuitive idea is the following.
The normalized wave function of a particle in a box of dimension $L^3$ is of order
$L^{-3/2}$.  If two such particles are put inside the box, and they interact, one
thus expects a shift in the two-particle energy of order the product of the
interaction strength time overlap of the
two one-particle wave functions, which is of order $1/L^3$.  Therefore, the
dependence of the
two-particle energy shifts on the volume gives us information about the interactions between the
two particles.  Unitarity plays an important role in the systematic analysis \cite{Lfv2},
making it not straightforward to generalize these ideas to PQQCD or QCD with mixed
actions.  However, ChPT gives us a handle on this problem, since, as we have seen,
it can be extended (in euclidean space) to the partially quenched and mixed-action
cases. The two-particle correlation functions, from which in the unitary case the
two-particle energy levels are determined, can thus be calculated in PQChPT
(or mixed-action ChPT).  One finds that
the double-hairpin contribution to flavor-neutral propagators gives rise to
unphysical phenomena such as enhanced finite-volume effects that can be estimated using ChPT
\cite{BGPQ,fvpq,SBMS,GIS}.  More study of these intricacies would be interesting.

%%####%%
\subsection{\label{eps-regime} $\e$-regime}
%%####%%
In this section, we will restrict ourselves to two flavors, under the assumption that the
strange quark is not in the $\e$-regime.  It can thus be taken into account for instance by the
method described in Sec.~\ref{kaon}.  Alternatively, one can start with $N_f=3$ ChPT,
but keep all fields describing mesons containing strange quarks in the $p$-regime.

Since in this section we will restrict ourselves to two flavors, $\S\in SU(2)$,
and instead of the parametrization~(\ref{Sinepsregime}) it is easier to use
\begin{eqnarray}
\label{Sdecomp}
\S&=&U\left(\s+i\;\frac{\sqrt{2}}{f}\;{\vec\t}\cdot{\vec\p}\right)\ ,\\
\s&=&\sqrt{1-\frac{2{\vec\p}\cdot{\vec\p}}{f^2}}\ ,\nonumber
\end{eqnarray}
with $\vec\t$ the Pauli matrices, and $\int d^4x\;\p_i(x)=0$.
With $m_u=m_d=m$, $V=L^4$, and
using Eq.~(\ref{Sdecomp}), we expand the chiral lagrangian to second order in the pion
field:
\begin{eqnarray}
\label{Leps}
\int d^4x\;\cl&=&-\frac{1}{4}f^2B_0mV\;\tr(U+U^\dagger)\\
&&+\frac{1}{2}\int d^4x\left[\partial_\m{\vec\p}\cdot\partial_\m{\vec\p}+
\frac{1}{2}\,B_0m\;{\vec\p}\cdot{\vec\p}\;\tr(U+U^\dagger)
\right]+O(\p^3)\ .\nonumber
\end{eqnarray}
In the $\e$-regime, as we have seen already,
the first term and the kinetic term are of order one.  This makes the
mass-dependent $O(\p^2)$ term a term of order $1/L^2$, and the whole
$O(p^4)$ lagrangian~(\ref{l4total}) of order $1/L^4$.  At this point, we observe several
differences between the $p$-regime and the $\e$-regime.  First, in the $p$-regime,
finite volume effects are exponentially suppressed, while in the $\e$-regime, they
are suppressed by powers of $1/(fL)$.   Then, we also see that the chiral
expansion in the $\e$-regime rearranges itself: the LECs $L_i$ only show up at
next-to-next-to-leading order in the expansion.  To leading order, we thus find for
the partition function, as a function of the quark mass:
\begin{equation}
\label{Zm}
Z(m)=\int[d\S]\;\exp\left(\frac{1}{4}f^2B_0mV\;\tr(U+U^\dagger)
-\frac{1}{2}\int d^4x\;\partial_\m{\vec\p}\cdot\partial_\m{\vec\p}\right)\ .
\end{equation}
Higher order terms in Eq.~(\ref{Leps}) can be expanded out from the exponent in the
integrand, and thus be taken into account systematically.  The notation $[d\S]$
indicates the $SU(2)$-invariant Haar measure for each $\S(x)$.

In order to proceed, we need to split the measure $[d\S]$ into an integration over
$U$ and an integration over $\vec\p$.
Following Ref.~\cite{HL}, we insert
\begin{eqnarray}
\label{one}
1&=&\int dH_0dH_1dH_2dH_3\;\d\left(H-\frac{1}{V}\int d^4x\;\S(x)\right)\\
&=&\int h^3dh\;dU\;\d\left(hU-\frac{1}{V}\int d^4x\;\S(x)\right)\ ,\nonumber\\
H&=&H_0+i{\vec\t}\cdot{\vec H}\ ,\nonumber
\end{eqnarray}
where, with $h=|H|\equiv\sqrt{\det(H)}$, we can write $H=hU$ with $U\in SU(2)$, and
with $dU$ the Haar measure on $SU(2)$.  Now we transform variables $\S=U\tS$,
so that
\begin{equation}
\label{Stildedecomp}
\tS=\s+i\;\frac{\sqrt{2}}{f}\;{\vec\t}\cdot{\vec\p}\ ,
\end{equation}
and use the fact that $d\S=d\tS$ to obtain
\begin{eqnarray}
\label{Zmagain}
Z(m)&=&\int dU\int [d\tS]\prod_{i=1}^3\d\left(\frac{\sqrt{2}}{fV}\int d^4x\;\p_i(x)\right)\\
&&\hspace{-15mm}\times\;\exp\left(\frac{1}{4}f^2B_0mV\;\tr(U+U^\dagger)
-\frac{1}{2}\int d^4x\;\partial_\m{\vec\p}\cdot\partial_\m{\vec\p}+3\log\left(\frac{1}{V}\int d^4x\;\s(x)\right)
\right)\nonumber\\
&=&\cn\int dU\;\exp\left(\frac{1}{4}f^2B_0mV\;\tr(U+U^\dagger)\right)\ ,\nonumber
\end{eqnarray}
where $\cn$ is independent of the quark mass $m$.  Carrying out the integral over
$U$ with
\begin{eqnarray}
\label{measure}
U&=&\cos{\theta}\;{\bf 1}+i\;\sin{\theta}\;{\vec n}\cdot{\vec\t}\ ,\ \ \ |{\vec n}|=1\ ,\ \ \ 0\le\theta\le\p\ ,\\
dU&=&\frac{1}{4\p^2}\;d\O({\vec n})\sin^2(\theta)\;d\theta\ ,\nonumber
\end{eqnarray}
we find
\begin{equation}
\label{Zmfinal}
Z(m)=\cn\frac{2}{f^2B_0mV}\;I_1\left(f^2B_0mV\right)\ ,
\end{equation}
where $I_1$ is the modified Bessel function of order one. From this, we find for the
chiral condensate
\begin{eqnarray}
\label{chcond}
\langle\bq q\rangle&=&-\frac{1}{ZV}\frac{dZ}{dm}=-f^2B_0\left(\frac{I_1'(f^2B_0mV)}{I_1(f^2B_0mV)}-\frac{1}{f^2B_0mV}\right)\\
&\stackrel{\null^{m\to 0}}{=}&-\frac{1}{4}f^4B_0^2mV
\ .\nonumber
\end{eqnarray}
Consistent with our previous discussion, the condensate vanishes for $m\to 0$.
For $m\ne 0$ fixed and $V\to\infty$, Eq.~(\ref{chcond}) reproduces the infinite-volume
result $\langle\bq q\rangle=-f^2B_0$, \seef\ Eq.~(\ref{psibarpsi}).  Note that the dimensionless
variable $f^2B_0mV$ can take any value; to be in the $\e$-regime, the
requirements are only that $f^2L^2\gg 1$ and $B_0mL^2\ll 1$.  This does
not restrict the size of their product.

Equation~(\ref{chcond}) provides an example of a ChPT calculation in the
$\e$-regime to lowest order.  Starting from Eq.~(\ref{Leps}), it is in principle
straightforward to go to higher orders in $1/(fL)$ \cite{HL,Letal}.  For recent work
discussing the ChPT calculation of the chiral condensate, see Ref.~\cite{DF}.

For instance, it is rather easy to see that the $m\;{\vec\p}\cdot{\vec\p}$
term in Eq.~(\ref{Leps}) leads to a renormalization of the first term in that
lagrangian by \cite{HL}
\begin{equation}
\label{renorm}
\r\equiv 1-\frac{1}{f^2V}\int d^4x\langle{\vec\p}(x)\cdot{\vec\p}(x)\rangle
=1-\frac{3}{f^2V}\sum_{p\ne 0}\frac{1}{p^2}=1-\frac{3\b_1}{f^2L^2}\ ,
\end{equation}
where $\b_1$ is a numerical constant \cite{HL},
and we used that $p_\m=2\p n_\m/L$ with $n_\m$ integer.\footnote{The sum
over $p$ is of course infinite, but we are interested here only in the
$1/L^2$ part, which is finite.  The volume-independent infinite part can be
absorbed into a multiplicative renormalization of $m$.  In dimensional
regularization, the ``infinite'' part actually vanishes.}

As we have noted before, the $O(p^4)$ LECs $L_i$ only come into play
at order $1/L^4$, \ie, at next-to-next-to-leading order in an expansion
in $1/(f^2L^2)$.  This is to be contrasted with the chiral expansion in
the $p$-regime, where they appear already at next-to-leading order.
This might be helpful for a precise determination of the leading-order
LECs ($f$ and $B_0$), because only these two parameters appear in fits
using next-to-leading order expressions in the $\e$-regime.  This is another
example in which we can extract physical quantities ($f$ and $B_0$)
from unphysical computations --- clearly, a small, euclidean, four-dimensional,
finite volume with
periodic boundary conditions is unphysical, but accessible to lattice QCD.

This concludes my brief overview of the application of ChPT to the study
of volume dependence.  For applications to the distribution of topological
charge, see Ref.~\cite{LStop}.
I have already mentioned the extension of the chiral expansion to the
partially quenched case with some quark
masses are in the $\e$-regime, with others in the $p$-regime (with the
obvious application to QCD with $2+1$ flavors with only the strange
quark mass in the $p$-regime), see Ref.~\cite{BDFH}.  For
a recent extension including lattice-spacing artifacts for Wilson fermions,
see Ref.~\cite{BNS}; for a recent higher-order study in a partially-quenched setting,
see Ref.~\cite{TW}.  These references contain fairly complete pointers to
earlier work in the $\e$-regime.

%%####%%
\section{\label{conclusion} Concluding remarks}
%%####%%

Clearly, ChPT is a very useful, and, in practice, indispensable tool for extracting
hadronic physical quantities from lattice QCD, which, in almost all cases can
only obtain correlation functions in some unphysical regime.   We have seen that
this includes nonzero lattice spacing and finite volume, necessarily, but also
``much more'' unphysical situations, such as independent choices for valence
and sea quarks on the lattice.   The chiral lagrangian, as well as other EFTs,
such as baryon ChPT, heavy-light ChPT and two-nucleon EFT, are local theories,
defined in terms of a number of LECs.  Once lattice QCD determines
the values of these LECs, even if this is by matching unphysical correlation
functions, they can be used to evaluate any hadronic quantity
accessible to the various EFTs.  While we have restricted ourselves in these
lectures to the physics of Nambu--Goldstone bosons, the conceptual points all carry
over to EFTs that include the interactions of other hadrons with Nambu--Goldstone bosons.

There are many applications of ChPT, relevant for lattice QCD, that were
left out of these lectures.  As already mentioned during these lectures,
one can extend ChPT to include the interactions of pions with heavier
hadrons.  While we briefly discussed this in the context of ``heavy'' kaons
in Sec.~\ref{kaon}, the methods described there can be extended to baryons
\cite{JM,BL} and hadrons containing heavy quarks \cite{MW,SZ}.  For a
recent review of hadron interactions, including $\p$-$\p$ and baryon-baryon
scattering, see Ref.~\cite{BOS}.

%%%%%%%%%%%%
%\newpage
\vspace{5ex}
\noindent {\large\bf  Some exercises}
\vspace{3ex}

Most explicit equations in these lecture notes can be derived with relative
ease.  None of them require more than a one-loop calculation, and in all
cases except Eq.~(\ref{integral}) the corresponding diagrams are tadpole
diagrams.  Here are a few more exercises:
\begin{itemize}
\item[1.]
Derive Eq.~(\ref{piscat}), which is, in fact, valid for an arbitrary number of flavors.
Simplify for $N_f=2$.
\item[2.]
Show that $\tr(L_\m L_\n L_\m L_\n)$ can be written as a linear combination
of the terms in Eq.~(\ref{l4}), using the Cayley--Hamilton theorem, which, for a
three-by-three matrix $A$ says that
\begin{equation}
\label{CH}
A^3-\tr(A)\;A^2+\frac{1}{2}\left(\tr(A)^2-\tr(A^2)\right)A-\det(A)\;\bf{1}=0\ .\nonumber
\end{equation}
[Hint: one way to proceed is to consider $\tr\Bigl((L_\m+L_\n)^4\Bigr)$.]  Simplify
Eq.~(\ref{l4}) for $N_f=2$.
\item[3.]
Show that the scattering lengths for pion scattering in the $O(N)$ linear
sigma model are discontinuous across the phase transition between the
symmetric phase and the phase in which $O(N)$ breaks down to $O(N-1)$
(\seef\ discussion after Eq.~(\ref{LCEpionmass})).
\item[4.]
Consider quenched ChPT (\ie, no sea quarks at all), with one valence quark
and ignore the anomaly. Show that using the naive (but wrong!) symmetry
group $SU(1|1)_L\otimes SU(1|1)_R$ for constructing the chiral lagrangian
leads to nonsense for the vacuum.  See Ref.~\cite{GSS} for more discussion.
\item[5.]
Verify cyclicity of the supertrace on $2\times 2$ graded matrices.
\item[6.]
Derive Eq.~(\ref{mixedmasses}).  Show that in the LCE regime all other terms
in the chiral lagrangian contribute at next-to-leading order.
\item[7.]
Investigate possible higher order terms, to order $p^4$, in the LCE regime
for Wilson ChPT (see Ref.~\cite{ABB}).
\item[8.]
Show that a solution of the equation $u_R=hu_L^\dagger h$ exists, for
given $u_R$ and $u_L$.  Here $u_{R,L}$ and $h$ are unitary matrices
with determinant one.
\item[9.]
Show that Eq.~(\ref{Kplag}) is equivalent to Eq.~(\ref{kaonfree}) plus a term
that can be absorbed into the $A_1$ term of Eq.~(\ref{kaon2}).
\item[10.]
Calculate the integral in Eq.~(\ref{integral}) and convince yourself of the
correctness of the discussion following Eq.~(\ref{denom}).
\item[11.]
Extend the discussion of Sec.~\ref{phase} to $N_f=3$ degenerate
light quarks (in the LCE regime).  I do not know the answer --- as far
as I know, you can publish the result!
\end{itemize}

%%%%%%%%%%%%
%\newpage
\vspace{5ex}
\noindent {\large\bf Acknowledgements}
\vspace{3ex}

First of all I would like to thank the organizers for inviting me to present
these lectures at this School.  I thank the students for the many questions they
asked, and all participants, students, lecturers and organizers alike, for  a lively and stimulating experience.

I would like to thank Oliver B\"ar, Claude Bernard, Santi Peris, Yigal Shamir, Steve Sharpe, and
Andr\'e Walker-Loud for  helping me understand many of the topics and concepts
covered in these lectures, as well as for useful comments on the manuscript.  I also would like to thank Claude
Bernard for providing me with a copy of his unpublished notes of the lectures on ChPT
he presented at the 2007 INT Summer School on ``Lattice QCD and its applications.''
I thank IFAE at the Universitat Aut\`onoma de Barcelona and the KITPC in Beijing,
where some of these lectures were written, for hospitality.
This work is supported in part by the US Department of Energy.

%%%%%%%%%%%%%%%%%%%%%%%%%%%%%%%%%%%%%%%%%%%%%%%%%%%%%%%%%%%%%%%%%%%%%%%%

\end{document}